\ifdefined\ForIEEE
\fi

\ifdefined\SingleColumnDraft
    \documentclass[12pt,draftclsnofoot,onecolumn]{IEEEtran}
\else
    \ifdefined\ForIEEE
        \ifdefined\SubmitToComSoc
            \documentclass[journal,comsoc,twoside]{IEEEtran}
        \else
            \documentclass[journal]{IEEEtran}
        \fi
    \else
        \documentclass[journal]{IEEEtran}
    \fi
\fi

\usepackage[T1]{fontenc}

\usepackage{cite}
\usepackage{amsmath,amsthm,mathtools}

\ifdefined\SubmitToComSoc
    \usepackage[cmintegrals]{newtxmath} 
\else
    \usepackage{amsfonts,amssymb,amsbsy} 
    \usepackage{esint} 
\fi

\usepackage{bm} 

\usepackage{array} 

\usepackage[caption=false,font=footnotesize]{subfig} 

\usepackage{url} 
\usepackage[bookmarks=false,hidelinks]{hyperref}

\usepackage{flushend} 

\usepackage{dsfont}
\usepackage{textcomp}

\usepackage{tikz}
\tikzset{every picture/.style={font issue=\footnotesize},
	font issue/.style={execute at begin picture={#1\selectfont}}
}
\usepackage{pgfplots}
\pgfplotsset{compat=newest}
\usetikzlibrary{backgrounds}
\usetikzlibrary{plotmarks}
\usetikzlibrary{positioning}
\usetikzlibrary{arrows,arrows.meta}
\usetikzlibrary{calc} 

\usepackage[capitalize]{cleveref}
	\Crefname{figure}{Fig.}{Fig.}
	\Crefname{section}{Sec.}{Sec.}
	\Crefname{subsection}{Sec.}{Sec.}
	\Crefname{prop}{Proposition}{Proposition}
	\Crefname{lemma}{Lemma}{Lemma}
	\Crefname{equation}{}{}
	\Crefname{footnote}{Footnote}{Footnote}

\input{macros}
\renewcommand{\d}{{\bf d}}
\newcommand{\cd}{c\hspace{0.3mm}}
\newcommand{\NObs}{N}
\newcommand{\KMin}{K_\text{min}}

\newcommand{\AssSync}{_{\text{sync}}}
\newcommand{\AssAsyn}{} 
\newcommand{\NoAssSync}{_{\text{sync,N/A}}}
\newcommand{\NoAssAsyn}{_{\text{N/A}}}

\newcommand{\E}{\mathbf{E}} 
\newcommand{\MLE}{{}^\text{\scriptsize{\,MLE}}}
\newcommand{\LSE}{{}^\text{\scriptsize{\,LSE}}}
\newcommand{\MVUE}{{}^\text{\scriptsize{\,MVUE}}}

\newcommand{\Hypo}[1]{\tilde{#1}} 

\newcommand{\AssViaTau}{_{\hspace{.15mm}\text{by}\hspace{.3mm}\tau}}
\newcommand{\AssViaTauSync}{_{\hspace{.15mm}\text{by}\hspace{.3mm}\tau,\text{sync}}}
\newcommand{\AssViaDiff}{_{\hspace{.15mm}\text{by}\hspace{.3mm}\Delta}}
\newcommand{\AssViaDiffPWA}{_{\hspace{.15mm}\text{by}\hspace{.3mm}\Delta,\text{PWA}}}

\newcommand{\AssViaDiffPWASync}{_{\hspace{.15mm}\text{by}\hspace{.3mm}\Delta,\text{PWA,sync}}}

\newcommand{\tauRMS}{\sigma_\tau } 
\newcommand{\errClock}{\epsilon}
\newcommand{\errClockA}[1][]{\errClock\AnnotateNodeA_{#1}}
\newcommand{\errClockB}[1][]{\errClock\AnnotateNodeB_{#1}}
\newcommand{\errClockAEstimate}[1][]{{\hat\errClock}\AnnotateNodeA_{#1}}

\newcommand{\errMeasSymbol}{n}
\newcommand{\errMeas}[1][]{\vect[#1]{\errMeasSymbol}}
\newcommand{\errMeasA}[1][]{\errMeas[#1]\AnnotateNodeA}
\newcommand{\errMeasB}[1][]{\errMeas[#1]\AnnotateNodeB}

\newcommand{\errClockO}[1][]{\vect[#1]{\errClock}} 

\newcommand{\AnnotateNode}[1]{^{\textnormal{\resizebox{3.3mm}{!}{\hspace{.2mm}(\hspace{.1mm}#1\hspace{.1mm})}}}}
\newcommand\AnnotateNodeA{\AnnotateNode{A}}
\newcommand\AnnotateNodeB{\AnnotateNode{B}}
\newcommand{\AnnotateNodeX}{^{\textnormal{\resizebox{2.7mm}{!}{\hspace{.2mm}\small(\hspace{.1mm}\textbullet\hspace{.1mm}\small)}}}}

\newcommand{\pos}{\vect{p}}
\newcommand{\posA}{\pos\AnnotateNodeA }
\newcommand{\posB}{\pos\AnnotateNodeB }

\newcommand{\delayMeasA}[1][]{\tau_{#1}\AnnotateNodeA }
\newcommand{\delayMeasB}[1][]{\tau_{#1}\AnnotateNodeB }

\newcommand{\delayTrueA}[1][]{\bar\tau_{#1}\AnnotateNodeA }
\newcommand{\delayTrueB}[1][]{\bar\tau_{#1}\AnnotateNodeB }
\newcommand{\delayTrueX}[1][]{\bar\tau_{#1}\AnnotateNodeX }

\newcommand{\delayDiff}[1][]{\vect[#1]{\Delta} }
\newcommand{\delayDiffTrue}[1][]{\vect[#1]{\bar\Delta} }

\newcommand{\cirA}{h\AnnotateNodeA}
\newcommand{\cirB}{h\AnnotateNodeB}
 
\newcommand{\mpc}{\psi}
\newcommand{\mpcA}{\mpc\AnnotateNodeA}
\newcommand{\mpcB}{\mpc\AnnotateNodeB}

\newcommand{\mpcSet}{\Psi}
\newcommand{\mpcSetA}{\mpcSet\AnnotateNodeA}
\newcommand{\mpcSetB}{\mpcSet\AnnotateNodeB}

\newcommand{\mpcSetAll}{\Omega}
\newcommand{\mpcSetAllA}{\mpcSetAll\AnnotateNodeA}
\newcommand{\mpcSetAllB}{\mpcSetAll\AnnotateNodeB}

\newcommand{\dirVectSymb}{\vect{e}}
\newcommand{\dirVect}[1][]{\dirVectSymb_{#1} } 
\newcommand{\dirVectA}[1][]{\vect{e}\AnnotateNodeA_{#1}}
\newcommand{\dirVectB}[1][]{\vect{e}\AnnotateNodeB_{#1} }
\newcommand{\dirVectX}[1][]{\vect{e}\AnnotateNodeX_{#1} }
\newcommand{\dirVectObs}[1][]{\vect{f}_{#1}} 
\newcommand{\dirVectObsA}[1][]{\vect{f}\AnnotateNodeA_{#1} }
\newcommand{\dirVectObsB}[1][]{\vect{f}\AnnotateNodeB_{#1} }

\newcommand{\permSmybol}{\pi}
\newcommand{\perm}{\permSmybol}
\newcommand{\permSetSymbol}{\Pi}
\newcommand{\permSet}[1][]{\ifthenelse{\isempty{#1}}{\permSetSymbol}{\permSetSymbol_{#1}}}
\newcommand{\permCount}{C}

\ifdefined\SubmitToComSoc
	\makeatletter
		\@ifpackageloaded{amssymb}{\PackageWarning{amssymb}{Discourraged package is loaded}}{}
		\@ifpackageloaded{amsfonts}{\PackageWarning{amsfonts}{Discourraged package is loaded}}{}
	\makeatother
\fi

\newcommand\OurPaperTitle{%
Pairwise Node Localization From Differences in Their UWB Channels to Observer Nodes}

\title{\OurPaperTitle}

\author{%
Gregor~Dumphart\ifdefined\ForIEEE\,\orcidlink{0000-0003-0253-5482}\fi,~\ifdefined\ForIEEE\IEEEmembership{Member,~IEEE,}
\fi
Robin~Kramer\ifdefined\ForIEEE\,\orcidlink{0000-0002-3875-8453}\fi,
Robert~Heyn\ifdefined\ForIEEE\,\orcidlink{0000-0002-1036-6490}\fi,~\ifdefined\ForIEEE\IEEEmembership{Student Member,~IEEE,}
\fi
Marc~Kuhn\ifdefined\ForIEEE\,\orcidlink{0000-0001-5885-4031}\fi,~\ifdefined\ForIEEE\IEEEmembership{Member,~IEEE,}
\fi		
and~Armin~Wittneben\ifdefined\ForIEEE\,\orcidlink{0000-0001-7768-1703},~\IEEEmembership{Member,~IEEE}
\fi%
\thanks{%
Accepted to appear in the IEEE Transactions on Signal Processing.
Manuscript received  August 20, 2021; revised January 6, 2022 and February 4, 2022; accepted February 6, 2022.
Date of publication February 14, 2022.
\ifdefined\ForIEEE%
The associate editor coordinating the review of this manuscript and approving it for publication was L. Venturino.
\fi
This work was partially supported by Innosuisse, Switzerland under project number 18723.2 PFES-ES.
This article was presented in part at the IEEE Conference on Communications (ICC),
Shanghai, China, May 2019, 
and the IEEE Global Communications Conference (GLOBECOM)
Workshops, Madrid, Spain, December 2021. 
(\textit{Corresponding author: Gregor Dumphart})}%
\thanks{G. Dumphart, R. Kramer, R. Heyn and A. Wittneben are with the Wireless Communications Group, D-ITET, ETH Zurich, Z\"urich, 8092 Switzerland, e-mail: dumphart@nari.ee.ethz.ch, robin.kramer@bluewin.ch, \{heyn,\,wittneben\}@nari.ee.ethz.ch.}%
\thanks{M. Kuhn is with the Zurich University of Applied Sciences (ZHAW), Winterthur, 8400 Switzerland, e-mail: kumn@zhaw.ch.}%
\thanks{Supplementary material (ray-tracing-based evaluations) and color versions of figures in this article are available online at \url{http://ieeexplore.ieee.org}.}%
\thanks{Digital Object Identifier 10.1109/TSP.2022.3150951}
}%

\begin{document}

\maketitle	

\ifdefined\ForIEEE
\fi

\begin{abstract}
We consider the problem of localization and distance estimation between a pair of wireless nodes in a multipath propagation environment, but not the usual way of processing a channel measurement between them. We propose a novel paradigm which compares the two nodes' ultra-wideband (UWB) channels to \textit{other} nodes, called \textit{observers}. The main idea is that the dissimilarity between the channel impulse responses (CIRs) increases with $d$ and allows for an estimate $\hat{d}$. Our approach relies on extracting common multipath components (MPCs) from the CIRs. This is realistic in indoor or urban scenarios and if $d$ is considerably smaller than the observer distances.

We present distance estimators which utilize the rich location information contained in MPC \textit{delay differences}. Likewise, we present estimators for the relative position vector which process both MPC delays and MPC directions.
We do so for various important cases: with and without time synchronization, delay measurement errors, and knowledge of the MPC association between the CIRs. The estimators exhibit great technological advantages: they do not require line-of-sight conditions, observer location knowledge, or environment knowledge.

We study the estimation accuracy with a numerical evaluation based on random sampling and, additionally, with an experimental evaluation based on measurements in an indoor environment. The proposal shows the potential for great accuracy in theory and practice. We describe how the paradigm could incorporate novel measurements into cooperative localization frameworks for spatio-temporal tracking. This could enable affordable wireless network localization in dynamic multipath settings.
\end{abstract}


\IEEEpubid{\begin{minipage}{\textwidth}\centering \ \\[12pt]
1053-587X~\copyright~2022 IEEE. Personal use is permitted, but republication/redistribution requires IEEE permission.\\
See \url{https://www.ieee.org/publications/rights/index.html} for more information.
\end{minipage}}
\IEEEpubidadjcol

\section{Introduction}
\label{sec:Intro}
\IEEEPARstart{W}{ireless}~
localization is the task of estimating a node's position from wireless channel measurements. 
It is a key requirement for many current or desired mobile applications in \VRev{cellular networks,} the Internet of Things, wireless sensor networks, and robotics. Thereby, great accuracy and reliability are desired. Important use cases concern indoor environments (retail stores \cite{Contigiani2016}, access gates \cite{HeynVTC2019}, assisted living \cite{WitrisalSPM2016}, public transport, warehouses), crowded urban settings and large events \cite{Gani2016}, autonomous vehicles \cite{AliCM2020}, and disaster sites.
The related problem of wireless ranging (i.e. distance estimation) has received a lot of interest in the context of access control \cite{HeynVTC2019}, keyless entry systems \cite{RanganathanCapkun2017}, and social distance monitoring \cite{AltuwaiyanICC2018} especially during the COVID-19 pandemic \cite{Leith2020}.


These applications concern dense and dynamic propagation environments, characterized by time-variant channels with frequent line-of-sight (LOS) obstruction \cite{VenusRADAR2021} and rich multipath propagation. This poses a great challenge to 
wireless localization and ranging.
Distance estimates obtained from the received signal strength (RSS) tend to have large relative error because shadowing, antenna patterns, and small-scale fading cause large RSS fluctuations \cite{SchultenVTC2019,GeziciSPM2005}.
Time of arrival (TOA) distance estimates can be obtained with wideband or ultra-wideband (UWB) systems. They often suffer from synchronization errors, processing delays, multipath interference, and non-line-of-sight (NLOS) bias \cite{DardariPIEEE2009,AlaviCL2006,JourdanTAES2008,YuPLANS2020,WangSYS2013}, which causes large relative error at short distances.
Naturally, trilateration of such inaccurate distance estimates will result in inaccurate position estimates. It is often infeasible to solve the problem by just adding infrastructure in order to ensure sufficient LOS anchors for most mobile positions \cite{WitrisalSPM2016}.
Location fingerprinting is also not an all-round alternative for accurate localization because the training data and the associated acquisition effort become obsolete quickly in time-varying settings \cite{HePC2016,SteinerTSP2010}.

\IEEEpubidadjcol

One approach to robust localization in NLOS multipath environments is the mitigation of NLOS-induced range biases \cite{QiTWC2006}. It has been tackled with machine learning \cite{WymeerschTC2012,Li2021DeepGEM} and adaptive probabilistic modeling \cite{PrietoTSP2012}.
%
%
%
%
%
Other promising approaches are soft information processing \cite{MazuelasTSP2018,ContiPIEEE2019} and temporal filtering \cite{BuehrerPIEEE2018,LeitingerTWC2019,HeynVTC2019} with the incorporation of inertial measurements \cite{PrietoTSP2016,HeynVTC2019}.
Furthermore, various recent work considers multipath as opportunity rather than interference
\cite{MeissnerWPNC2010,WitrisalSPM2016,LeitingerJSAC2015,LeitingerTWC2019,WinPIEEE2018,UlmschneiderACCESS2020,YuPLANS2020,VenusRADAR2021}.
Multipath-assisted localization yields improved accuracy and robustness if knowledge about the propagation environment is either available a-priori \cite{LeitingerJSAC2015} or obtained with mapping \cite{LeitingerTWC2019,UlmschneiderACCESS2020}. Thereby, UWB operation is crucial for resolving the delays of individual multipath components (MPCs) of the propagation channel \cite{MolischPIEEE2009,WitrisalSPM2016,KulmerWCL2019}.
Further rich location information is held by the MPC directions (e.g., the direction of arrival) which can be resolved with wideband antenna arrays (e.g., by millimeter-wave massive-MIMO systems) \cite{WenTWC2020,BuehrerPIEEE2018,HanTIT2015}.
Another promising approach is cooperative (a.k.a. collaborative) network localization \cite{WinPIEEE2018,BuehrerPIEEE2018,PatwariSPM2005,WinCM2011,ShenJSAC2012,LiJSAC2015,LiuTSP2018,MeyerIOT2018,MeyerPIEEE2018} which uses distance estimates between mobiles in the computation of an improved joint position estimate.
These developments demonstrate the importance of relative location information between pairs of nodes and of embracing both NLOS and multipath propagation in the signal processing.

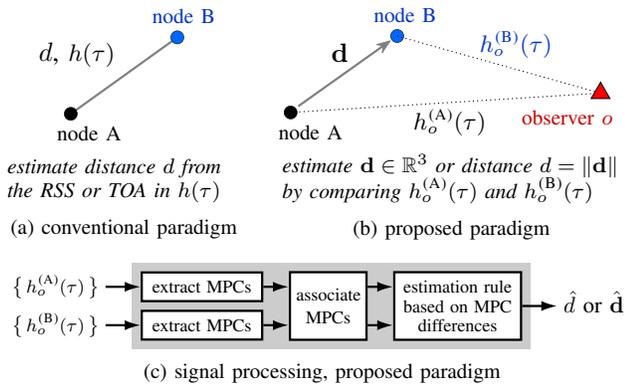
\begin{figure}[!ht]
\centering
\vspace{-4mm}
\subfloat[conventional paradigm]{
\begin{tabular}{c}
\includegraphics[height=15mm,trim=0 -7mm 0 0]{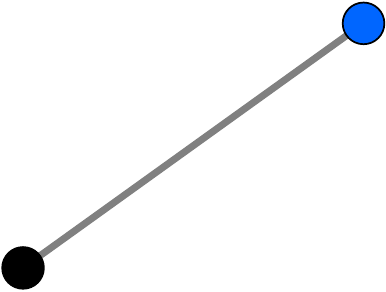}
\put(-55,31){$d$, \small{$h(\tau)$}}
\put(-48,0.75){\footnotesize{node A}}
\put(-12,45.25){\footnotesize\textcolor[rgb]{0,.2,.8}{node B}}
\\[-2.5mm]
\parbox[l]{31mm}{\textit{\footnotesize{\begin{flushleft}
estimate distance $d$ from the RSS or TOA in $h(\tau)$
\end{flushleft}}}}
\\[-1.5mm]
\end{tabular}
\label{fig:Concept_Conventional}}\!\!\!\!
\subfloat[proposed paradigm]{
\centering
\begin{tabular}{c}
\includegraphics[height=15mm,trim=0 -7mm -3mm 0]{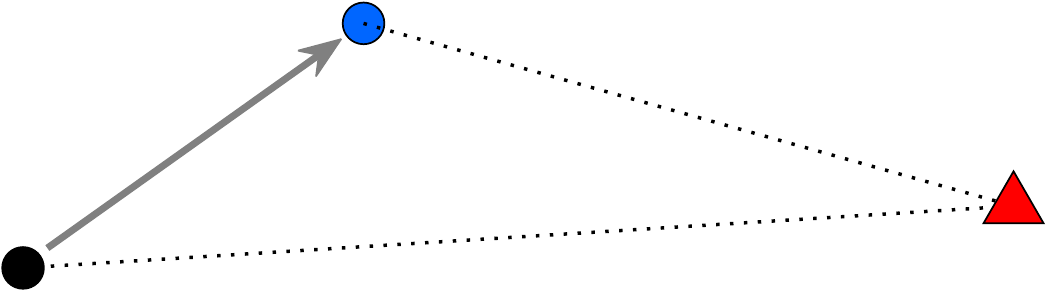}
\put(-109,30){$\d$}
\put(-78,4.5){\small{$\cirA_o(\tau)$}}
\put(-129,0.75){\footnotesize{node A}}
\put(-53,34){\textcolor[rgb]{0,.2,.8}{\small{$\cirB_o(\tau)$}}}
\put(-93,45.25){\footnotesize\textcolor[rgb]{0,.2,.8}{node B}}
\put(-37,6.5){\footnotesize{\textcolor[rgb]{.8,0,0}{observer $o$}}}
\\[-2.5mm]
\parbox[l]{45mm}{\textit{\footnotesize{\begin{flushleft}
estimate $\d \in \bbR^3$ or distance $d = \|\d\|$ by comparing $\cirA_o(\tau)$ and $\cirB_o(\tau)$
\end{flushleft}}}}
\\[-1.5mm]
\end{tabular}
\label{fig:Concept_Proposed}}
\ifdefined\SingleColumnDraft\else
\\
\fi
\subfloat[signal processing, proposed paradigm]{
\centering
\begin{tabular}{c}
\ \\[-5mm]
\resizebox{84mm}{!}{\begingroup
\begin{tikzpicture}[%
box/.style={draw,text width=9mm,minimum height=7mm,align=center,line width=1pt}]
\node[box,minimum height=10.6mm,minimum width=12mm,text width=10mm,fill=white] (AssocBox) {\baselineskip=10pt associate MPCs \par};
\node[box,right=4mm of AssocBox,minimum height=10.6mm,minimum width=19mm,text width=18mm,fill=white] (CompBox) {\baselineskip=8.5pt  estimation rule based on MPC differences \par}; 
\node[right=-1mm of AssocBox.154] (MpcOutA) {};
\node[right=-1mm of AssocBox.206] (MpcOutB) {};
\node[box,left=3mm of MpcOutA,minimum height=4.5mm,minimum width=17mm,text width=17mm,fill=white](MpcA) {extract MPCs \par};
\node[box,left=3mm of MpcOutB,minimum height=4.5mm,minimum width=17mm,text width=17mm,fill=white](MpcB) {extract MPCs \par};
\node[right=3mm of AssocBox.26] (AssocOutA) {};
\node[right=3mm of AssocBox.334] (AssocOutB) {};
\draw[-{Latex[length=2.5mm,width=1.6mm]},line width=1pt] (AssocBox.26) -- (AssocOutA.center);
\draw[-{Latex[length=2.5mm,width=1.6mm]},line width=1pt] (AssocBox.334) -- (AssocOutB.center);
\node[left=5.5mm of MpcA] (CirsA) {$\big\{\, \cirA_o(\tau) \,\big\}$};
\node[left=5.5mm of MpcB] (CirsB) {$\big\{\, \cirB_o(\tau) \,\big\}$};
\node[right=5.5mm of CompBox] (MyOutput) {\raisebox{1.5mm}{\normalsize{$\hat d$ or $\hat\d$}}};
\begin{scope}[on background layer]
\fill[gray,thick,dotted,fill={rgb:black,1;white,4}] ($(MpcA.north west)+(-0.13,0.13)$)  rectangle ($(CompBox.south east)+(0.13,-0.13)$);
\end{scope}
\draw[-{Latex[length=2.5mm,width=1.6mm]},line width=1pt] (CirsA) -- (MpcA.180);
\draw[-{Latex[length=2.5mm,width=1.6mm]},line width=1pt] (CirsB) -- (MpcB.180);
\draw[-{Latex[length=2.5mm,width=1.6mm]},line width=1pt] (MpcA.0) -- (MpcOutA.center);
\draw[-{Latex[length=2.5mm,width=1.6mm]},line width=1pt] (MpcB.0) -- (MpcOutB.center);
\draw[-{Latex[length=2.5mm,width=1.6mm]},line width=1pt] (CompBox) -- (MyOutput);
\end{tikzpicture}
\endgroup}
\\[-1mm]
\end{tabular}
\label{fig:SigProcBlackBox}}
\caption{
Paradigms for relative localization of two wireless nodes.}
\label{fig:IntroConcepts}
\end{figure}

This paper concerns specifically the acquisition of such pairwise relative location information in multipath environments with possible LOS obstruction.
In particular, we propose and study an \textit{alternative paradigm} for obtaining estimates of the distance $d = \|\d\|$ or the relative position vector $\d$ between two nodes.
We abandon the conventional notion (\Cref{fig:Concept_Conventional}) that an estimate of the distance $d$ between two nodes A and B should be based on a direct measurement between them, e.g. on the TOA or RSS.
Instead, we consider the presence of one or more \textit{other} nodes, 
henceforth referred to as \textit{observers} $o \in \{ 1, 2, \ldots, \NObs \}$ (\Cref{fig:Concept_Proposed}). We consider the channel impulse responses (CIRs) $\cirA_o(\tau)$ between node A and the observers as well as the CIRs $\cirB_o(\tau)$ between node B and the observers.
The paradigm is intended for \VRev{environments with distinct MPCs and} node distances $d$ considerably smaller than the link distances to the observers.
The CIRs can be obtained via channel estimation at the observers after transmitting training sequences at A and B, or vice versa \cite{MolischPIEEE2009}.
If measured with sufficient bandwidth, the CIRs are descriptive signatures of the multipath environment \cite{WitrisalSPM2016}.
Our starting point is the observation that the CIRs $\cirA_o(\tau)$ and $\cirB_o(\tau)$ are similar for small $d$ but \textit{differ increasingly and systematically} with increasing $d$.
A good metric for CIR dissimilarity could give rise to an accurate estimate of $d$ or even of the relative position vector $\d$, with the prospect of particularly good accuracy at short distances
and no requirements for LOS connections.

Our main technological motive is to provide localization systems with a new method for acquiring pairwise estimates, with the aim of overcoming the outlined problems in dynamic NLOS multipath settings. We also want to enable novel opportunities for systems that lack precise time-synchronization and the training data required for machine learning. 
The proposed paradigm is intended to complement and enhance existing localization approaches, not to outright beat and replace them.

A naturally arising question is how to implement the proposed paradigm with a specific estimation rule.
Since our setup in \Cref{fig:Concept_Proposed} is similar to the location-fingerprinting setup \cite{SteinerTSP2010}, one option is a regression model (e.g., a neural network). This would require a large set of CIR training data and associated ground-truth values $d$ or $\d$ for supervised machine learning. This onerous approach would withhold analytical insights and struggle with time-varying channels.

\subsubsection*{Contribution}
Using estimation theory and geometric principles, we derive estimators that process the geometric information in MPCs according to the proposed paradigm (delineated in \Cref{fig:SigProcBlackBox}). Specifically, we state closed-form distance estimators $\hat d$ from the difference of MPC delays and, furthermore, relative position estimators $\hat\d$ which additionally process the MPC directions at the nodes A and/or B.
To the best of our knowledge, neither the proposed paradigm nor any of the presented associated estimators are covered by existing work. Our specific contributions are:
\begin{itemize}
\item Regarding \textit{distance estimation}, we derive the \textit{maximum-likelihood} estimate (MLE) under random delay measurement errors and random MPC directions (assuming a uniform distribution in 3D). We also derive the MLE for the case of unknown MPC association, the minimum-variance unbiased estimate (MVUE) for the zero-measurement-error case, and its scaling behavior.
\item Regarding \textit{relative position estimation} when the MPC directions are observable, we derive the MLE and the least-squares estimate (LSE) for various cases. We append a tailored scheme for MPC association.
\item We evaluate the estimation accuracy:
\begin{itemize}
\item By simulation, using random sampling of MPCs.
\item In practice, based on UWB channel measurements from a compliant environment (a large empty room).
\end{itemize}
\item We identify \textit{technological advantages} of the estimators: they do not require LOS conditions, precise time-synchronization, pairwise interaction, or knowledge about the environment or observer locations. 
\item We identify \textit{important use cases and applications}. 
\end{itemize}

\subsubsection*{Paper Structure}
In \Cref{sec:SystemModel} we state the employed system model, identified geometric properties, and assumptions on synchronization, MPC selection and association.
\Cref{sec:EstimateDist,sec:EstimateRelLoc} present the derived estimators for distance and relative position, respectively.
\Cref{sec:EvalSim} presents a numerical evaluation of the estimation accuracy and major influences. \Cref{sec:EvalMeas} presents a practical proof of concept. \Cref{sec:TechComparison} discusses the technological 
potential. 
\Cref{sec:Summary} concludes the paper.

\subsubsection*{Notation}
Scalars $x$ are written lowercase italic, vectors $\bf x$ lowercase boldface, and matrices $\bf X$ uppercase boldface. $\eye_K$ is the $K \times K$ identity matrix. All vectors are column vectors unless transposed explicitly. $\|{\bf x}\|$ is the Euclidean norm $\|{\bf x}\|_2$. For a random variable $x$, the probability density function (PDF) is denoted $f_x(x)$.
For simplicity we do not use distinct random variable notation.
$\EVSymb[x]$ is the expected value. 



\section{System Model and Key Principles}
\label{sec:SystemModel}
We consider an ultra-wideband (UWB) wireless system operating in a multipath propagation environment with distinct objects that reflect, scatter, or diffract the propagated waves. This gives rise to multipath components (MPCs). The MPCs can be resolved from UWB channel measurements owing to the high delay resolution.
The approximate minimum bandwidth is $500\unit{MHz}$ (indoor) \cite{MolischPIEEE2009}.
The objects may or may not obstruct the LOS path of a link.
Suitable application scenarios are indoor, urban, or industrial. For a thorough introduction to UWB multipath channels, we refer to \cite{MolischPIEEE2009,WitrisalSPM2016}.

We consider the nodes A and B whose positions are written in Cartesian coordinates $\posA, \, \posB \in \bbR^3$ in an arbitrary reference frame. Their arrangement is characterized by:
\begin{align}
&\text{relative position:} &&
\d = \posB - \posA \, , &&
 \d \in \bbR^3 \, ,
 \label{eq:DisplacementDef} \\
&\text{node distance:} &&
d  = \|\posB - \posA\| \, , &&
d  \in \bbR_+ \, .
\label{eq:DistanceDef}
\end{align}
In a temporal tracking use of the paradigm, the positions $\posA$ and $\posB$ would simply relate to the same node at different times A and B. They could even relate to different nodes at different times.

We consider the observers $o \in \{1, \ldots, \NObs\}$ with $\NObs \geq 1$. The multipath channels between the observers and node A are characterized by the CIRs
$\cirA_o(\tau)$
and those between the observers and node B by
$\cirB_o(\tau)$.

\begin{figure}[!ht]
\centering
\subfloat[multipath propagation from observer $o$ to A and B]{\centering
	\includegraphics[height=38mm]{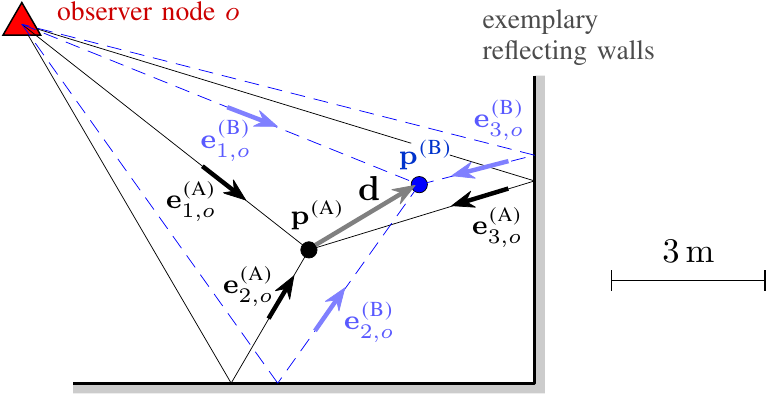}
	\label{fig:InterUserPaths}}
	\ \\[.85mm]
\subfloat[CIRs between $o$ and A, between $o$ and B]{\centering
	\includegraphics[width=84mm]{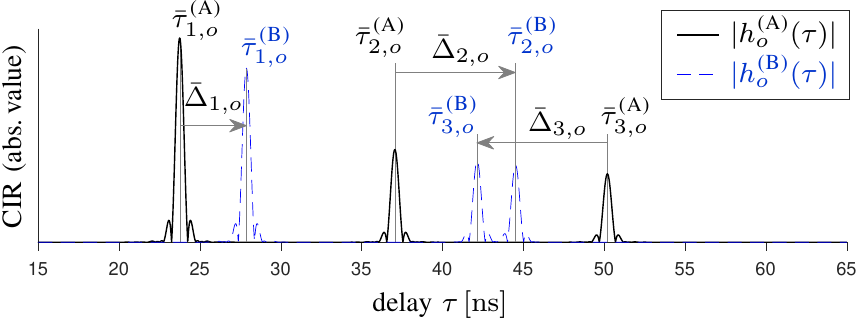}
	\label{fig:InterUserSignals}}
\caption{Considered wireless setup in an exemplary indoor environment with two walls, one observer ($\NObs = 1$), and $K_o = 3$ propagation paths. Here, $k=1$ is the LOS path and $k \in \{2,3\}$ are reflections. The CIR illustration assumes a raised-cosine pulse of $2\,\mathrm{GHz}$ bandwidth and no noise or interference.}
\label{fig:InterUserIntro}
\end{figure}

The specific requirements of the proposed paradigm are:
\begin{enumerate}
\item \label{li:reqMulti}
Wireless operation in a multipath propagation environment where distinct MPCs arise.
\item \label{li:reqUWB}
Sufficient system bandwidth for the resolution of MPCs.
\item \label{li:reqSelect}
An MPC selection step that selects a total of $K_o$ MPCs from both $\cirA_o(\tau)$ and $\cirB_o(\tau)$. The scheme must exclude alien MPCs, i.e. MPCs with no counterpart in the other CIR in terms of matching propagation path.
\item \label{li:reqK}
The number of selected MPCs $K = \sum_{o=1}^\NObs K_o$ fulfills an estimator-specific criterion $K \geq \KMin$.
\end{enumerate}
Some estimators will furthermore require:
\begin{enumerate}
\item[6)] \label{li:reqMatch}
An MPC association step that pairs those MPCs in $\cirA_o(\tau)$ and $\cirB_o(\tau)$ with matching propagation paths (e.g., both are reflections off the same wall).
\end{enumerate}
\Cref{apdx:MpcSetTheory} provides formal definitions of the terms MPC selection, MPC association, matching paths, alien MPC, correct association, and association error.
The absence of reliable MPC association is referred to as unknown association.
The specific schemes for MPC selection and MPC association are left unspecified in this system model. These delicate steps will be discussed at the end of the section. The MPCs may or may not comprise the LOS path.
%
%
The employed indices are:
\begin{align} 
& \text{observer index:} &
o &\in \{ 1, 2, \ldots, \NObs \}
\, , \\
& \text{MPC index:} &
k &\in \{ 1, 2, \ldots, K_o \}
\, , \\
& \text{total MPC count:} &
K \! &= K_1 + \ldots + K_\NObs
\, .
\end{align}
We consider the following MPC parameters:
\begin{align}
&\text{MPC delay, true value:} &&
\delayTrueA[k,o], \delayTrueB[k,o] \in \mathbb{R}_+ \\
&\text{MPC delay, measured:} &&
\delayMeasA[k,o], \delayMeasB[k,o] \in \mathbb{R}_+ \\
&\text{delay difference, true value:} &&
\delayDiffTrue[k,o] = \delayTrueB[k,o] - \delayTrueA[k,o]
\label{eq:DelayShiftDef_True} \\
&\text{delay difference, measured:} &&
\delayDiff[k,o]     = \delayMeasB[k,o] - \delayMeasA[k,o] 
\label{eq:DelayShiftDef_Meas} \\
&\text{MPC direction (unit vectors):} \!\! &&
\dirVectA[k,o], \dirVectB[k,o] \in \bbR^3
\end{align}
Specifically, $\dirVectA[k,o], \dirVectB[k,o]$ are the directions of arrival at $\posA\! , \posB$ if an observer is transmitting (see \Cref{fig:InterUserPaths}). Vice versa, $-\dirVectA[k,o]$ and $-\dirVectB[k,o]$ would be the directions of departure at $\posA\! , \posB$ if A, B were transmitting. We do not define the transmitter and receiver roles due to channel reciprocity.


We recall that a delay $\tau$ is caused by having traveled a path length $\cd\tau$ at the wave propagation velocity $c \approx 3 \cdot 10^8\unit{m/s}$ (the speed of light). Therefrom, we establish crucial \textit{geometric properties} regarding MPC dissimilarity between the two sets of CIRs. Each MPC $k,o$ fulfills the delay-difference bounds
\begin{align}
-d \leq \cd\delayDiffTrue[k,o] \leq d
\label{eq:DelayDiffBounds}
\end{align}
and, furthermore, two equalities on the relative position vector:
\begin{align}
\d &= \cd\delayTrueB[k,o] \dirVectB[k,o] - \cd\delayTrueA[k,o] \dirVectA[k,o]
\label{eq:VectorEquality}
\, , \\[1mm]
( \hspace{.1mm} \dirVectA[k,o] + \dirVectB[k,o] \hspace{.1mm} )\Tr \d
&= \cd\delayDiffTrue[k,o] \big(1 + (\dirVectA[k,o])\Tr \dirVectB[k,o] \,\big)
\, . 
\label{eq:ProjectionEquality}
\end{align}
In \Cref{apdx:PropagationGeometry} we give a simple proof based on the triangle between $\posA$, $\posB$, and the virtual source of the MPC $k,o$.
If
$d \ll \cd\delayTrueA[k,o]$,
$d \ll \cd\delayTrueB[k,o]$, and the MPC is not caused by a scatterer near A or B, then $\dirVectA[k,o] \approx \dirVectB[k,o]$ holds. In consequence
\begin{align}
(\dirVectA[k,o])\Tr \d \ \approx \
(\dirVectB[k,o])\Tr \d \ \approx \ \cd\delayDiffTrue[k,o]
\label{eq:ProjectionApproximation}
\end{align}
holds in good approximation. This is essentially a plane-wave assumption (PWA) in the vicinity of node A and B. It constitutes a simpler version of \Cref{eq:ProjectionEquality}.

We believe that the mathematical structure in
\Cref{eq:VectorEquality,eq:ProjectionEquality,eq:ProjectionApproximation,eq:ProjectionApproximation} is fundamentally responsible for successes in wideband MIMO location fingerprinting, e.g. reported in \cite{VieiraPIMRC2017,LiSENS2021}. In this paper we will utilize these properties systematically to formulate estimators. A key strength is the formal absence of the observer positions and environment specifics in the expressions.
The bounds \Cref{eq:DelayDiffBounds} show that the value range of $\{ \cd\delayDiff[k,o] \}$ is expressive of $d$. This forms the basis for the distance estimators in \Cref{sec:EstimateDist}.
Likewise, \Cref{eq:VectorEquality,eq:ProjectionEquality,eq:ProjectionApproximation} will be utilized by the relative position estimators in \Cref{sec:EstimateRelLoc}. We note that equation \Cref{eq:VectorEquality} readily provides an estimation rule for vector $\d$ from the delay and direction of a single MPC, if accurate measurements thereof can actually be obtained.


For the measured MPC delays we consider the error model
\begin{align}
\delayMeasA[k,o] &= \delayTrueA[k,o] + \errMeasA[k,o] + \errClockA[o]
\, , \label{eq:SignalModelTauA} \\
\delayMeasB[k,o] &= \delayTrueB[k,o] + \errMeasB[k,o] + \errClockB[o]
\label{eq:SignalModelTauB}
\end{align}
where
$\errMeasA[k,o] , \errMeasB[k,o]$
are measurement errors due to noise, interference, clock jitter, limited bandwidth, and limited receiver resolution \cite{DardariPIEEE2009,MolischPIEEE2009}.
The clock offsets
$\errClockA[o]$ between A and $o$ and 
$\errClockB[o]$ between B and $o$ occur because we do not assume precise time-synchronization, neither between A and B nor between the observers. 
However, we assume that the setup is able to conduct the necessary channel estimation steps in quick succession, such that the clock drift is negligible over the duration of the entire process (i.e. while recording all the received signals).
A consequence is the property
\footnote{Even without this assumption regarding negligible clock drift, the estimator results of this paper apply with straightforward adaptations. These adapted estimators are stated in \Cref{apdx:EstimatorsAsyncObs}.}
\begin{align}
\errClockB[o] - \errClockA[o] = \errClock
\label{eq:QuickChannelEstimationAssumption}
\end{align}
for the clock offset $\errClock$ between A and B, which specifically does not depend on the observer index $o$.
This property yields a particularly simple \textit{signal model} for the measured delay differences
$\delayDiff[k,o]$ 
from \Cref{eq:DelayShiftDef_Meas},
\begin{align}
\delayDiff[k,o] 
&= \delayDiffTrue[k,o] + \errMeas[k,o] + \errClock \, .
\label{eq:SignalModel}
\end{align}
The measurement error
$\errMeas[k,o] = \errMeas[k,o]\AnnotateNode{B} - \errMeas[k,o]\AnnotateNode{A},
\errMeas[k,o] \in \bbR$
is considered as a random variable. We note that potential biases are compensated by the difference. It could thus be reasonable to assume $\EV{\errMeas[k,o]} = 0$. Likewise, a Gaussian-distributed $\errMeas[k,o]$ could be a reasonable assumption because of the many different influences in
$\errMeas[k,o]\AnnotateNode{B} - \errMeas[k,o]\AnnotateNode{A}$
(central limit theorem).
The clock offset $\errClock \in \bbR$ is considered as non-random variable. We distinguish between the asynchronous case where $\errClock$ is unknown and the synchronous case where $\errClock$ is known a-priori.


We have yet to discuss the intricacies of MPC selection and association.
Methods for resolving MPCs from channel measurements are given in \cite{FleuryJSAC1999,KulmerWCL2019,LeitingerASILOMAR2020}. 
The MPC selection and association processes are complicated by the fact that descriptive identifiers of the MPC propagation paths are unavailable in most every application. In this case one has to resort to measured MPC parameters like delays and possibly  also amplitudes and directions.
Suitable data association schemes are addressed in \cite{WitrisalSPM2016,LeitingerTWC2019} (temporal tracking) and \cite{DokmanicPNAS2013,MeyerICASSP2017} (single temporal snapshot).

A simple MPC association scheme is given by sorting the measured MPC delays of either CIR in ascending order. 
This has a high chance of finding the correct association if $d/c$ is much smaller than the channel delay spread. Because in this case, the two CIRs will likely exhibit the same delay order.
In general however, sorting is prone to association errors because $\delayMeasA[k,o] < \delayMeasA[l,o]$ does not guarantee $\delayMeasB[k,o] < \delayMeasB[l,o]$ (an example is seen in \Cref{fig:InterUserSignals}).
Furthermore, alien MPCs can occur due to selective fading or shadowing, even for very small $d$.
%
%
%

\ifdefined\AddedContent
We always assume the correct association of a measured CIR to the involved node and observer. This can easily be established with an appropriate protocol.
\fi

The number of selected MPCs $K_o$ is left as an unspecified design parameter. At the design stage it must be noted that MPC selection has two opposing goals:
(i) establish a large $K_o$ because the estimators will rely thereon, (ii) exclude alien MPCs as they are useless and detrimental to the estimators.
A smart selection scheme will adapt $K_o$ to the channel conditions and accuracy requirements.

Obstruction of the LOS path (i.e. NLOS) will usually decrease $K_o$ by $1$, so in this case, one may have to resolve and select an additional MPC in order to meet accuracy targets.
The risk of selecting alien MPCs increases with $K_o$ and also with $d$, as the MPCs in $\cirA_o(\tau)$ and $\cirB_o(\tau)$ become less likely to stem from the same propagation paths. This aspect will determine the maximum usable distance. However, the distance threshold can not easily be stated: it is a complicated function of the accuracy targets, technical parameters, node arrangement, environment geometry, and the selection scheme.

\section{Distance Estimators}
\label{sec:EstimateDist}
This section presents estimators of the inter-node distance $d$ from measured delay differences $\delayDiff[k,o] = \delayDiffTrue[k,o] + \errMeas[k,o] + \errClock$ as defined in \Cref{eq:SignalModel}. The estimators do not use or require the MPC directions.
In order to derive estimators with desirable estimation-theoretic properties, we have to establish an adequate statistical model for $\delayDiff[k,o]$. We achieve this with the following assumptions regarding rich multipath propagation:
\newcommand{\Ass}[1]{\uppercase\expandafter{\romannumeral#1}}
\begin{enumerate}[\IEEEsetlabelwidth{\Ass{3}}]
\item[\Ass{1}:] Each MPC direction $\dirVectA[k,o] \in \bbR^3$ is random and has uniform distribution on the 3D unit sphere.
\item[\Ass{2}:] The random directions $\dirVectA[k,o]$ are statistically independent for different MPCs $k,o$.
\item[\Ass{3}:] The plane-wave approximation \Cref{eq:ProjectionApproximation} is assumed to be exact, i.e. $\dirVectA[k,o] = \dirVectB[k,o]$ and $
(\dirVectA[k,o])\Tr \d =
(\dirVectB[k,o])\Tr \d = 
\cd\delayDiffTrue[k,o]$.
\end{enumerate}
These assumptions actually result in i.i.d. uniform distributions
$\cd\delayDiffTrue[k,o] \sim \ \mathcal{U}(-d,d)$.
Details can be found in
\cite[Lemma 4.1]{Dumphart2020}.

%
For the measurement errors $\errMeas[k,o]$ we assume the distribution to be known. For mathematical simplicity we assume statistical independence between the $\errMeas[k,o]$ of different MPCs $k,o$.

\Cref{sec:EstimateDistWithAssoc} assumes correct MPC association while  \Cref{sec:EstimateDistNoAssoc} assumes unknown MPC association.



\subsection{With Known MPC Association}
\label{sec:EstimateDistWithAssoc}
Given the delay differences $\delayDiff[k,o]$ subject to measurement errors $\errMeas[k,o]$ and unknown clock offset $\errClock$, we show in \Cref{apdx:DeriveEstDistanceWithAssoc} that the joint maximum-likelihood estimate (MLE) of distance $d$ and clock offset $\errClock$ is given by the maximization problem
\begin{align}
& ( \hat d \MLE\AssAsyn , \hat \errClock \MLE\AssAsyn )
\in \argmax_{\Hypo{d} \in \bbR_+, \Hypo{\errClock} \in \bbR} \,\f{1}{\Hypo{d}^K \!} \prod_{o=1}^{\NObs} \prod_{k=1}^{K_o}
I_{k o}\big( \delayDiff[k,o] \!-\! \Hypo{\errClock}, \Hypo{d} \, \big)
, \label{eq:distMLE_general} \\
& I_{k o}\big(\, \bullet \, ,\Hypo{d} \,\big) =
F_{\errMeas[k,o]}\big(\bullet + \tfrac{\Hypo{d}}{c} \,\big) -
F_{\errMeas[k,o]}\big(\bullet - \tfrac{\Hypo{d}}{c} \,\big) .
\label{eq:SoftIndicFunc}
\end{align}
The free variables $\Hypo{d}$ and $\Hypo{\errClock}$ represent hypothesis values and $F_{\errMeas[k,o]}$ is the cumulative distribution function (CDF) of the measurement error $\errMeas[k,o]$.
The clock offset $\errClock$ is necessarily included as a nuisance parameter because it affects the statistics of $\delayDiff[k,o]$.

The term $I_{k o}$ can be regarded as soft indicator function that evaluates the set membership $\cd (\delayDiff[k,o] - \errClock) \in [-\Hypo{d},\Hypo{d}\,]$.
For the case of Gaussian errors $\errMeas[k,o] \sim \mathcal{N}(0 , \sigma_{k o}^2)$, the CDF is described by the \mbox{$Q$-function}, $F_{\errMeas[k,o]}(x) = 1-Q(x/\sigma_{k o})$.

\Cref{fig:LHF_Asyn,fig:LHF_Asyn_Gauss} show examples of the two-dimensional likelihood function, i.e. of the maximization objective function in \Cref{eq:distMLE_general}. They concern the setup in \Cref{fig:InterUserPaths}.

\newcommand{\myFigHeight}{28.5mm}
\begin{figure}[t]\centering
\subfloat[known assoc., $\sigma = 0$]{
\includegraphics[height=\myFigHeight,trim=0 4 0 2]{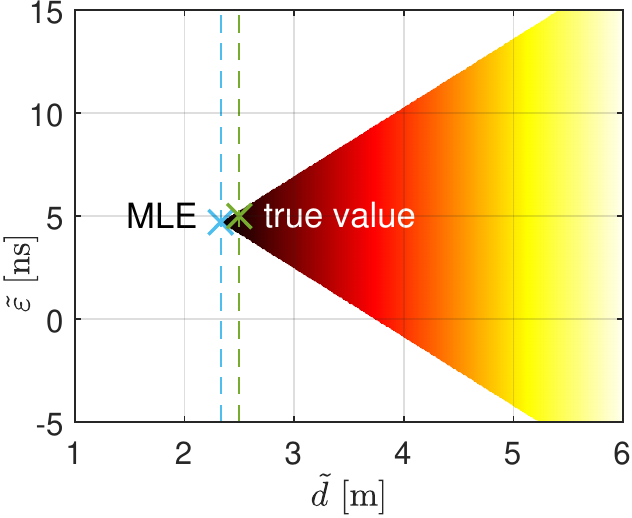}
\label{fig:LHF_Asyn}}
\subfloat[known assoc., $\sigma = 1\unit{ns}$]{
\includegraphics[height=\myFigHeight,trim=0 4 0 2]{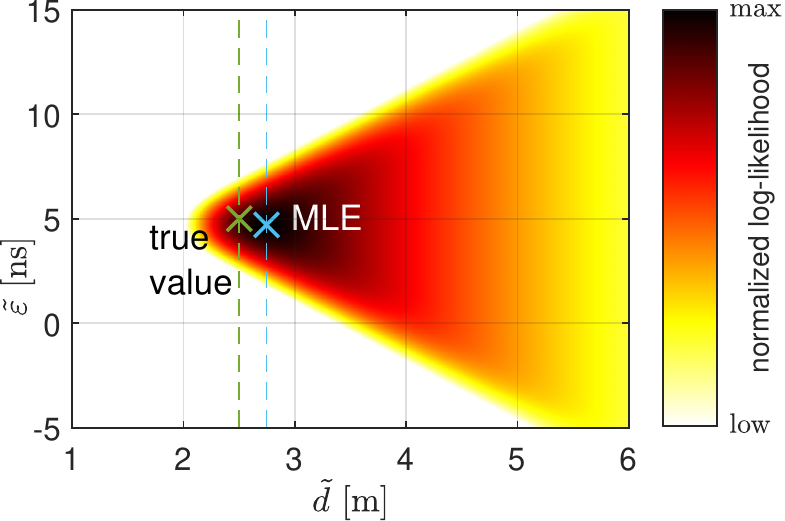}
\label{fig:LHF_Asyn_Gauss}}
\\
\subfloat[unknown assoc., $\sigma = 0$]{
\includegraphics[height=\myFigHeight,trim=0 4 0 2]{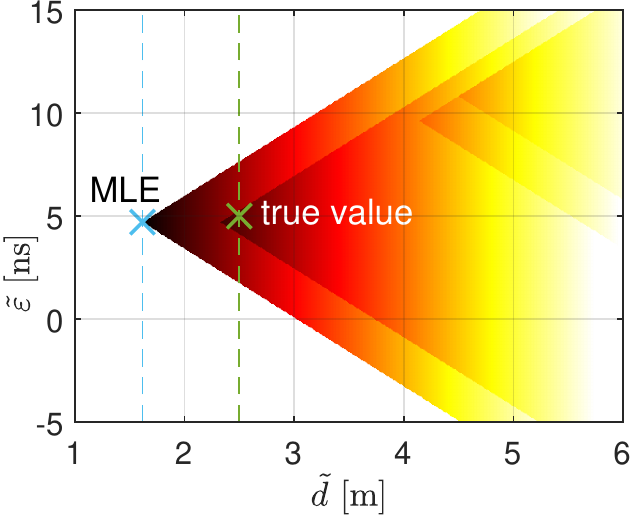}
\label{fig:LHF_Asyn_NoAssoc}}
\subfloat[unknown assoc., $\sigma = 1\unit{ns}$]{
\includegraphics[height=\myFigHeight,trim=0 4 0 2]{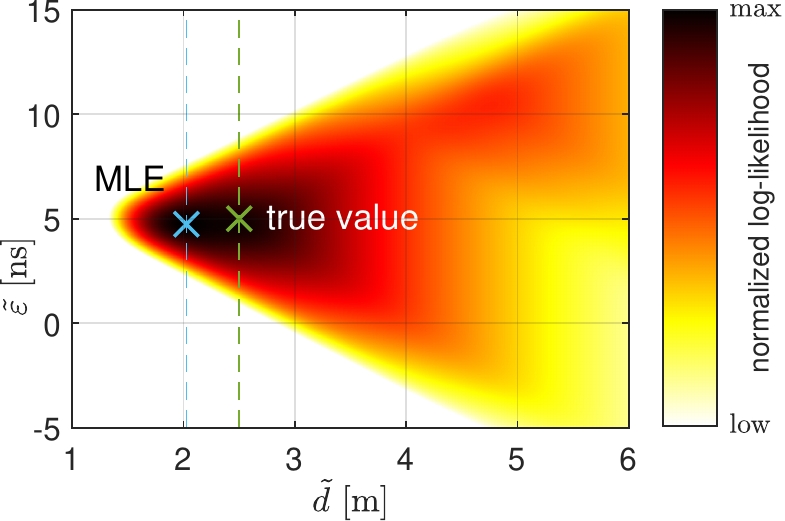}
\label{fig:LHF_Asyn_NoAssoc_Gauss}}
\caption{Likelihood function of distance (abscissa) and clock offset (ordinate) given delay differences $\delayDiff[k,o]$, cf. \Cref{eq:distMLE_general}.
Shown are cases with and without Gaussian measurement errors and known MPC association. The evaluation assumes the setup and CIRs from \Cref{fig:InterUserIntro} ($K = 3, d = 2.5\unit{m}$) and $\errClock = 5\unit{ns}$.}
\label{fig:LHF_Asyn_AllCases}
\centering
\vspace{2mm}
\resizebox{!}{36mm}{\input{LHF_Sync}}
\vspace{-2mm}
\caption{Distance likelihood function when precise time-synchronization between A and B is given a-priori.
The evaluation assumes the setup in \Cref{fig:InterUserIntro}.}
\label{fig:LHF_Sync}
\end{figure}

A general closed-form solution of \Cref{eq:distMLE_general} is unavailable and the properties of the optimization problem depend on the specific error statistics. The likelihood function is non-concave in general. This applies also to the Gaussian case because the $Q$-function is neither convex nor concave. Yet, the Gaussian case exhibits a unimodal likelihood function that is infinitely differentiable (i.e. smooth) and, thus, \Cref{eq:distMLE_general} can easily be solved with a few iterations of a gradient-based numerical solver.%

In the time-synchronous case where the clock offset $\errClock$ is a-priori known, the MLE
$\hat d \MLE\AssSync$
is found by maximizing a now univariate likelihood function of distance, which is obtained by fixing $\Hypo{\errClock} = \errClock$ in \Cref{eq:distMLE_general}. Examples thereof are given by the black graphs in \Cref{fig:LHF_Sync}.



Henceforth we consider the special case of zero measurement errors ($\errMeas[k,o] \equiv 0$).
Here the actual indicator function
$I_{k o}(\bullet \, ,\Hypo{d} \,) = \IndFunc_{[-\Hypo{d}/c,\Hypo{d}/c]}(\bullet)$
applies and, in consequence, the likelihood function attains a distinct structure with discontinuities. This can be seen in \Cref{fig:LHF_Asyn,fig:LHF_Sync}.
The MLE problem can now be solved in closed form, which is conducted in detail in \Cref{apdx:DeriveEstDistanceWithAssoc}. The solutions for the different cases constitute very simple formulas for distance estimation:
\begin{align}
& \hat d \MLE\AssAsyn
= \f{c}{2} \Big( \max_{k,o} \delayDiff[k,o] - \min_{k,o} \delayDiff[k,o] \Big)
\, , \label{eq:distMLE} \\
& \hat d \MLE\AssSync
= c \cdot \max_{k,o} |\delayDiff[k,o] - \errClock|
\, . \label{eq:distMLESync}
\end{align}
Both are underestimates with probability $1$ and thus biased. An underestimate $\hat d \MLE\AssAsyn < d$ occurs unless two of the random MPC directions happen to coincide with the directions of $\d$ and $-\d$, respectively. Only then do the values ${\cd\delayDiffTrue[k,o]}$ attain their minimum and maximum possible values $\pm d$. An underestimate $\hat d \MLE\AssSync < d$ occurs unless one MPC direction coincides with the direction of either $\d$ or $-\d$.
\footnote{The estimators \Cref{eq:distMLE}, \Cref{eq:distMLESync}, \Cref{eq:EpsMVUE} have the MLE property also for a two-dimensional system model where $\dirVectA[k,o]$ has uniform distribution on the unit circle. The MVUE rules however do not transfer. The details are omitted.

Another fortunate aspect is that the plane-wave assumption $\dirVectA[k,o] \approx \dirVectB[k,o]$ is practically superfluous here; the MPCs relevant to estimation fulfill it quite naturally. To see this, note that in \Cref{eq:distMLE}, \Cref{eq:distMLESync}, \Cref{eq:EpsMVUE} these MPCs have $\cd\delayDiffTrue[k,o]$ near $-d$ or $d$, so their $\dirVectA[k,o]$ is near-colinear with $\pm\d$. Thus, these MPC directions remain near-constant when moving from $\posA$ to $\posB$.}


In \Cref{apdx:DeriveEstDistanceWithAssoc_NoErrorMVUE} we derive bias-corrected versions of \Cref{eq:distMLE,eq:distMLESync} and furthermore show that each result is in fact the minimum-variance unbiased estimate (MVUE) of the respective case. They are given by
\begin{align}
&\hat d \MVUE\AssAsyn = \f{K+1}{K-1}\cdot \f{c}{2} \Big( \max_{k,o} \delayDiff[k,o] - \min_{k,o} \delayDiff[k,o] \Big)
\, , \label{eq:distMVUE} \\
& \hat \errClock\MVUE\AssAsyn = \hat \errClock\MLE\AssAsyn = \f{1}{2} \Big( \max_{k,o} \delayDiff[k,o] + \min_{k,o} \delayDiff[k,o] \Big)
\, , \label{eq:EpsMVUE} \\
&\hat d \MVUE\AssSync = \f{K+1}{K}\ c \cdot \max_{k,o} |\delayDiff[k,o] - \errClock|
\, . \label{eq:distMVUESync} 
\end{align}
The clock offset estimator \Cref{eq:EpsMVUE} could be useful in its own right, e.g. for distributed synchronization in dense multipath. It is both the MLE and the MVUE.

Despite $\errMeas[k,o] \equiv 0$ and the MVUE property, the estimators \Cref{eq:distMVUE,eq:EpsMVUE,eq:distMVUESync} exhibit a non-zero estimation error with probability $1$. This is caused by the random effect of the unknown MPC directions. The simple statistics
$\cd \delayDiffTrue[k,o] \overset{\text{i.i.d.}}{\sim} \mathcal{U}(-d,d)$
allow for an analytic study of the estimation-error statistics. By applying properties of order statistics \cite[Cpt.~12 \& 13]{Ahsanullah2013} to \Cref{eq:distMVUE,eq:EpsMVUE,eq:distMVUESync} we obtain for the estimators' root-mean-squared error (RMSE)
\begin{align}
\sqrt{\EV{ ( \hat d \MVUE\AssAsyn - d )^2 }} 
&= d\ \f{\sqrt{2}}{\sqrt{(K-1)(K+2)} }
\ , \label{eq:stdMVUE} \\
\sqrt{\EV{ (  \hat \errClock \MVUE\AssAsyn - \errClock )^2 }}
&= \f{d}{c}\ \f{\sqrt{2}}{\sqrt{(K+1)(K+2)}}
\ , \label{eq:stdEpsMVUE} \\
\sqrt{\EV{ ( \hat d \MVUE\AssSync - d )^2 }}
&= d\ \f{1}{\sqrt{K(K+2)}}
\ . \label{eq:stdMVUESync}
\end{align}
We note that each RMSE is asymptotically proportional to $d/K$, thus exhibiting a linear increase with distance. The formulas suggest that considering more MPCs is an efficient means to reduce the RMSE (it may however jeopardize the requirement of a correct MPC association). We also find that, asymptotically, the distance RMSE reduces by a factor $\sqrt{2}$ through precise time-synchronization between A and B. The reason is that the synchronous-case estimator \Cref{eq:distMVUESync} is less dependent on diverse MPC directions than \Cref{eq:distMVUE}.



\subsection{With Unknown MPC Association}
\label{sec:EstimateDistNoAssoc}
We now assume an unknown association between the MPC delays
$\{ \delayMeasA[1,o] \, , \ldots \, , \, \delayMeasA[K_o,o] \}$
and
$\{ \delayMeasB[1,o] \, , \ldots \, , \, \delayMeasB[K_o,o] \}$
for all $o$.
A distance estimator now faces the problem that, without any prior knowledge, any MPC association is eligible. Hence, any conceivable delay-difference $\delayMeasB[\perm(k),o] - \delayMeasA[k,o]$ with any choice of permutation $k' = \perm(k)$ is eligible a-priori. We regard the permutation $\perm$, a bijective map from and to $\{1,\ldots,K_o\}$, as a formal representation of an MPC association.

We adopt the statistical assumptions from the beginning of this section.
In \Cref{apdx:DeriveEstDistanceNoAssoc} we use tools from order statistics to show that the joint MLE of distance $d$ and clock offset $\errClock$ with unknown MPC association is given by
\begin{multline}
\left( \hat d \MLE\NoAssAsyn, \hat\errClock \MLE\NoAssAsyn \right) 
\in
\argmax_{\Hypo{d} \in \R_+, \, \Hypo{\errClock} \in \R}
\\
\frac{1}{\Hypo{d}^K} \prod_{o=1}^\NObs \sum_{\perm \in \permSet[K_o]} \prod_{k=1}^{K_o}
I_{ko}\big( \delayMeasB[\perm(k),o] - \delayMeasA[k,o] - \Hypo{\errClock}, \Hypo{d} \,\big) .
\label{eq:distMLENoAss_}
\end{multline}
There $I_{ko}$ is the soft indicator function from \Cref{eq:SoftIndicFunc}. 
\Cref{fig:LHF_Asyn_NoAssoc,fig:LHF_Asyn_NoAssoc_Gauss} show examples of the likelihood function, which now has a multimodal structure: it is a superposition of likelihood functions of the type in \Cref{fig:LHF_Asyn,fig:LHF_Asyn_Gauss} for different $\perm$.
The estimates can be computed by attempting to find the global solution of the optimization problem \Cref{eq:distMLENoAss_} with a numerical solver, e.g., an iterative gradient-based algorithm with a multistart approach. We note that each likelihood evaluation has combinatorial time complexity, which can become prohibitive for large $K_o$.

Consider the case without measurement errors ($\errMeas[k,o] \equiv 0$). The likelihood function in \Cref{eq:distMLENoAss_} attains the distinct structure in \Cref{fig:LHF_Asyn_NoAssoc}. As shown in \Cref{app:lhfEvalNoAssDerviation}, the simpler MLE rule
\begin{align}
\left( \hat d \MLE\NoAssAsyn, \hat\errClock \MLE\NoAssAsyn \right)
\in \argmax_{(\Hypo{d}, \Hypo{\errClock}) \in \calH} \
\f{1}{\Hypo{d}^K} \prod_{o=1}^\NObs \permCount_o(\Hypo{d}, \Hypo{\errClock})
\label{eq:distMLENoAss}
\end{align}
now applies. Relating to observer $o$, $C_o \in \bbN_0$ is the number of permutations for which the hypotheses $\Hypo{d}$ and $\Hypo{\errClock}$ do not contradict the observed delay differences. Formally,
\begin{align}
\permCount_o = \Big|\Big\{
\perm \in \permSet[K_o] \, \Big| \
c\cdot\max_k |\delayMeasB[\perm(k),o]\! - \delayMeasA[k,o]\! - \Hypo{\errClock}| \leq \Hypo{d}
\,\Big\}\Big| .
\label{eq:distMLENoAss_PermCount}
\end{align}
Due to the specific likelihood function structure, it suffices to evaluate the likelihood for a finite set of candidate hypotheses
$(\Hypo{d}, \Hypo{\errClock}) \in \calH$,
$\calH =  \{ ( c\,\f{L-S}{2}, \f{L+S}{2} ) \, | \,
(L,S) \in \bigcup_{o=1}^\NObs \calL_o \times \calS_o \} $
with the delay differences
$\calL_o = \bigcup_{\perm \in \permSet[K_o]} \max_k \delayMeasB[\perm(k),o]\! - \delayMeasA[k,o]$
and
$\calS_o = \bigcup_{\perm \in \permSet[K_o]} \min_k \delayMeasB[\perm(k),o]\! - \delayMeasA[k,o]$.


In the synchronous case, the MLE becomes the one-dimensional problem  
$\hat d \MLE\NoAssSync = \hat d \MLE\NoAssAsyn \big|_{\Hypo{\errClock} = \errClock}$.
The red graphs in \Cref{fig:LHF_Sync} show examples of the associated likelihood function, which again shows a superposition of known-association-type likelihoods for different permutations. If furthermore $\errMeas[k,o] \equiv 0$, then the candidates for maximum-likelihood distance reduce to the finite set
$\Hypo{d} \in \bigcup_{o=1}^\NObs \bigcup_{\perm \in \permSet[K_o]}\! c\cdot
\max_k |\delayMeasB[\perm(k),o] - \delayMeasA[k,o] - \errClock|$.

	
	

\section{Relative Position Estimators}
\label{sec:EstimateRelLoc}
This section presents estimators for the relative position vector $\d = \posB - \posA$.
These estimators do use the MPC directions $\dirVectA[k,o], \dirVectB[k,o]$ and thus require their availability, e.g., by measuring them during channel estimation with the use of antenna arrays at A and B.
The estimators in \Cref{sec:EstimateRelLocFromDelta} furthermore use the MPC delay differences $\delayDiff[k,o]$ while those in \Cref{sec:EstimateRelLocFromTau} use the MPC delays $\delayMeasA[k,o]$ and $\delayMeasB[k,o]$ directly. The two methods will results in characteristic differences.

Please note that inaccurately measured $\dirVectA[k,o], \dirVectB[k,o]$ can dominate the estimation error, but this is not captured by the following formalism. We will however investigate the effect numerically later in \Cref{sec:EvalSim}.

The estimators assume that the MPC association was established correctly by a preceding signal processing step, e.g., by the scheme presented later in \Cref{sec:EstimateRelLocNoAssoc}.

\subsection{Position Estimates From Delay Differences}
\label{sec:EstimateRelLocFromDelta}
The considered approach is based on the projection property \Cref{eq:ProjectionEquality} and the processing of delay differences $\delayDiff[k,o]$. As preparation, we define stacked vector and matrix quantities:
\begin{align}
\delayDiff
&= [ \, \delayDiff[1,1] \ldots \delayDiff[K_1,1] \ \delayDiff[1,2] \ldots \delayDiff[K_\NObs,\NObs] ]\Tr
\! \! \! 
&& \in \bbR^{K \times 1}
, \label{eq:DelayShiftStackVector} \\
\errMeas
&= [ \ \, \errMeas[1,1] \ldots \ \errMeas[K_1,1] \ \ \errMeas[1,2] \ldots \, \,\errMeas[K_\NObs,\NObs] ]\Tr
&& \in \bbR^{K \times 1}
, \label{eq:ErrorStackVector} \\
\E &= \hspace{-.5mm}
\mtx{cc}{
\! {\bf s}_{1,1} \,\ldots\, {\bf s}_{K_1,1} \!\! & \!\! {\bf s}_{1,2} \,\ldots\, {\bf s}_{K_\NObs, \NObs} \! \! \\
\!\!\! 1 \,\ \ldots \, \ \ 1 \!\! & \!\!\!\!\!\!\!\!\!\! 1 \ \ \ \ldots \ \ 1
} &&\in \bbR^{4 \times K}
, \label{eq:EMatrix} \\
{\bf s}_{ko} &= 
\f{1}{1 + (\dirVectA[k,o])\Tr \dirVectB[k,o]}
\left(\, \dirVectA[k,o] + \dirVectB[k,o] \,\right)
&& \in \bbR^{3 \times 1} .
\label{eq:sVectorForEMatrix}
\end{align}
The vector ${\bf s}_{ko}$ fulfills ${\bf s}_{ko}\Tr \d = \cd\delayDiffTrue[k,o]$, which is a convenient restatement of \Cref{eq:ProjectionEquality}. Therewith, the delay-differences signal model \Cref{eq:SignalModel} can now handily be written as
$\delayDiff
= \delayDiffTrue + \mathbf{1}\errClock + \errMeas$
or rather as linear equation system
$\cd\delayDiff
= \E\Tr[\d\Tr , \cd\errClock]\Tr + \cd\errMeas$.

We consider the joint estimation problem of $\d$ and $\errClock$ after observing $\delayDiff$ subject to measurement error $\errMeas$ and observing the MPC directions $\dirVectA[k,o], \dirVectB[k,o]$ without error (by assumption).
In \Cref{apdx:DeriveRelLocMLE_GeneralMLE} we show that the MLE is given by the unconstrained four-dimensional optimization problem
\begin{align}
\big( \, \hat\d\MLE\AssViaDiff \, , \hat\errClock\MLE\AssViaDiff \,\big)
\in \argmax_{\Hypo{\d} \in \bbR^3, \Hypo{\errClock} \in \bbR}
f_{\errMeas} \!\left( \delayDiff - \f{1}{c} \, \E\Tr \! \mtx{c}{\Hypo{\d} \\ \! c\cdot\Hypo{\errClock} \!} \right)
\label{eq:displMLEViaDiff}
\end{align}
where $f_{\errMeas}$ is the joint PDF of the measurement errors.
This could be tackled with established numerical methods such as iterative gradient search.
The properties of the likelihood function (e.g., concavity) depend on the specifics of $f_{\errMeas}$.

The least-squares estimate (LSE) is given by the formula
\begin{align}
\mtx{c}{\hat\d\LSE\AssViaDiff \\[1mm] \!
c \cdot \hat\errClock\LSE\AssViaDiff \!}
&= 
\left(\E\E\Tr \right)^{-1} \E\, (\cd\delayDiff)
\, . \label{eq:displLSEAssViaDiff}
\end{align}
Conveniently, the LSE can be computed without knowledge of the statistics of $\errMeas$.
For the special case $\errMeas \sim \calN({\bf 0}, \sigma^2 \,\eye_K)$, the LSE \Cref{eq:displLSEAssViaDiff} is also the MLE and the MVUE. For a general Gaussian distribution $\errMeas \sim \calN(\boldsymbol\mu, \boldsymbol\Sigma)$, the MLE and MVUE is instead given by $( \E\,\boldsymbol\Sigma^{-1} \E\Tr )^{-1} \E \, \boldsymbol\Sigma^{-1} (\cd\delayDiff - c\boldsymbol\mu)$, which deviates from the LSE in general. \cite[Thm~4.2 and Thm~7.5]{Kay1993}.

If the clock offset $\errClock$ between A and B is known (synchronous case) then $\hat\d\LSE\AssViaDiff$ is obtained by first considering a system of equations ${\bf s}_{ko}\Tr \d = \cd(\delayDiff[k,o] - \errClock)$ for all MPCs $k,o$ to then solve it for $\d$ with the Moore-Penrose inverse of $[{\bf s}_{1,1}\ {\bf s}_{2,1} \ldots]\Tr$.
Vice versa, when $\d$ is known, then the clock offset LSE is 
the mean of the values $\delayDiff[k,o] - \f{1}{c} {\bf s}_{ko}\Tr \d$ over all $k,o$.

Under the employed assumptions, the estimation error
$\hat\d\LSE\AssViaDiff - \d = (\E\E{}\Tr)^{-1} \E\, (\cd\errMeas)$
applies, however a more intuitive description of the estimation accuracy is desirable.
In \Cref{apdx:DeriveApproxRMSE} we show that the RMSE approximation
\begin{align}
\sqrt{\EV{ \| \hat\d\LSE\AssViaDiff \! - \d\|^2 }}
\approx \f{3\cd\sigma}{\sqrt{K}}
\label{eq:RMSEdisplLSEViaDiff}
\end{align}
is accurate for the special case $\errMeas \sim \mathcal{N}({\bf 0}, \sigma^2 \eye_K)$ with the plane-wave assumption, large $K$, and random $\dirVectA[k,o]$ with i.i.d. uniform distributions on the 3D unit sphere. 
In comparison to the RMSE of the MVUE distance estimator \Cref{eq:stdMVUE},
the expression \Cref{eq:RMSEdisplLSEViaDiff} shows no systematic increase with $d$. The decay with increasing $K$ is rather slow. The same RMSE expression \Cref{eq:RMSEdisplLSEViaDiff} applies in the synchronous case. From this perspective, a-priori synchronization is inessential to the accuracy.

Finally, we point out a technically interesting adaptation.
When only the directions $\dirVectA[k,o]$ but not $\dirVectB[k,o]$ are available, then the plane-wave assumption (PWA)
$\dirVectA[k,o] \approx \dirVectB[k,o] \ \Rightarrow\ {\bf s}_{ko} \approx \dirVectA[k,o]$
allows to still use the estimators \Cref{eq:displMLEViaDiff,eq:displLSEAssViaDiff} with hardly any accuracy loss for small $d$; cf. \Cref{eq:ProjectionEquality} versus \Cref{eq:ProjectionApproximation}. We define an estimator which utilizes this technological advantage:
\begin{align}
\hat\d\LSE\AssViaDiffPWA := \hat\d\LSE\AssViaDiff \Big|_{{\bf s}_{ko} = \dirVectA[k,o]}
\, . \label{eq:displLSEAssViaDiffPWA}
\end{align}

\subsection{Position Estimates Directly From Delays}
\label{sec:EstimateRelLocFromTau}
Based on the property $\d = \cd\delayTrueB[k,o] \dirVectB[k,o] - \cd\delayTrueA[k,o] \dirVectA[k,o]$ from \Cref{eq:VectorEquality},
we study an alternative scheme which directly uses the delays instead of their differences.
We consider the estimation of $\d \in \bbR^3$ from measured delays
$\delayMeasA[k,o]$ and $\delayMeasB[k,o]$ subject to 
clock offsets $\errClockA[o]$ and $\errClockB[o] = \errClockA[o] - \errClock$, respectively, and measurement errors $\errMeasA[k,o]$ and $\errMeasB[k,o]$.
Again, the processing relies on accurate knowledge of the MPC directions $\dirVectA[k,o]$ and $\dirVectB[k,o]$.

In \Cref{apdx:DeriveRelLocLSE_FromTau} we show that the joint LSE of $\d$ and all relevant clock offsets is given by
\begin{align}
& \left[
(\hat\d\LSE\AssViaTau)\Tr \! ,
\cd\hat\errClock ,
\cd\errClockAEstimate[1] ,
\ldots ,
\cd\errClockAEstimate[\NObs]
\right]\Tr
\! = 
( {\bf G}\Tr {\bf G} )^{-1} {\bf G}\Tr {\bf t}
\label{eq:EstimateRelLocViaTau}
\end{align}
whereby ${\bf G} \in \bbR^{(3K) \times (4 + \NObs)}$ and ${\bf t} \in \bbR^{(3K) \times 1}$ are defined as
\newcommand\myPartOne{
{\bf G} = &
\mtx{llccc}{
\eye_3 & \dirVectB[1,1]   & \dirVectB[ 1 ,1]\!-\dirVectA[ 1 ,1] \\[-2mm]
\vdots & \,\vdots         &    \vdots                           & & {\bf 0} \\
\eye_3 & \dirVectB[K_1,1] & \dirVectB[K_1,1]\!-\dirVectA[K_1,1] \\[-2mm]
\vdots & \,\vdots & & \!\!\!\!\ddots\!\!\!\! \\
\eye_3 & \dirVectB[1,\NObs] & & & \dirVectB[ 1 ,\NObs]\!-\dirVectA[ 1 ,\NObs] \\[-2mm]
\vdots & \,\vdots       & {\bf 0} & & \vdots \\
\eye_3 & \dirVectB[K_\NObs,\NObs] & & & \!\!\!\dirVectB[K_\NObs,\NObs]\!-\dirVectA[K_\NObs,\NObs]\!\!\! } 
,}
\newcommand\myPartTwo[2]{
{\bf t} = &
\mtx{c}{
\cd\delayMeasB[1,1] \dirVectB[1,1] - \cd\delayMeasA[1,1] \dirVectA[1,1] \\[#1]
\vdots \\[#2]
\!\!
\cd\delayMeasB[K_\NObs,\NObs] \dirVectB[K_\NObs,\NObs] - \cd\delayMeasA[K_\NObs,\NObs] \dirVectA[K_\NObs,\NObs]
\!\!}
. \label{eq:EstimateRelLocViaTauDetails}
}
\ifdefined\SingleColumnDraft
\begin{align} \myPartOne & \myPartTwo{4mm}{4mm} \end{align}
\else
\begin{align} \myPartOne \nonumber \\ \myPartTwo{-1.5mm}{-1mm} \end{align}
\fi
The MLE is omitted as its formulation requires the joint PDF of $\errMeasA,\errMeasB$.
When all clock offsets between the nodes and all observers are a-priori known,
then the LSE reduces to
\begin{align}
\hat\d\LSE\AssViaTauSync \! = \!
\f{c}{K} \!\sum_{o=1}^\NObs \sum_{k=1}^{K_o} 
(\delayMeasB[k,o]\! - \!\errClockB[o]) \dirVectB[k,o] \! - \!
(\delayMeasA[k,o]\! - \!\errClockA[o]) \dirVectA[k,o]
.
\label{eq:EstimateRelLocViaTauSync}
\end{align}
%
It is apparent that the estimators \Cref{eq:EstimateRelLocViaTau,eq:EstimateRelLocViaTauSync} rely on the MPC directions $\dirVectA[k,o]$, $\dirVectB[k,o]$ being measured with high accuracy.

This approach is fundamentally incompatible with the plane-wave assumption \Cref{eq:ProjectionApproximation} due to the nature of the underlying property \Cref{eq:VectorEquality}. This can be seen best in \Cref{apdx:PropagationGeometry}.

\subsection{Establishing the MPC Association}
\label{sec:EstimateRelLocNoAssoc}
The relative position estimators stated in \Cref{sec:EstimateRelLocFromDelta,sec:EstimateRelLocFromTau}
assumed knowledge of the MPC directions $\dirVectA[k,o]$ and $\dirVectB[k,o]$. Such direction knowledge is particularly useful for reconstructing the MPC association in case it is a-priori unknown, which is the topic of this subsection.

We assume that the MPC directions $\dirVectA[k,o]$ and $\dirVectB[k,o]$ are stated within the same frame of reference (this could be enforced by solving an orthogonal Procrustes problem).
The MPC association relating to observer $o$ is formalized in terms of a permutation $k' = \pi_o(k)$ with $k,k' \in \{1,\ldots,K_o\}$.
We propose the following geometry-inspired data-fitting rule to reconstruct the MPC association given the MPC directions and delays:
\begin{align}
\hat\perm_o &=
\argmin_{\perm \in \permSet[K_o]}
\sum_{k=1}^{K_o}
J_o(k,\perm(k))
\, , \label{eq:AssocEstim} \\
J_o(k,l) &=
\big\| {\dirVectB[{l,o}]} \! - \dirVectA[k,o] \big\|^2
+ \lambda^2 \,
\big| \delayMeasB[l,o]\! - \mu_o\AnnotateNode{B}\! - \delayMeasA[k,o]\! - \mu_o\AnnotateNode{A} \big|^2 .
\nonumber 
\end{align}
This search is done individually per observer $o \in \{1, \ldots, \NObs\}$.
It associates the MPCs $\bullet_{k,o}\AnnotateNode{A}$ and $\bullet_{\hat\perm_o{(k)},o}\AnnotateNode{B}$. Afterwards, any estimation rule from \Cref{sec:EstimateDistWithAssoc,sec:EstimateRelLocFromDelta,sec:EstimateRelLocFromTau} can be used.

The expression uses the averages
$\mu_o\AnnotateNode{\textbullet} = \f{1}{K_o} \sum_{k=1}^{K_o} \tau_{k,o}\AnnotateNode{\textbullet}$. They can be replaced by $\mu_o\AnnotateNode{\textbullet} = \errClock\AnnotateNode{\textbullet}$ in the synchronous case.
The regularization constant $\lambda^2$ balances cost contributions by MPC directions and delays. 
If $\lambda$ is large then \Cref{eq:AssocEstim} tends to associate the MPCs in the order of their delay.
A sensible choice is given by $\lambda = 1/\tauRMS$ where $\tauRMS$ is the channel delay spread (or a coarse estimate thereof).


If the MPC directions at the observer (denoted $\dirVectObs[k,o]\AnnotateNode{\textbullet}$) are also known, then they can be included in the scheme by adding the summand
$\| \dirVectObsB[\perm(k),o] - \dirVectObsA[k,o] \|^2$
to $J_o(k,l)$ in \Cref{eq:AssocEstim}. Likewise, if the $\dirVect[k,o]\AnnotateNode{\textbullet}$ were unavailable, then $\dirVectObs[k,o]\AnnotateNode{\textbullet}$ could replace them in \Cref{eq:AssocEstim}, e.g., for distance estimation between low-complexity nodes A and B with high-complexity observers with antenna arrays.

The optimization problem \Cref{eq:AssocEstim} is a linear assignment problem and is thus solved efficiently by the Hungarian method \cite{SchuhmacherTSP2008}. 
This avoids the tedious evaluation of all $K_o!$ possible permutations. The cost $\min \sum_k J_o(k,\pi(k))$ is related to the optimal subpattern assignment (OSPA) metric \cite[Eq.~(3)]{SchuhmacherTSP2008}.

The framework of linear assignment problems can handle MPCs $\bullet_{k,o}\AnnotateNode{A}$ without a corresponding $\bullet_{k,o}\AnnotateNode{B}$. It is furthermore able to detect and reject bad associations: large angles between $\dirVectA[k,o]$ and $\dirVectB[l,o]$ indicate an incorrect association. We implement this notion by setting the cost $J_o(k,l) = \infty$ if the angle exceeds a threshold of $30^\circ$.


\section{Numerical Accuracy Evaluation}
\label{sec:EvalSim}
\newcommand\SimPlotsInnerWidth{.5\columnwidth}  
\newcommand\SimPlotsInnerHeight{.425\columnwidth} 

\newcommand\SimPlotsSyncColor{0.0,0.8,0.8}
\newcommand\SimPlotsSyncColorDarker{0.0,0.7,0.7}
\newcommand\SimPlotsLineWidth{0.5pt}
\newcommand\SimPlotsYMaxDist{1.2}
\newcommand\SimPlotsYMaxPos{1.2}

\newcommand\SimPlotsGrid{false}
\newcommand\SimPlotsGridLineWidth{0.5pt}
\newcommand\SimPlotsGridColor{gray!18}

\newcommand\SimPlotsLegendMVUE{MVUE} 
\newcommand\SimPlotsLegendMLE{MLE} 
\newcommand\SimPlotsLegendNA{N/A} 
\newcommand\SimPlotsLegendSORT{SORT} 
\newcommand\SimPlotsLegendDD{DD} 
\newcommand\SimPlotsLegendPWA{PWA} 
\newcommand\SimPlotsLegendDDN{DDN} 
\newcommand\SimPlotsLegendTAU{TAU} 
\newcommand\SimPlotsLegendTNA{TNA} 
\newcommand\SimPlotsTextMVUE{$\hat d\MVUE$ from \Cref{eq:distMVUE} (designed for $\sigma_{k,o} \equiv 0$)}
\newcommand\SimPlotsTextMLE{$\hat d\MLE$ from \Cref{eq:distMLE_general}}
\newcommand\SimPlotsTextNA{$\hat d\MLE\NoAssAsyn$ from \Cref{eq:distMLENoAss_} (unknown MPC association )}
\newcommand\SimPlotsTextSORT{$\hat d\MVUE$ from \Cref{eq:distMVUE}, assoc. estimated via $\tau$-sorting}
\newcommand\SimPlotsTextDD{$\hat\d\LSE\AssViaDiff$ from \Cref{eq:displLSEAssViaDiff} (uses delay differences)}
\newcommand\SimPlotsTextPWA{$\hat\d\LSE\AssViaDiffPWA$ from \Cref{eq:displLSEAssViaDiffPWA} (plane-wave assumption)}
\newcommand\SimPlotsTextDDN{$\hat\d\LSE\AssViaDiff$ from \Cref{eq:displLSEAssViaDiff}, assoc. estimated via \Cref{sec:EstimateRelLocNoAssoc}}
\newcommand\SimPlotsTextTAU{$\hat\d\LSE\AssViaTau$ from \Cref{eq:EstimateRelLocViaTau} (directly uses delays)}
\newcommand\SimPlotsTextTNA{$\hat\d\LSE\AssViaTau$ from \Cref{eq:EstimateRelLocViaTau}, assoc. estimated via \Cref{sec:EstimateRelLocNoAssoc}}

\newcommand\SimPlotsMVUELineStyle{solid}
\newcommand\SimPlotsMVUEMarkSize{2.25pt}
\newcommand\SimPlotsMVUEMarker{square}
\newcommand\SimPlotsMVUERotate{0}

\newcommand\SimPlotsMLELineStyle{dotted}
\newcommand\SimPlotsMLEMarkSize{3.0pt}
\newcommand\SimPlotsMLEMarker{asterisk}

\newcommand\SimPlotsNALineStyle{dotted}
\newcommand\SimPlotsNAMarkSize{3.5pt}
\newcommand\SimPlotsNAMarker{triangle*}

\newcommand\SimPlotsSORTLineStyle{dashed}
\newcommand\SimPlotsSORTMarkSize{1.25pt}
\newcommand\SimPlotsSORTMarker{square*}

\newcommand\SimPlotsDDLineStyle{solid}
\newcommand\SimPlotsDDMarkSize{3.5pt}
\newcommand\SimPlotsDDMarker{triangle}

\newcommand\SimPlotsPWALineStyle{dotted}
\newcommand\SimPlotsPWAMarkSize{4.0pt}
\newcommand\SimPlotsPWAMarker{|}

\newcommand\SimPlotsDDNLineStyle{dotted}
\newcommand\SimPlotsDDNMarkSize{1.25pt}
\newcommand\SimPlotsDDNMarker{*}

\newcommand\SimPlotsTAULineStyle{solid}
\newcommand\SimPlotsTAUMarkSize{3.5pt}
\newcommand\SimPlotsTAUMarker{diamond}

\newcommand\SimPlotsTNALineStyle{dashed}
\newcommand\SimPlotsTNAMarkSize{4.5pt}
\newcommand\SimPlotsTNAMarker{x}

\newcommand\LegendSymb[4]{\resizebox{6mm}{!}{\begin{tikzpicture}
\begin{axis}[%
width=23mm, height=20mm, at={(0,0)},
axis line style={draw=none}, xtick=\empty, ytick=\empty,
xmin=0, xmax=2, ymin=-1, ymax=1,
axis background/.style={fill=white},
axis x line*=bottom, axis y line*=left,
]
\addplot [color=black, #1, line width=\SimPlotsLineWidth]
  table[row sep=crcr]{0 0\\2 0\\};
\addplot [color=black, line width=\SimPlotsLineWidth, mark size=#2, mark=#3, mark options={solid, black,rotate=#4}]
  table[row sep=crcr]{1 0\\};
\end{axis}
\end{tikzpicture}}}

This section presents a numerical evaluation of the estimators' accuracy as a function of $d$, $K$, and the main technical parameters and conditions. The employed methodology is random sampling of MPC parameters, with statistical assumptions characteristic of dense indoor multipath channels.

The assumed basic setup parameters are as follows.
All observers are assumed at random positions at $5\unit{m}$ distance from $\posA$. The minimum delay is thus $\tau_\mathrm{min} = 16.7\unit{ns} \leq \delayTrueA[k,o]$, which is attained by an eventual LOS path.
The LOS path to an observer occurs with probability $p_\mathrm{LOS}$.
We use the parameter values $d = 2.5\unit{m}$, $p_\mathrm{LOS} = 0.5$, $\NObs = 3$, and $K_o = 4\,\forall o$ (i.e. $K = 3 \cdot 4 = 12$) unless they are redefined explicitly.


For all NLOS paths in $\cirA(\tau)$ the excess delays are iid sampled from a PDF that is proportional to a power-delay profile (PDP) $S_\nu(\tau)$ with double-exponential shape \cite[Eq.~(9)]{KaredalTWC2007},\cite[Eq.~(22)]{MolischUWB2005}. This channel model is characteristic of office and industrial environments.
Guided by \cite[Tab.~I]{LeitingerJSAC2015}, \cite[Tab.~II]{KaredalTWC2007}, \cite[Sec.~III-E]{MolischUWB2005} we choose the following PDP parameter values:
rise time $\gamma_\mathrm{rise} = 10 \unit{ns}$,
fall time $\gamma_1 = 30 \unit{ns}$,
normalized power
$\Omega_1 = 1.5 \cdot 10^{-6}$
(yielding $T_\mathrm{p} \f{\Omega_1}{\gamma_1} = 5 \cdot N_0$),
and shape parameter $\chi = 0.9$.
This results in
a mean excess delay of $36.5\unit{ns}$
and an RMS delay spread of $\tau_\mathrm{RMS} = 30.3\unit{ns}$.

The MPC directions $\dirVectA[k,o]$ are independently drawn from a uniform distribution over the 3D unit sphere (rich scattering assumption).
We set $\posB = \posA + \d$ with $\d = [d,0,0]\Tr$ and then $\delayTrueB[k,o] = \f{1}{c} \|{\bf b}\|$ and $\dirVectB[k,o] = \f{\bf b}{\|{\bf b}\|}$ with ${\bf b} := \cd\delayTrueA[k,o]\dirVectA[k,o] + \d$. This calculation is based on the triangle in \Cref{apdx:PropagationGeometry}.

The delay measurement errors
$\errMeasA[k,o]$ and $\errMeasB[k,o]$
are independently drawn from zero-mean Gaussian distributions with standard deviations
$\sigma\AnnotateNodeA_{k,o}$ and $\sigma\AnnotateNodeB_{k,o}$, respectively.
The delay-difference error variance is thus
$\sigma_{k,o}^2 =
(\sigma\AnnotateNodeA_{k,o})^2 +
(\sigma\AnnotateNodeB_{k,o})^2$.
In order to describe those statistics, we assume a transmit pulse with a block spectrum (no roll-off) and $B_\mathrm{N} = 2\unit{GHz}$ Nyquist bandwidth. The pulse duration is $T_\mathrm{p} = 1 / B_\mathrm{N} = 0.5\unit{ns}$ and the effective mean-square bandwidth
$\beta = B_\mathrm{N} / \sqrt{12}$, cf. \cite{LeitingerJSAC2015}.
We describe $\sigma\AnnotateNodeA_{k,o}$ and $\sigma\AnnotateNodeB_{k,o}$ with the formula \cite[Eq.~(13)]{LeitingerJSAC2015}
\begin{align}
\sigma\AnnotateNodeX_{k,o} &=
\Big( \pi\beta \sqrt{8\cdot\mathrm{SINR}\AnnotateNodeX_{k,o} } \,\Big)^{-1}
\label{eq:CRLB}
\end{align}
that stems from the Cram\'{e}r-Rao lower bound.
In this model, the signal-to-interference-plus-noise ratio (SINR) of delay measurements is limited by receiver noise and interference from non-resolvable diffuse multipath \cite[Eq.~(14)]{LeitingerJSAC2015}:
\begin{align}
\mathrm{SINR}\AnnotateNodeX_{k,o} &=
\f{|\alpha\AnnotateNodeX_{k,o}|^2}{N_0 + T_\mathrm{p}\, S_\nu(\delayTrueX[k,o])} .
\end{align}
For the noise spectral density we assume $N_0 = 5 \cdot 10^{-9}$ with unit $\f{\mathrm{mW}}{\mathrm{GHz}} = \mathrm{pJ}$.
Like in \cite{LeitingerJSAC2015} we describe the MPC path loss due to the traveled path length $\cd\delayTrueX[k,o]$ with a Friis-type formula
$|\alpha\AnnotateNodeX_{k,o}|^2 = \xi_{k,o}  E_1 ( \cd\delayTrueX[k,o] )^{-2}$.
It assumes isotropic antennas. The factor $\xi_{k,o}$ describes eventual attenuation due to lossy reflection or scattering. For simplicity we set $\xi_{k,o} = -5\dB$ for all NLOS paths, which is in decent agreement with measurements \cite{KaredalTWC2007}, and $\xi_{k,o} = 1$ for a LOS path.
The term $E_1$ is the squared path amplitude over a $1\unit{m}$ LOS link; we assume $E_1 = 2.5 \cdot 10^{-5}$.
This yields a LOS-path SNR of
$(\f{1\unit{m}}{5\unit{m}})^2 E_1 / N_0 = 23\dB$
for the $\unit{5m}$ link between an observer and node A (the SINR is $20.8\dB$).
For each NLOS MPC, the model results in $\mathrm{SINR}\AnnotateNodeX_{k,o} < 10\dB$ with $93\%$ chance and in $\mathrm{SINR}\AnnotateNodeX_{k,o} < 0\dB$ with $4\%$ chance. A LOS path has $\cd\sigma\AnnotateNodeA_{k,o} = 5.3\unit{mm}$ while most NLOS paths exhibit $10$ to $50 \unit{mm}$.
The value $\cd\sigma_{k,o} \approx \sqrt{2} \cdot \cd\sigma\AnnotateNodeA_{k,o}$ of $\delayDiff[k,o]$-measurements is thus between $14$ and $71 \unit{mm}$ for most NLOS paths.

\Cref{tab:SimPlotColors,tab:SimPlotCases} describe the considered synchronization assumptions and the evaluated estimators, respectively.

\begin{table}[!hb]
\centering
\begin{tabular}{c|l}
black & No synchronization assumption other than \Cref{eq:QuickChannelEstimationAssumption} \\
{\color[rgb]{\SimPlotsSyncColor}cyan} & Also assumes precise a-priori time sync. between A, B
\end{tabular}
\vspace{1mm}
\caption{Legend of synchronization assumptions (color codes).}
\label{tab:SimPlotColors}
\newcommand\myDist{-.15mm}
\newcolumntype{L}[1]{>{\arraybackslash}m{#1}}
\begin{tabular}{cc|L{.65\columnwidth}}
\hline
\ \\[-2.7mm]
\multicolumn{3}{c}{abbreviations for distance estimators} \\[.5mm]\hline
& \\[-2.7mm]
\LegendSymb{\SimPlotsMVUELineStyle}{\SimPlotsMVUEMarkSize}{\SimPlotsMVUEMarker}{\SimPlotsMVUERotate}
& \SimPlotsLegendMVUE & \SimPlotsTextMVUE \\[\myDist]
\LegendSymb{\SimPlotsMLELineStyle}{\SimPlotsMLEMarkSize}{\SimPlotsMLEMarker}{0}
& \SimPlotsLegendMLE & \SimPlotsTextMLE \\[\myDist]
\LegendSymb{\SimPlotsSORTLineStyle}{\SimPlotsSORTMarkSize}{\SimPlotsSORTMarker}{0}
& \SimPlotsLegendSORT & \SimPlotsTextSORT \\[\myDist]
\LegendSymb{\SimPlotsNALineStyle}{\SimPlotsNAMarkSize}{\SimPlotsNAMarker}{0}
& \SimPlotsLegendNA & \SimPlotsTextNA \\[.8mm]\hline
\ \\[-2.7mm]
\multicolumn{3}{c}{abbreviations for position estimators} \\[.5mm]\hline
& \\[-2.7mm]
\LegendSymb{\SimPlotsDDLineStyle}{\SimPlotsDDMarkSize}{\SimPlotsDDMarker}{0}
& \SimPlotsLegendDD & \SimPlotsTextDD \\[\myDist]
\LegendSymb{\SimPlotsPWALineStyle}{\SimPlotsPWAMarkSize}{\SimPlotsPWAMarker}{0}
& \SimPlotsLegendPWA & \SimPlotsTextPWA \\[\myDist]
\LegendSymb{\SimPlotsDDNLineStyle}{\SimPlotsDDNMarkSize}{\SimPlotsDDNMarker}{0}
& \SimPlotsLegendDDN & \SimPlotsTextDDN \\[\myDist]
\LegendSymb{\SimPlotsTAULineStyle}{\SimPlotsTAUMarkSize}{\SimPlotsTAUMarker}{0}
& \SimPlotsLegendTAU & \SimPlotsTextTAU \\[\myDist]
\LegendSymb{\SimPlotsTNALineStyle}{\SimPlotsTNAMarkSize}{\SimPlotsTNAMarker}{0}
& \SimPlotsLegendTNA & \SimPlotsTextTNA
\end{tabular}
\vspace{.35mm}
\caption{Legend of estimators for simulation results.}
\label{tab:SimPlotCases}
\end{table}


\begin{figure}[t]
\centering\!\!\!\!\!\!
\subfloat[distance estimation]{\centering\label{fig:RMSE_vs_d_SyncAspects_dEst}
\resizebox{.5\columnwidth}{!}{
%
%
\definecolor{colorSync}{rgb}{\SimPlotsSyncColor}%
\begin{tikzpicture}

\pgfplotsset{every x tick label/.append style={font=\tiny, yshift=0.3ex}}
\pgfplotsset{every y tick label/.append style={font=\tiny, xshift=0.4ex}}

\newcommand\MyMarkRepeat{4}

\begin{axis}[%
width=\SimPlotsInnerWidth,
height=\SimPlotsInnerHeight,
at={(0,0)},
scale only axis,
xmode=log,
ymode=log,
xmin=.1,
xmax=10,
xtick={.1, 1, 10},
xlabel={\footnotesize{inter-node distance $d$ [m]}},
ymin=.01,
ymax=3,
ytick={.01, .1, 1, 10},
ylabel={\footnotesize{distance RMSE [m]}},
x label style={at={(axis description cs:0.5,-0.09)},anchor=north},
y label style={at={(axis description cs:-0.135,0.5)},anchor=south},
xmajorgrids=\SimPlotsGrid,
ymajorgrids=\SimPlotsGrid,
grid style={line width=\SimPlotsGridLineWidth, draw=\SimPlotsGridColor, solid},
legend style={at={(.12,.9)}, anchor=north west, legend cell align=center, draw=white!15!black},
legend columns=2,
transpose legend
]

\addplot [color=colorSync, \SimPlotsMVUELineStyle, line width=\SimPlotsLineWidth, mark size=\SimPlotsMVUEMarkSize, mark=\SimPlotsMVUEMarker, mark options={solid, rotate=\SimPlotsMVUERotate, colorSync},
forget plot, mark repeat=\MyMarkRepeat, mark phase=1]
  table[row sep=crcr]{%
0.1	0.089019899529269\\
0.125892541179417	0.0840292878533894\\
0.158489319246111	0.0791319699260849\\
0.199526231496888	0.0739048425594127\\
0.251188643150958	0.0726381214241619\\
0.316227766016838	0.0706185782349792\\
0.398107170553497	0.071009714090409\\
0.501187233627272	0.074775821644571\\
0.630957344480193	0.0784233345892771\\
0.794328234724281	0.0863956765778507\\
1	0.0987641923929703\\
1.25892541179417	0.117335988541001\\
1.58489319246111	0.136080891095425\\
1.99526231496888	0.164207754173953\\
2.51188643150958	0.200224452797181\\
3.16227766016838	0.257371385282481\\
3.98107170553497	0.31657494372047\\
5.01187233627272	0.412315392599412\\
6.30957344480193	0.486630903627422\\
7.94328234724282	0.594339774897164\\
10	0.713924592544784\\
};

\addplot [color=colorSync, \SimPlotsMLELineStyle, line width=\SimPlotsLineWidth, mark size=\SimPlotsMLEMarkSize, mark=\SimPlotsMLEMarker, mark options={solid, colorSync},
forget plot, mark repeat=\MyMarkRepeat, mark phase=3]
  table[row sep=crcr]{%
0.1	0.0402403556457259\\
0.125892541179417	0.0397809648604885\\
0.158489319246111	0.0399427759195961\\
0.199526231496888	0.0413471471258316\\
0.251188643150958	0.0440729869226267\\
0.316227766016838	0.0479692103671389\\
0.398107170553497	0.0529950771269636\\
0.501187233627272	0.0615010246654653\\
0.630957344480193	0.0699474975688697\\
0.794328234724281	0.0805147125449853\\
1	0.0971033953578267\\
1.25892541179417	0.11711022403416\\
1.58489319246111	0.145913072252759\\
1.99526231496888	0.183520207269769\\
2.51188643150958	0.235606529514664\\
3.16227766016838	0.287354831058482\\
3.98107170553497	0.37779924746312\\
5.01187233627272	0.473266810632124\\
6.30957344480193	0.581604291373746\\
7.94328234724282	0.719430176252707\\
10	0.90142250981244\\
};

\addplot [color=black, \SimPlotsMVUELineStyle, line width=\SimPlotsLineWidth, mark size=\SimPlotsMVUEMarkSize, mark=\SimPlotsMVUEMarker, mark options={solid, rotate=\SimPlotsMVUERotate, black},
mark repeat=\MyMarkRepeat, mark phase=3]
  table[row sep=crcr]{%
0.1	0.0720431710400428\\
0.125892541179417	0.0669702967812523\\
0.158489319246111	0.0629502703658703\\
0.199526231496888	0.0598167183690343\\
0.251188643150958	0.0601366260590916\\
0.316227766016838	0.0630239591290086\\
0.398107170553497	0.0663877788791542\\
0.501187233627272	0.0755331250078776\\
0.630957344480193	0.0853756344402675\\
0.794328234724281	0.100995072598569\\
1	0.123632770510689\\
1.25892541179417	0.15515929558411\\
1.58489319246111	0.191122012710974\\
1.99526231496888	0.236582090317288\\
2.51188643150958	0.300734205050987\\
3.16227766016838	0.381501037578764\\
3.98107170553497	0.491434296143033\\
5.01187233627272	0.679310295782928\\
6.30957344480193	0.925554878926245\\
7.94328234724282	1.32529728596517\\
10	1.9091102837069\\
};
\addlegendentry{\SimPlotsLegendMVUE}

\addplot [color=black, \SimPlotsMLELineStyle, line width=\SimPlotsLineWidth, mark size=\SimPlotsMLEMarkSize, mark=\SimPlotsMLEMarker, mark options={solid, black},
mark repeat=\MyMarkRepeat, mark phase=1]
  table[row sep=crcr]{%
0.1	0.0453335682631101\\
0.125892541179417	0.0453231178102718\\
0.158489319246111	0.045565417757392\\
0.199526231496888	0.0471707053287204\\
0.251188643150958	0.0494994956490006\\
0.316227766016838	0.0563795983735626\\
0.398107170553497	0.0631410348154113\\
0.501187233627272	0.077788309187549\\
0.630957344480193	0.0931986337731031\\
0.794328234724281	0.114294456508292\\
1	0.147250093174775\\
1.25892541179417	0.181753732771135\\
1.58489319246111	0.240111679421074\\
1.99526231496888	0.310353935800937\\
2.51188643150958	0.414976330713325\\
3.16227766016838	0.532325418885747\\
3.98107170553497	0.714920644022238\\
5.01187233627272	0.96119031994794\\
6.30957344480193	1.33019763780357\\
7.94328234724282	1.84239581078741\\
10	2.65258315532714\\
};
\addlegendentry{\SimPlotsLegendMLE}

\end{axis}
\end{tikzpicture}%
}}\!\!\!\!\!
\subfloat[position estimation]{\centering\label{fig:RMSE_vs_d_SyncAspects_vEst}
\resizebox{.5\columnwidth}{!}{
%
%
\definecolor{colorSync}{rgb}{\SimPlotsSyncColor}%
\begin{tikzpicture}

\pgfplotsset{every x tick label/.append style={font=\tiny, yshift=0.3ex}}
\pgfplotsset{every y tick label/.append style={font=\tiny, xshift=0.4ex}}

\newcommand\MyMarkRepeat{4}

\begin{axis}[%
width=\SimPlotsInnerWidth,
height=\SimPlotsInnerHeight,
at={(0,0)},
scale only axis,
xmode=log,
ymode=log,
xmin=.1,
xmax=10,
xtick={.1, 1, 10},
xlabel={\footnotesize{inter-node distance $d$ [m]}},
ymin=.01,
ymax=3,
ylabel={\footnotesize{position RMSE [m]}},
x label style={at={(axis description cs:0.5,-0.09)},anchor=north},
y label style={at={(axis description cs:-0.135,0.5)},anchor=south},
xmajorgrids=\SimPlotsGrid,
ymajorgrids=\SimPlotsGrid,
grid style={line width=\SimPlotsGridLineWidth, draw=\SimPlotsGridColor, solid},
legend style={at={(.12,.9)}, anchor=north west, legend cell align=center, draw=white!15!black}
]

\addplot [color=colorSync, \SimPlotsDDLineStyle, line width=\SimPlotsLineWidth, mark size=\SimPlotsDDMarkSize, mark=\SimPlotsDDMarker, mark options={solid, colorSync},
forget plot, mark repeat=\MyMarkRepeat, mark phase=1]
  table[row sep=crcr]{%
0.1	0.0583208350221823\\
0.125892541179417	0.0582143568363382\\
0.158489319246111	0.0583895412666307\\
0.199526231496888	0.0584340870114936\\
0.251188643150958	0.0579500808152773\\
0.316227766016838	0.0585561799366097\\
0.398107170553497	0.0583862226977865\\
0.501187233627272	0.0584977349045947\\
0.630957344480193	0.0587867895150245\\
0.794328234724281	0.0576555970748503\\
1	0.0582664424538674\\
1.25892541179417	0.0588697188908174\\
1.58489319246111	0.0591049990749522\\
1.99526231496888	0.0590077757850898\\
2.51188643150958	0.0589795786124016\\
3.16227766016838	0.0583536195215578\\
3.98107170553497	0.0588355328552575\\
5.01187233627272	0.0585614275163513\\
6.30957344480193	0.0574579778076974\\
7.94328234724282	0.0572912718204066\\
10	0.0581534596221786\\
};

\addplot [color=colorSync, \SimPlotsPWALineStyle, line width=\SimPlotsLineWidth, mark size=\SimPlotsPWAMarkSize, mark=\SimPlotsPWAMarker, mark options={solid, colorSync},
forget plot, mark repeat=\MyMarkRepeat, mark phase=1]
  table[row sep=crcr]{%
0.1	0.0583282160452522\\
0.125892541179417	0.0582127148338085\\
0.158489319246111	0.0584101577173279\\
0.199526231496888	0.0584618460176516\\
0.251188643150958	0.0580081138795803\\
0.316227766016838	0.0586517967025961\\
0.398107170553497	0.0587069942723535\\
0.501187233627272	0.0594356063194676\\
0.630957344480193	0.0608145603787133\\
0.794328234724281	0.0627614996276253\\
1	0.0704443045197999\\
1.25892541179417	0.0851525728502267\\
1.58489319246111	0.113830904772656\\
1.99526231496888	0.16855631700506\\
2.51188643150958	0.256676592128621\\
3.16227766016838	0.396489815057274\\
3.98107170553497	0.6319499976876\\
5.01187233627272	1.01648191223715\\
6.30957344480193	1.63487522105633\\
7.94328234724282	2.58107542538266\\
10	4.05304203241497\\
};

\addplot [color=colorSync, \SimPlotsTAULineStyle, line width=\SimPlotsLineWidth, mark size=\SimPlotsTAUMarkSize, mark=\SimPlotsTAUMarker, mark options={solid, colorSync},
forget plot, mark repeat=\MyMarkRepeat, mark phase=1]
  table[row sep=crcr]{%
0.1	0.0175093500482343\\
0.125892541179417	0.0176195360574257\\
0.158489319246111	0.017677344224762\\
0.199526231496888	0.0174779588042419\\
0.251188643150958	0.0174074740511637\\
0.316227766016838	0.0176288704592767\\
0.398107170553497	0.017474295058957\\
0.501187233627272	0.0176222402671174\\
0.630957344480193	0.0177074044054144\\
0.794328234724281	0.0174787249704072\\
1	0.0175702607052513\\
1.25892541179417	0.0178222584836754\\
1.58489319246111	0.0177645303277815\\
1.99526231496888	0.0177872189354151\\
2.51188643150958	0.017865149580109\\
3.16227766016838	0.0178980596052217\\
3.98107170553497	0.0178630366832831\\
5.01187233627272	0.0180562909691889\\
6.30957344480193	0.0180990605450354\\
7.94328234724282	0.0186733484607908\\
10	0.0191791479759765\\
};

\addplot [color=black, \SimPlotsDDLineStyle, line width=\SimPlotsLineWidth, mark size=\SimPlotsDDMarkSize, mark=\SimPlotsDDMarker, mark options={solid, black},
mark repeat=\MyMarkRepeat, mark phase=3]
  table[row sep=crcr]{%
0.1	0.0626756343175576\\
0.125892541179417	0.0624329022857151\\
0.158489319246111	0.0626106271117253\\
0.199526231496888	0.061865562695877\\
0.251188643150958	0.0628105040496404\\
0.316227766016838	0.0631231422965785\\
0.398107170553497	0.0626585336264435\\
0.501187233627272	0.0629471549597483\\
0.630957344480193	0.0631343141905639\\
0.794328234724281	0.0620734053443119\\
1	0.0624514940119277\\
1.25892541179417	0.0634575892157286\\
1.58489319246111	0.0636376668612395\\
1.99526231496888	0.0635139535056496\\
2.51188643150958	0.0633386223867407\\
3.16227766016838	0.0637236815338339\\
3.98107170553497	0.0639379697325471\\
5.01187233627272	0.0651516293411097\\
6.30957344480193	0.0650791765353898\\
7.94328234724282	0.0671864741494272\\
10	0.0709566753749985\\
};
\addlegendentry{\SimPlotsLegendDD}

\addplot [color=black, \SimPlotsPWALineStyle, line width=\SimPlotsLineWidth, mark size=\SimPlotsPWAMarkSize, mark=\SimPlotsPWAMarker, mark options={solid, black},
mark repeat=\MyMarkRepeat, mark phase=3]
  table[row sep=crcr]{%
0.1	0.0626831757667116\\
0.125892541179417	0.0624405994796463\\
0.158489319246111	0.0625997402934159\\
0.199526231496888	0.0619015987700549\\
0.251188643150958	0.0628431020753237\\
0.316227766016838	0.0631495399101271\\
0.398107170553497	0.0627882056128971\\
0.501187233627272	0.0632646877634033\\
0.630957344480193	0.0639240674783606\\
0.794328234724281	0.0643652150544446\\
1	0.0679879847117881\\
1.25892541179417	0.0754336145808155\\
1.58489319246111	0.0922048670324079\\
1.99526231496888	0.12561368598692\\
2.51188643150958	0.180869343760652\\
3.16227766016838	0.276233294790559\\
3.98107170553497	0.432618083677182\\
5.01187233627272	0.704449406305999\\
6.30957344480193	1.13489253449459\\
7.94328234724282	1.82532309697287\\
10	2.88771974137183\\
};
\addlegendentry{\SimPlotsLegendPWA}

\addplot [color=black, \SimPlotsTAULineStyle, line width=\SimPlotsLineWidth, mark size=\SimPlotsTAUMarkSize, mark=\SimPlotsTAUMarker, mark options={solid, black},
mark repeat=\MyMarkRepeat, mark phase=3]
  table[row sep=crcr]{%
0.1	0.0271186290921377\\
0.125892541179417	0.0268798366273928\\
0.158489319246111	0.0272794125168939\\
0.199526231496888	0.0268964583967539\\
0.251188643150958	0.0269743487326721\\
0.316227766016838	0.0273384612546708\\
0.398107170553497	0.0266482755023575\\
0.501187233627272	0.0269374274302137\\
0.630957344480193	0.0273731906886412\\
0.794328234724281	0.0274868535237601\\
1	0.0273451027270414\\
1.25892541179417	0.0271654925425195\\
1.58489319246111	0.0277212169805774\\
1.99526231496888	0.0277218176778904\\
2.51188643150958	0.0276779598602726\\
3.16227766016838	0.0279724979558206\\
3.98107170553497	0.0281369870024853\\
5.01187233627272	0.0285664586425639\\
6.30957344480193	0.0293586929639166\\
7.94328234724282	0.0315843692133318\\
10	0.03382488519776\\
};
\addlegendentry{\SimPlotsLegendTAU}

\end{axis}
\end{tikzpicture}%
}}
\caption{Estimation RMSE versus inter-node distance $d$. The experiment assumes $\NObs = 3$ observers and $K_1 = K_2 = K_3 = 4$ MPCs. The light-colored (cyan) graphs relate to cases with precise a-priori time synchronization.}
\label{fig:RMSE_vs_d_SyncAspects}
\centering\!\!\!
\subfloat[distance estimation]{\centering\label{fig:RMSE_vs_d_AssocAspects_dEst}
\resizebox{.48\columnwidth}{!}{
%
%
\definecolor{colorSync}{rgb}{\SimPlotsSyncColor}%
\begin{tikzpicture}

\pgfplotsset{every x tick label/.append style={font=\tiny, yshift=0.3ex}}
\pgfplotsset{every y tick label/.append style={font=\tiny, xshift=0.4ex}}

\newcommand\MyMarkRepeat{4}
\newcommand\MyMarkPhase{3}

\begin{axis}[%
width=\SimPlotsInnerWidth,
height=\SimPlotsInnerHeight,
at={(0,0)},
scale only axis,
xmode=log,
ymode=log,
xmin=.1,
xmax=10,
xtick={.1, 1, 10},
xlabel={\small{inter-node distance $d$ [m]}},
ymin=.01,
ymax=2,
ytick={.01, .1, 1, 10},
ylabel={\footnotesize{distance RMSE [m]}},
x label style={at={(axis description cs:0.5,-0.09)},anchor=north},
y label style={at={(axis description cs:-0.135,0.5)},anchor=south},
xmajorgrids=\SimPlotsGrid,
ymajorgrids=\SimPlotsGrid,
grid style={line width=\SimPlotsGridLineWidth, draw=\SimPlotsGridColor, solid},
legend style={at={(.95,.05)}, anchor=south east, legend cell align=center, draw=white!15!black},
legend columns=1,
transpose legend
]

\addplot [color=black, \SimPlotsMVUELineStyle, line width=\SimPlotsLineWidth, mark size=\SimPlotsMVUEMarkSize, mark=\SimPlotsMVUEMarker, mark options={solid, rotate=\SimPlotsMVUERotate, black},
mark repeat=\MyMarkRepeat, mark phase=\MyMarkPhase]
  table[row sep=crcr]{%
0.1	0.0720431710400428\\
0.125892541179417	0.0669702967812523\\
0.158489319246111	0.0629502703658703\\
0.199526231496888	0.0598167183690343\\
0.251188643150958	0.0601366260590916\\
0.316227766016838	0.0630239591290086\\
0.398107170553497	0.0663877788791542\\
0.501187233627272	0.0755331250078776\\
0.630957344480193	0.0853756344402675\\
0.794328234724281	0.100995072598569\\
1	0.123632770510689\\
1.25892541179417	0.15515929558411\\
1.58489319246111	0.191122012710974\\
1.99526231496888	0.236582090317288\\
2.51188643150958	0.300734205050987\\
3.16227766016838	0.381501037578764\\
3.98107170553497	0.491434296143033\\
5.01187233627272	0.679310295782928\\
6.30957344480193	0.925554878926245\\
7.94328234724282	1.32529728596517\\
10	1.9091102837069\\
};
\addlegendentry{\SimPlotsLegendMVUE}

\addplot [color=black, \SimPlotsSORTLineStyle, line width=\SimPlotsLineWidth, mark size=\SimPlotsSORTMarkSize, mark=\SimPlotsSORTMarker, mark options={solid, black},
mark repeat=\MyMarkRepeat, mark phase=1]
  table[row sep=crcr]{%
0.1	0.0718819888074352\\
0.125892541179417	0.0668028361799694\\
0.158489319246111	0.0627732910302149\\
0.199526231496888	0.0597674526881637\\
0.251188643150958	0.0597665085195755\\
0.316227766016838	0.0627227153083097\\
0.398107170553497	0.06641704238742\\
0.501187233627272	0.0752322368549054\\
0.630957344480193	0.0868074111890829\\
0.794328234724281	0.103558749612379\\
1	0.128398120652383\\
1.25892541179417	0.165478735022488\\
1.58489319246111	0.208517078133353\\
1.99526231496888	0.272008980643105\\
2.51188643150958	0.354961060785114\\
3.16227766016838	0.470960106302944\\
3.98107170553497	0.659650009271651\\
5.01187233627272	0.969578015504514\\
6.30957344480193	1.42505396889669\\
7.94328234724282	2.08932265072787\\
10	3.10679929560437\\
};
\addlegendentry{\SimPlotsLegendSORT}

\addplot [color=black, \SimPlotsNALineStyle, line width=\SimPlotsLineWidth, mark size=\SimPlotsNAMarkSize, mark=\SimPlotsNAMarker, mark options={solid, black},
mark repeat=\MyMarkRepeat, mark phase=1]
  table[row sep=crcr]{%
0.1	0.0460886949924966\\
0.125892541179417	0.0454472455859701\\
0.158489319246111	0.0447195927294841\\
0.199526231496888	0.0459080196327769\\
0.251188643150958	0.0502900449185892\\
0.316227766016838	0.0573402976180379\\
0.398107170553497	0.0651998875015067\\
0.501187233627272	0.0799295825523364\\
0.630957344480193	0.0961582007601956\\
0.794328234724281	0.121176618273445\\
1	0.155357109534009\\
1.25892541179417	0.2043205916866\\
1.58489319246111	0.264302717234237\\
1.99526231496888	0.353942341706585\\
2.51188643150958	0.466505689128714\\
3.16227766016838	0.620241277901979\\
3.98107170553497	0.848083522275528\\
5.01187233627272	1.21666533675164\\
6.30957344480193	1.72250089852215\\
7.94328234724282	2.43472830357887\\
10	3.49928998452746\\
};
\addlegendentry{\SimPlotsLegendNA}

\end{axis}
\end{tikzpicture}%
}}\!\!\!
\subfloat[position estimation]{\centering\label{fig:RMSE_vs_d_AssocAspects_vEst}
\resizebox{.48\columnwidth}{!}{
%
%
\definecolor{colorSync}{rgb}{\SimPlotsSyncColor}%
\begin{tikzpicture}

\pgfplotsset{every x tick label/.append style={font=\tiny, yshift=0.3ex}}
\pgfplotsset{every y tick label/.append style={font=\tiny, xshift=0.4ex}}

\newcommand\MyMarkRepeat{4}
\newcommand\MyMarkPhase{3}

\begin{axis}[%
width=\SimPlotsInnerWidth,
height=\SimPlotsInnerHeight,
at={(0,0)},
scale only axis,
xmode=log,
ymode=log,
xmin=.1,
xmax=10,
xtick={.1, 1, 10},
xlabel={\small{inter-node distance $d$ [m]}},
ymin=.01,
ymax=2,
ytick={.01, .1, 1, 10},
ylabel={\footnotesize{position RMSE [m]}},
x label style={at={(axis description cs:0.5,-0.09)},anchor=north},
y label style={at={(axis description cs:-0.135,0.5)},anchor=south},
xmajorgrids=\SimPlotsGrid,
ymajorgrids=\SimPlotsGrid,
grid style={line width=\SimPlotsGridLineWidth, draw=\SimPlotsGridColor, solid},
legend style={at={(.2,.93)}, anchor=north west, legend cell align=center, draw=white!15!black}
]
\addplot [color=black, \SimPlotsDDLineStyle, line width=\SimPlotsLineWidth, mark size=\SimPlotsDDMarkSize, mark=\SimPlotsDDMarker, mark options={solid, black},
mark repeat=\MyMarkRepeat, mark phase=\MyMarkPhase]
  table[row sep=crcr]{%
0.1	0.0626756343175576\\
0.125892541179417	0.0624329022857151\\
0.158489319246111	0.0626106271117253\\
0.199526231496888	0.061865562695877\\
0.251188643150958	0.0628105040496404\\
0.316227766016838	0.0631231422965785\\
0.398107170553497	0.0626585336264435\\
0.501187233627272	0.0629471549597483\\
0.630957344480193	0.0631343141905639\\
0.794328234724281	0.0620734053443119\\
1	0.0624514940119277\\
1.25892541179417	0.0634575892157286\\
1.58489319246111	0.0636376668612395\\
1.99526231496888	0.0635139535056496\\
2.51188643150958	0.0633386223867407\\
3.16227766016838	0.0637236815338339\\
3.98107170553497	0.0639379697325471\\
5.01187233627272	0.0651516293411097\\
6.30957344480193	0.0650791765353898\\
7.94328234724282	0.0671864741494272\\
10	0.0709566753749985\\
};
\addlegendentry{\SimPlotsLegendDD}

\addplot [color=black, \SimPlotsDDNLineStyle, line width=\SimPlotsLineWidth, mark size=\SimPlotsDDNMarkSize, mark=\SimPlotsDDNMarker, mark options={solid, black},
mark repeat=\MyMarkRepeat, mark phase=\MyMarkPhase]
  table[row sep=crcr]{%
0.1	0.0626756343175576\\
0.125892541179417	0.0624329022857151\\
0.158489319246111	0.0626106271117253\\
0.199526231496888	0.061865562695877\\
0.251188643150958	0.0628105040496404\\
0.316227766016838	0.0631231422965785\\
0.398107170553497	0.0626585336264435\\
0.501187233627272	0.0629471549597483\\
0.630957344480193	0.0631343141905639\\
0.794328234724281	0.0620734053443119\\
1	0.0624514940119277\\
1.25892541179417	0.0634575892157286\\
1.58489319246111	0.0636376668612395\\
1.99526231496888	0.0716989477807067\\
2.51188643150958	0.134854634546674\\
3.16227766016838	0.312989960503565\\
3.98107170553497	0.486876778596048\\
5.01187233627272	1.08910713979828\\
6.30957344480193	195667265141050\\
7.94328234724282	84.7184133138888\\
10	136628799191049\\
};
\addlegendentry{\SimPlotsLegendDDN}

\addplot [color=black, \SimPlotsTAULineStyle, line width=\SimPlotsLineWidth, mark size=\SimPlotsTAUMarkSize, mark=\SimPlotsTAUMarker, mark options={solid, black},
mark repeat=\MyMarkRepeat, mark phase=1]
  table[row sep=crcr]{%
0.1	0.0271186290921377\\
0.125892541179417	0.0268798366273928\\
0.158489319246111	0.0272794125168939\\
0.199526231496888	0.0268964583967539\\
0.251188643150958	0.0269743487326721\\
0.316227766016838	0.0273384612546708\\
0.398107170553497	0.0266482755023575\\
0.501187233627272	0.0269374274302137\\
0.630957344480193	0.0273731906886412\\
0.794328234724281	0.0274868535237601\\
1	0.0273451027270414\\
1.25892541179417	0.0271654925425195\\
1.58489319246111	0.0277212169805774\\
1.99526231496888	0.0277218176778904\\
2.51188643150958	0.0276779598602726\\
3.16227766016838	0.0279724979558206\\
3.98107170553497	0.0281369870024853\\
5.01187233627272	0.0285664586425639\\
6.30957344480193	0.0293586929639166\\
7.94328234724282	0.0315843692133318\\
10	0.03382488519776\\
};
\addlegendentry{\SimPlotsLegendTAU}

\addplot [color=black, \SimPlotsTNALineStyle, line width=\SimPlotsLineWidth, mark size=\SimPlotsTNAMarkSize, mark=\SimPlotsTNAMarker, mark options={solid, black},
mark repeat=\MyMarkRepeat, mark phase=1]
  table[row sep=crcr]{%
0.1	0.0271186290921377\\
0.125892541179417	0.0268798366273928\\
0.158489319246111	0.0272794125168939\\
0.199526231496888	0.0268964583967539\\
0.251188643150958	0.0269743487326721\\
0.316227766016838	0.0273384612546708\\
0.398107170553497	0.0266482755023575\\
0.501187233627272	0.0269374274302137\\
0.630957344480193	0.0273731906886412\\
0.794328234724281	0.0274868535237601\\
1	0.0273451027270414\\
1.25892541179417	0.0271654925425195\\
1.58489319246111	0.0277212169805774\\
1.99526231496888	0.0430735122401897\\
2.51188643150958	0.102081854226323\\
3.16227766016838	0.219618473200556\\
3.98107170553497	0.58851326144163\\
5.01187233627272	1.73998599390415\\
6.30957344480193	4.06669651098493\\
7.94328234724282	7.07355175815538\\
10	9.79218108818038\\
};
\addlegendentry{\SimPlotsLegendTNA}

\end{axis}
\end{tikzpicture}%
}}
\caption{Estimation RMSE versus $d$; known versus unknown MPC association.}
\label{fig:RMSE_vs_d_AssocAspects}
\centering\!\!\!\!
\subfloat[distance estimation]{\centering\label{fig:RMSEvsK_dEst}
\resizebox{.48\columnwidth}{!}{
%
%
\definecolor{colorSync}{rgb}{\SimPlotsSyncColor}%
\begin{tikzpicture}

\pgfplotsset{every x tick label/.append style={font=\footnotesize, yshift=0.1ex}}
\pgfplotsset{every y tick label/.append style={font=\footnotesize, xshift=0.1ex}}

\newcommand\MyMarkRepeat{2}

\begin{axis}[%
width=\SimPlotsInnerWidth,
height=\SimPlotsInnerHeight,
at={(0,0)},
scale only axis,
xmin=1,
xmax=12,
xtick={1, 2, 3, 4, 5, 6, 7, 8, 9, 10, 11, 12},
xlabel style={},
xlabel={\small{number of MPCs $K$}},
ymin=0,
ymax=2,
ytick={0, .5, 1, 1.5, 2},
yticklabels={0, .5, 1, 1.5, 2},
x label style={at={(axis description cs:0.5,-0.09)},anchor=north},
y label style={at={(axis description cs:-.1,0.5)},anchor=south},
ylabel style={},
ylabel={\footnotesize{distance RMSE [m]}},
xmajorgrids=\SimPlotsGrid,
ymajorgrids=\SimPlotsGrid,
grid style={line width=\SimPlotsGridLineWidth, draw=\SimPlotsGridColor, solid},
legend style={at={(.9,.9)}, anchor=north east, legend cell align=center, draw=white!15!black},
legend columns=2,
transpose legend
]

\addplot [color=colorSync, \SimPlotsMVUELineStyle, line width=\SimPlotsLineWidth, mark size=\SimPlotsMVUEMarkSize, mark=\SimPlotsMVUEMarker, mark options={solid, rotate=\SimPlotsMVUERotate, colorSync},
forget plot, mark repeat=\MyMarkRepeat, mark phase=1]
  table[row sep=crcr]{%
1	1.43746695778434\\
2	0.888654355213048\\
3	0.651311176346261\\
4	0.517716261164868\\
5	0.432300884712479\\
6	0.36483044197617\\
7	0.320711528935393\\
8	0.282453507062262\\
9	0.260563966680597\\
10	0.237686304134487\\
11	0.225135136803974\\
12	0.202696084064544\\
};

\addplot [color=colorSync, \SimPlotsSORTLineStyle, line width=\SimPlotsLineWidth, mark size=\SimPlotsSORTMarkSize, mark=\SimPlotsSORTMarker, mark options={solid, colorSync},
forget plot, mark repeat=\MyMarkRepeat, mark phase=2]
  table[row sep=crcr]{%
1	1.43746695778434\\
2	0.903282679720185\\
3	0.69485013878796\\
4	0.581682319845569\\
5	0.520142286958601\\
6	0.478853416171167\\
7	0.453800954292196\\
8	0.428669585703806\\
9	0.419611406511936\\
10	0.411302701229322\\
11	0.406423298329213\\
12	0.403827193773262\\
};


\addplot [color=black, \SimPlotsMVUELineStyle, line width=\SimPlotsLineWidth, mark size=\SimPlotsMVUEMarkSize, mark=\SimPlotsMVUEMarker, mark options={solid, rotate=\SimPlotsMVUERotate, black},
mark repeat=\MyMarkRepeat, mark phase=1]
  table[row sep=crcr]{%
2	1.77742239013657\\
3	1.12220412929655\\
4	0.842023109175635\\
5	0.67997483718691\\
6	0.568341738852694\\
7	0.494802787500211\\
8	0.432113072932829\\
9	0.386228450366072\\
10	0.355232867604254\\
11	0.32700115012686\\
12	0.295984814942279\\
};
\addlegendentry{\SimPlotsLegendMVUE}

\addplot [color=black, \SimPlotsSORTLineStyle, line width=\SimPlotsLineWidth, mark size=\SimPlotsSORTMarkSize, mark=\SimPlotsSORTMarker, mark options={solid, black},
mark repeat=\MyMarkRepeat, mark phase=2]
  table[row sep=crcr]{%
2	1.74479929771351\\
3	1.12363049969177\\
4	0.873097299953449\\
5	0.74818621769587\\
6	0.673177892666733\\
7	0.628961547995792\\
8	0.592802884469639\\
9	0.569970156401917\\
10	0.565069232637426\\
11	0.557015425440892\\
12	0.551524975918593\\
};
\addlegendentry{\SimPlotsLegendSORT}


\end{axis}
\end{tikzpicture}%
}}\!
\subfloat[position estimation]{\centering\label{fig:RMSEvsK_vEst}
\resizebox{.48\columnwidth}{!}{
%
%
\definecolor{colorSync}{rgb}{\SimPlotsSyncColor}%
\begin{tikzpicture}

\pgfplotsset{every x tick label/.append style={font=\footnotesize, yshift=0.1ex}}
\pgfplotsset{every y tick label/.append style={font=\tiny, xshift=0.4ex}}

\newcommand\MyMarkRepeat{2}

\begin{axis}[%
width=\SimPlotsInnerWidth,
height=\SimPlotsInnerHeight,
at={(0,0)},
scale only axis,
ymode=log,
xmin=1,
xmax=12,
xtick={1, 2, 3, 4, 5, 6, 7, 8, 9, 10, 11, 12},
xlabel={\small{number of MPCs $K$}},
ymin=.01,
ymax=3,
ylabel={\footnotesize{position RMSE [m]}},
x label style={at={(axis description cs:0.5,-0.09)},anchor=north},
y label style={at={(axis description cs:-0.135,0.5)},anchor=south},
ylabel style={},
ylabel={\footnotesize{position RMSE [m]}},
axis background/.style={fill=white},
xmajorgrids=\SimPlotsGrid,
ymajorgrids=\SimPlotsGrid,
grid style={line width=\SimPlotsGridLineWidth, draw=\SimPlotsGridColor, solid},
legend style={at={(.9,.9)}, anchor=north east, legend cell align=center, draw=white!15!black}
]

\addplot [color=colorSync, \SimPlotsDDLineStyle, line width=\SimPlotsLineWidth, mark size=\SimPlotsDDMarkSize, mark=\SimPlotsDDMarker, mark options={solid, colorSync},
forget plot, mark repeat=\MyMarkRepeat, mark phase=1]
  table[row sep=crcr]{%
3	1.83186454747658\\
4	0.280632783577881\\
5	0.137115321110086\\
6	0.109704660238369\\
7	0.0946612402355909\\
8	0.0862666891428081\\
9	0.0796426195470509\\
10	0.0746144215172485\\
11	0.0703321631393936\\
12	0.0668903281587872\\
};

\addplot [color=colorSync, \SimPlotsTAULineStyle, line width=\SimPlotsLineWidth, mark size=\SimPlotsTAUMarkSize, mark=\SimPlotsTAUMarker, mark options={solid, colorSync},
forget plot, mark repeat=\MyMarkRepeat, mark phase=2]
  table[row sep=crcr]{%
1	0.0393734650040037\\
2	0.0382880908781679\\
3	0.0336636744317978\\
4	0.0307575198282548\\
5	0.0283692427899378\\
6	0.0268924181233614\\
7	0.025079497282518\\
8	0.0239872238866397\\
9	0.022770459868963\\
10	0.0218200265274829\\
11	0.0211513294292161\\
12	0.0203105702102717\\
};

\addplot [color=black, \SimPlotsDDLineStyle, line width=\SimPlotsLineWidth, mark size=\SimPlotsDDMarkSize, mark=\SimPlotsDDMarker, mark options={solid, black},
mark repeat=\MyMarkRepeat, mark phase=1]
  table[row sep=crcr]{%
4	2.25078635456067\\
5	0.388020362932769\\
6	0.162085360719827\\
7	0.11834360944767\\
8	0.102176406047825\\
9	0.0900297839157447\\
10	0.0827793013647685\\
11	0.0766992225850969\\
12	0.0720362920340396\\
};
\addlegendentry{\SimPlotsLegendDD}

\addplot [color=black, \SimPlotsTAULineStyle, line width=\SimPlotsLineWidth, mark size=\SimPlotsTAUMarkSize, mark=\SimPlotsTAUMarker, mark options={solid, black},
mark repeat=\MyMarkRepeat, mark phase=1]
  table[row sep=crcr]{%
5	0.0438013931490329\\
6	0.0407662701605981\\
7	0.0378095314563031\\
8	0.0361947112762177\\
9	0.0344026624340758\\
10	0.0330074171091227\\
11	0.0318308410551324\\
12	0.0304638604254529\\
};
\addlegendentry{\SimPlotsLegendTAU}

\end{axis}
\end{tikzpicture}%
}}
\caption{Estimation RMSE versus the number of MPCs $K$ for $\NObs = 1$ observers and $d = 2.5\unit{m}$ distance.
Cyan graphs: precise a-priori time synchronization.}
\label{fig:RMSEvsK_d2m}
\centering
\vspace{2mm}
\resizebox{.75\columnwidth}{!}{
%
%
\definecolor{colorSync}{rgb}{\SimPlotsSyncColor}%
\begin{tikzpicture}

\pgfplotsset{every x tick label/.append style={font=\footnotesize, yshift=0.1ex}}
\pgfplotsset{every y tick label/.append style={font=\footnotesize, xshift=0.1ex}}

\newcommand\MyMarkRepeat{2}

\begin{axis}[%
width=.92\columnwidth,
height=.4\columnwidth,
at={(0,0)},
scale only axis,
xmin=0,
xmax=18,
xtick distance=3,
xlabel={\normalsize{direction error standard deviation $\sigma_{\mathrm{dir}}$ [degree]}},
ymin=0,
ymax=1,
ytick={0, .2, .4, .6, .8, 1},
ylabel={\small{position RMSE [m]}},
axis background/.style={fill=white},
xmajorgrids=\SimPlotsGrid,
ymajorgrids=\SimPlotsGrid,
grid style={line width=\SimPlotsGridLineWidth, draw=\SimPlotsGridColor, solid},
legend style={at={(.35,.73)}, anchor=north west, legend cell align=center, draw=white!15!black}
]
\addplot [color=black, \SimPlotsDDLineStyle, line width=\SimPlotsLineWidth, mark size=\SimPlotsDDMarkSize, mark=\SimPlotsDDMarker, mark options={solid, black},
mark repeat=\MyMarkRepeat, mark phase=2]
  table[row sep=crcr]{%
0	0.0629001009753723\\
0.5	0.063079924562694\\
1	0.0638500147392608\\
1.5	0.0638520828537768\\
2	0.0644799774814773\\
3	0.0666294084125857\\
4	0.0693067637733523\\
5	0.0728496098758858\\
6	0.077166375584886\\
7	0.0812998204835806\\
8	0.0863413566731083\\
9	0.0912596798486168\\
10	0.0981302919930769\\
11	0.103740478999269\\
12	0.110087587342489\\
13	0.117101206622563\\
14	0.123834827122089\\
15	0.130687677937129\\
16	0.137238368404758\\
17	0.145654142625755\\
18	0.153533788922082\\
};
\addlegendentry{\SimPlotsLegendDD}

\addplot [color=black, \SimPlotsDDNLineStyle, line width=\SimPlotsLineWidth, mark size=\SimPlotsDDNMarkSize, mark=\SimPlotsDDNMarker, mark options={solid, black},
mark repeat=\MyMarkRepeat, mark phase=2]
  table[row sep=crcr]{%
0	0.0629001009753723\\
0.5	0.0630798636208141\\
1	0.0638499111963174\\
1.5	0.0638521792907865\\
2	0.064479966139432\\
3	0.0667492927788638\\
4	0.0694393106483852\\
5	0.0737468595232807\\
6	0.0828807407744028\\
7	0.0946264228900855\\
8	0.136686199059787\\
9	0.171965789280476\\
10	0.210543604538951\\
11	0.259194339077922\\
12	0.379093548498264\\
13	0.413460825629461\\
14	0.512805420744582\\
15	0.61778440274616\\
16	1.43883629864188\\
17	1.02018908807205\\
18	1.74954133605103\\
};
\addlegendentry{\SimPlotsLegendDDN}

\addplot [color=black, \SimPlotsTAULineStyle, line width=\SimPlotsLineWidth, mark size=\SimPlotsTAUMarkSize, mark=\SimPlotsTAUMarker, mark options={solid, black},
mark repeat=\MyMarkRepeat, mark phase=1]
  table[row sep=crcr]{%
0	0.0272482471530772\\
0.5	0.0938691888734578\\
1	0.211797213160785\\
1.5	0.334238204135749\\
2	0.443505007079557\\
3	0.608875410387709\\
4	0.716085622408154\\
5	0.790990136990126\\
6	0.843330242843899\\
7	0.889349183579765\\
8	0.935124595668272\\
9	0.977131761507324\\
10	1.01558349336221\\
11	1.05757708996893\\
12	1.09444401037953\\
13	1.13600485894576\\
14	1.17348845643732\\
15	1.22038333308299\\
16	1.25767528091682\\
17	1.28831744407088\\
18	1.33209899214906\\
};
\addlegendentry{\SimPlotsLegendTAU}


\end{axis}
\end{tikzpicture}%
}
\vspace{-2mm}
\caption{Position estimation RMSE versus the standard deviation of angular measurement errors on the MPC directions.}
\label{fig:RMSE_vs_sigmaAngular}
\end{figure}

We proceed with the primary numerical performance results.
We study the distance estimation accuracy versus the true inter-node distance $d$ by means of \Cref{fig:RMSE_vs_d_SyncAspects_dEst}.
For very small $d$, the RMSEs are dominated by the largest $\cd\sigma_{k,o}$-values, i.e. the RMSE is bandwidth- and SINR-limited. In this regime the \SimPlotsLegendMLE{} \Cref{eq:distMLE_general}, which does account for these error statistics, beats the \SimPlotsLegendMVUE{}. For $d \gg \max \cd\sigma_{k,o}$ on the other hand, the RMSE is instead dominated by the unknown (and randomly modeled) MPC directions. The RMSE becomes linear in $d$ and agrees very well with the closed-form expressions \eqref{eq:stdMVUE},\eqref{eq:stdMVUESync}. In this regime the unbiased \SimPlotsLegendMVUE{} beats the biased \SimPlotsLegendMLE{}.

In \Cref{fig:RMSE_vs_d_SyncAspects_vEst}, the position estimator $\d\LSE\AssViaDiff$ (\SimPlotsLegendDD{}) exhibits a near-constant RMSE of about $60\unit{mm}$, dominated by the largest $\cd\sigma_{k,o}$-values. The behavior is in accordance with the analytical prediction \Cref{eq:RMSEdisplLSEViaDiff}, e.g., $\f{3 \cdot 71\unit{mm}}{\sqrt{12}} \approx 61.5\unit{mm}$.
As expected, the PWA-induced error is insignificant if and only if $d$ is much smaller than the traveled path lengths.
The \SimPlotsLegendTAU{} scheme achieves an RMSE of about $27\unit{mm}$ (sync.: $18\unit{mm}$) and thus beats the DD scheme. This is due to processing the information in both $\delayMeasB[ko]$ and $\delayMeasA[ko]$ instead of just $\delayMeasB[ko]-\delayMeasA[ko]$, which improves the handling of measurement errors. We will however find that \SimPlotsLegendTAU{} has serious problems in less ideal conditions.

We expect that the schemes with unknown MPC association (\SimPlotsLegendSORT{}, \SimPlotsLegendNA{}, \SimPlotsLegendDDN{}) will run into problems unless $d \ll \cd\tau_\mathrm{RMS} \approx 9\unit{m}$.
This behavior is evaluated in \Cref{fig:RMSE_vs_d_AssocAspects_dEst} for distance estimation.
The simple \SimPlotsLegendSORT{} scheme, which just applies $\hat d\MVUE$ after associating the MPCs by ascending delay order, surprisingly outperforms the sophisticated \SimPlotsLegendNA{}. One reason is the lack of bias-correction in $\hat d\MLE\NoAssAsyn$, which could be addressed by future work.
Regarding position estimation, we find in \Cref{fig:RMSE_vs_d_AssocAspects_vEst} that the association scheme from \Cref{sec:EstimateRelLocNoAssoc}, implemented by \SimPlotsLegendDDN{} and \SimPlotsLegendTNA{}, works flawlessly up to about half the observer distance (and then breaks down).

The experiment in \Cref{fig:RMSEvsK_d2m} studies the effect of the number of MPCs $K$ on the estimation accuracy. It considers only one observer ($\NObs=1$). Clearly, all estimators benefit from an increasing $K$. 
The MVUE is in accordance with the closed-form expressions \eqref{eq:stdMVUE},\eqref{eq:stdMVUESync}.
The gap between SORT and MVUE widens with increasing $K$ because of the decreasing probability that delay-sorting gives the correct association. SORT has a negative bias (it tends to underestimate) which becomes significant for large $K$ or large $d$.
For $\d\LSE\AssViaDiff$ (\SimPlotsLegendDD{}) it seems particularly fruitful to exceed $\KMin$ by a little margin, to ensure that $\E\E\Tr$ in \Cref{eq:displLSEAssViaDiff} is well-conditioned.


So far we assumed that the position estimators have perfect knowledge of the MPC directions. We will now consider measurements that deviate from the true value by an angle $\alpha \sim \calN(0,\sigma_\text{dir}^2)$. Each unit vector $\dirVectX[k o] \in \bbR^3$ is sampled uniformly from the cone defined by $\alpha$. The performance implications are shown in \Cref{fig:RMSE_vs_sigmaAngular}. 
We find that the \SimPlotsLegendTAU{} scheme deteriorates heavily. This is because the entries $\dirVectB[k,o]\!-\dirVectA[k,o]$ of matrix ${\bf G}$ in \Cref{eq:EstimateRelLocViaTauDetails} are extremely susceptible to inaccurate directions. In simpler terms, it is intuitive that the underlying property $\d = \cd\delayTrueB[k,o] \dirVectB[k,o] - \cd\delayTrueA[k,o] \dirVectA[k,o]$
from \Cref{eq:VectorEquality} relies on very accurate knowledge of $\dirVectA[k,o]$ and $\dirVectB[k,o]$.
A very important observation is that the \SimPlotsLegendDD{} scheme performs robustly even with vastly inaccurate direction measurements.
It is based on the projection property \Cref{eq:ProjectionEquality} and thus avoids the above problem.
Another pleasant observation is that the MPC association scheme of \SimPlotsLegendDDN{} performs flawlessly up to $6^\circ$ error level.
This advantage together with the large usable $d$ (cf. \Cref{fig:RMSE_vs_d_AssocAspects_vEst}) makes the \SimPlotsLegendDDN{} scheme an important cornerstone for the use of the proposed paradigm in non-idealistic conditions.


While the presented estimators do not rely on LOS conditions, it is still helpful to have LOS paths to the observers. The positive effects are: (i) additional paths increase $K$ and (ii) the LOS delays can be measured with high accuracy due to their high SINR. Both effects improve the estimation accuracy. A numerical evaluation is presented in \Cref{fig:EffectNLOS}. We find that the RMSE reduction from effect (i) is significant while that from (ii) is not. The reason for latter is that, in the SINR-limited regime, the RMSE is still dominated by the NLOS-path MPCs.

\renewcommand\SimPlotsInnerHeight{.35\columnwidth} 
\renewcommand\SimPlotsInnerWidth{.5\columnwidth}  

\begin{figure}[!ht]
\vspace{-2mm}
\centering\!\!
\subfloat[distance MVUE]{\centering\label{fig:EffectNLOS_dEst}
\resizebox{.48\columnwidth}{!}{
%
%
\begin{tikzpicture}

\pgfplotsset{every x tick label/.append style={font=\footnotesize}}
\pgfplotsset{every y tick label/.append style={font=\footnotesize, xshift=0.3ex}}

\begin{axis}[%
width=\SimPlotsInnerWidth,
height=\SimPlotsInnerHeight,
at={(0,0)},
scale only axis,
xmin=0.1,
xmax=1,
xtick={0.1, 0.4, 0.7, 1},
xticklabels={0.1, 0.4, 0.7, 1},
xlabel={\small{inter-node distance $d$ [m]}},
ylabel={\footnotesize{distance RMSE [m]}},
ymin=0.04,
ymax=0.16,
ytick={.05, .1, .15},
yticklabels={.05, .1, .15},
axis background/.style={fill=white},
x label style={at={(axis description cs:0.5,-0.1)},anchor=north},
y label style={at={(axis description cs:-0.08,0.5)},anchor=south},
legend style={at={(0.09,1)}, anchor=north west, legend cell align=left, align=left, draw=white!15!black}
]

\addplot [color=black, line width=1.0pt]
  table[row sep=crcr]{%
0.1	0.0738607550244412\\
0.125	0.0704585932308808\\
0.15	0.0684503069180795\\
0.175	0.0681508351346355\\
0.2	0.0683411966664272\\
0.225	0.0686716861994616\\
0.25	0.0704992759098584\\
0.275	0.0724555184108879\\
0.3	0.0742118333997921\\
0.325	0.0762588650027344\\
0.35	0.0787339359807862\\
0.375	0.081448569279419\\
0.4	0.0837001988504827\\
0.5	0.0960375236712792\\
0.6	0.109707531924705\\
0.7	0.1232875427498\\
0.8	0.13754000099297\\
0.9	0.152658724957337\\
1	0.166455340525\\
};
\addlegendentry{$K_o = 3$, NLOS}

\addplot [color=black, dashed, line width=1.0pt]
  table[row sep=crcr]{%
0.1	0.0748253742204674\\
0.125	0.0702661431677012\\
0.15	0.0669662136794472\\
0.175	0.0646822051450085\\
0.2	0.0632230471899263\\
0.225	0.063005236794797\\
0.25	0.0627929529935536\\
0.275	0.0626038062372462\\
0.3	0.0641360567524804\\
0.325	0.0646869220010921\\
0.35	0.0651475078013962\\
0.375	0.0669210300948446\\
0.4	0.0684174558624749\\
0.5	0.0762703538125408\\
0.6	0.0837298727237397\\
0.7	0.0938166840284984\\
0.8	0.103073669482028\\
0.9	0.113266885655705\\
1	0.124064313086867\\
};
\addlegendentry{$K_o = 4$, NLOS}

\addplot [color=green, line width=1.5pt]
  table[row sep=crcr]{%
0.1	0.0679928972899276\\
0.125	0.063898932376978\\
0.15	0.0610130492330165\\
0.175	0.0595645774433765\\
0.2	0.0583834676870143\\
0.225	0.0579427291150617\\
0.25	0.0585937757466939\\
0.275	0.0589294376948816\\
0.3	0.0597086970132327\\
0.325	0.060378595676281\\
0.35	0.0622292432772452\\
0.375	0.0632503213579688\\
0.4	0.0650339298445757\\
0.5	0.0728489219792008\\
0.6	0.0820017209964226\\
0.7	0.0915717446069305\\
0.8	0.101317921654274\\
0.9	0.111509041317652\\
1	0.123026468081036\\
};
\addlegendentry{$K_o = 4$, LOS}

\end{axis}
\end{tikzpicture}%
}}\!\!
\subfloat[position LSE]{\centering\label{fig:EffectNLOS_vEst}
\resizebox{.48\columnwidth}{!}{
%
%
\begin{tikzpicture}

\pgfplotsset{every x tick label/.append style={font=\footnotesize}}
\pgfplotsset{every y tick label/.append style={font=\footnotesize, xshift=0.3ex}}

\begin{axis}[%
width=\SimPlotsInnerWidth,
height=\SimPlotsInnerHeight,
at={(0,0)},
scale only axis,
xmin=0.1,
xmax=1,
xtick={0.1, 0.4, 0.7, 1},
xticklabels={0.1, 0.4, 0.7, 1},
xlabel={\small{inter-node distance $d$ [m]}},
ymin=0.04,
ymax=0.16,
ytick={.05, .1, .15},
yticklabels={.05, .1, .15},
ylabel={\footnotesize{position RMSE [m]}},
axis background/.style={fill=white},
x label style={at={(axis description cs:0.5,-0.1)},anchor=north},
y label style={at={(axis description cs:-0.08,0.5)},anchor=south},
legend style={at={(0.5,1)}, anchor=north, legend cell align=left, align=left, draw=white!15!black}
]

\addplot [color=black, line width=1.0pt]
  table[row sep=crcr]{%
0.1	0.0807370529411263\\
0.125	0.0812917439496118\\
0.15	0.0812369852318478\\
0.175	0.0807205072943237\\
0.2	0.0812675034546045\\
0.225	0.0811959097176013\\
0.25	0.0813755786552457\\
0.275	0.0814834364771898\\
0.3	0.0811719700538164\\
0.325	0.0811674280807172\\
0.35	0.081462905773295\\
0.375	0.0811934623868002\\
0.4	0.0811123886080659\\
0.5	0.0812650813039371\\
0.6	0.0812972487857768\\
0.7	0.082099618104336\\
0.8	0.0813882922149329\\
0.9	0.0813724171673778\\
1	0.0813466764721659\\
};
\addlegendentry{$K_o = 3$, NLOS}

\addplot [color=black, dashed, line width=1.0pt]
  table[row sep=crcr]{%
0.1	0.0652878599853995\\
0.125	0.0654985217117979\\
0.15	0.0654244268514657\\
0.175	0.0652115215114196\\
0.2	0.0654219731232963\\
0.225	0.06545546894895\\
0.25	0.0653162859065771\\
0.275	0.06536612083723\\
0.3	0.0656512041528209\\
0.325	0.0652695139020411\\
0.35	0.0655689209346802\\
0.375	0.065265769558714\\
0.4	0.0654582704191104\\
0.5	0.0652397234813123\\
0.6	0.0652177861796891\\
0.7	0.0655739795249917\\
0.8	0.0655684640923335\\
0.9	0.065599380366225\\
1	0.0654971872471066\\
};
\addlegendentry{$K_o = 4$, NLOS}

\addplot [color=green, line width=1.5pt]
  table[row sep=crcr]{%
0.1	0.0605507496377417\\
0.125	0.0602766476858508\\
0.15	0.0605958518621586\\
0.175	0.060640277140287\\
0.2	0.0605097856317019\\
0.225	0.0602684771650356\\
0.25	0.0605638378436488\\
0.275	0.0605567711897088\\
0.3	0.0604724898745731\\
0.325	0.0605656029550999\\
0.35	0.0602428860327434\\
0.375	0.0604167242436942\\
0.4	0.0606330375762162\\
0.5	0.0603033541177115\\
0.6	0.0602682272592702\\
0.7	0.0602627592861015\\
0.8	0.0605960033347421\\
0.9	0.0604658199669311\\
1	0.060274088456877\\
};
\addlegendentry{$K_o = 4$, LOS}

\end{axis}

\end{tikzpicture}%
}}
\caption{Comparison between LOS and NLOS conditions to the observers.}
\label{fig:EffectNLOS}
%
\centering\!\!
\subfloat[distance MVUE]{\centering\label{fig:EffectSNR_dEst}
\resizebox{.48\columnwidth}{!}{
%
%
\definecolor{mycolor1}{rgb}{1.00000,0.00000,1.00000}%
\begin{tikzpicture}

\pgfplotsset{every x tick label/.append style={font=\scriptsize, yshift=0.3ex}}
\pgfplotsset{every y tick label/.append style={font=\scriptsize, xshift=0.3ex}}

\begin{axis}[%
width=\SimPlotsInnerWidth,
height=\SimPlotsInnerHeight,
at={(0,0)},
scale only axis,
xmode=log,
xmin=0.1,
xmax=10,
xminorticks=true,
xlabel={\small{inter-node distance $d$ [m]}},
ylabel={\footnotesize{distance RMSE [m]}},
ymode=log,
ymin=0.01,
ymax=2,
yminorticks=true,
axis background/.style={fill=white},
x label style={at={(axis description cs:0.5,-0.09)},anchor=north},
y label style={at={(axis description cs:-0.135,0.5)},anchor=south},
legend style={at={(1,0.03)}, font=\scriptsize, anchor=south east, legend cell align=left, align=left, draw=white!15!black}
]

\addlegendimage{empty legend}
\addlegendentry{\hspace{-6mm}$\SNR_\mathrm{LOS}$:}

\addplot [color=red, dashed, line width=1.0pt]
  table[row sep=crcr]{%
0.1	0.852211693085833\\
0.115478198468946	0.835064691289803\\
0.133352143216332	0.822358521589775\\
0.153992652605949	0.815411051871984\\
0.177827941003892	0.807774339190735\\
0.205352502645715	0.782488206138487\\
0.237137370566166	0.758432506731315\\
0.273841963426436	0.733399963543663\\
0.316227766016838	0.710478510537653\\
0.365174127254838	0.69018282726665\\
0.421696503428582	0.652530492070831\\
0.486967525165863	0.618703883272015\\
0.562341325190349	0.605201470538104\\
0.649381631576211	0.581850339038777\\
0.749894209332456	0.560219702335343\\
0.865964323360065	0.53803291537694\\
1	0.517620332961492\\
1.33352143216332	0.466438220070537\\
1.77827941003892	0.461922036907662\\
2.37137370566166	0.485462936632068\\
3.16227766016838	0.517554418483048\\
4.21696503428582	0.62874438476142\\
5.62341325190349	0.850736903219666\\
7.49894209332456	1.30327857076155\\
10	2.06143498143885\\
};
\addlegendentry{$\,\ 3\dB$}

\addplot [color=mycolor1, dotted, line width=2.0pt]
  table[row sep=crcr]{%
0.1	0.228734984301515\\
0.115478198468946	0.224815157158723\\
0.133352143216332	0.214048201054258\\
0.153992652605949	0.20542686117716\\
0.177827941003892	0.19666179771278\\
0.205352502645715	0.190650745421772\\
0.237137370566166	0.176800902571182\\
0.273841963426436	0.179176370827875\\
0.316227766016838	0.168549401833648\\
0.365174127254838	0.15784144508386\\
0.421696503428582	0.157925035929146\\
0.486967525165863	0.147976872215077\\
0.562341325190349	0.147799723681279\\
0.649381631576211	0.151303102127413\\
0.749894209332456	0.151087497839903\\
0.865964323360065	0.158232257608852\\
1	0.169186617502679\\
1.33352143216332	0.186976535593214\\
1.77827941003892	0.234291981272869\\
2.37137370566166	0.307857673777106\\
3.16227766016838	0.412442276755478\\
4.21696503428582	0.564549516370796\\
5.62341325190349	0.852549729607119\\
7.49894209332456	1.33053122357613\\
10	2.03458508578242\\
};
\addlegendentry{$13\dB$}

\addplot [color=black, line width=1.0pt]
  table[row sep=crcr]{%
0.1	0.0696849120803761\\
0.115478198468946	0.0650541591635369\\
0.133352143216332	0.0615410276329903\\
0.153992652605949	0.0604445961052921\\
0.177827941003892	0.0590352367358124\\
0.205352502645715	0.0579432024666916\\
0.237137370566166	0.0584315586053843\\
0.273841963426436	0.0598895579927147\\
0.316227766016838	0.0608189667122973\\
0.365174127254838	0.0628003911714719\\
0.421696503428582	0.0667292169088795\\
0.486967525165863	0.0733427766469561\\
0.562341325190349	0.0791501121113384\\
0.649381631576211	0.0863848478195702\\
0.749894209332456	0.0963646535717493\\
0.865964323360065	0.106714007828045\\
1	0.12268523848104\\
1.33352143216332	0.158249859333158\\
1.77827941003892	0.217384865804767\\
2.37137370566166	0.281209968805405\\
3.16227766016838	0.405261042394551\\
4.21696503428582	0.552313197541473\\
5.62341325190349	0.831296974604804\\
7.49894209332456	1.31862313303234\\
10	2.04244298974809\\
};
\addlegendentry{$23\dB$}

\addplot [color=blue, dashdotted, line width=1.0pt]
  table[row sep=crcr]{%
0.1	0.0411748968049255\\
0.115478198468946	0.0405425952430535\\
0.133352143216332	0.0393400994812522\\
0.153992652605949	0.0392174146331999\\
0.177827941003892	0.0405101357326319\\
0.205352502645715	0.0410396503497293\\
0.237137370566166	0.0419028846497179\\
0.273841963426436	0.0432651363001183\\
0.316227766016838	0.0476727641012768\\
0.365174127254838	0.0504561175301238\\
0.421696503428582	0.0562585024682629\\
0.486967525165863	0.0638430088057495\\
0.562341325190349	0.0707412795631328\\
0.649381631576211	0.0808812965422878\\
0.749894209332456	0.0921770313373835\\
0.865964323360065	0.104153854743011\\
1	0.117314101882356\\
1.33352143216332	0.158127123984604\\
1.77827941003892	0.21007782525296\\
2.37137370566166	0.283219560068167\\
3.16227766016838	0.388668637894161\\
4.21696503428582	0.554261709402897\\
5.62341325190349	0.849764500953498\\
7.49894209332456	1.28947568411415\\
10	2.06776683311345\\
};
\addlegendentry{$33\dB$}

\end{axis}

\end{tikzpicture}%
}}\!\!
\subfloat[position LSE]{\centering\label{fig:EffectSNR_vEst}
\resizebox{.48\columnwidth}{!}{
%
%
\definecolor{mycolor1}{rgb}{1.00000,0.00000,1.00000}%
\begin{tikzpicture}

\pgfplotsset{every x tick label/.append style={font=\scriptsize, yshift=0.3ex}}
\pgfplotsset{every y tick label/.append style={font=\scriptsize, xshift=0.3ex}}

\begin{axis}[%
width=\SimPlotsInnerWidth,
height=\SimPlotsInnerHeight,
at={(0,0)},
scale only axis,
xmode=log,
xmin=0.1,
xmax=10,
xminorticks=true,
xlabel={\small{inter-node distance $d$ [m]}},
ylabel={\footnotesize{position RMSE [m]}},
ymode=log,
ymin=0.01,
ymax=2,
yminorticks=true,
axis background/.style={fill=white},
x label style={at={(axis description cs:0.5,-0.09)},anchor=north},
y label style={at={(axis description cs:-0.135,0.5)},anchor=south},
legend style={at={(0.82,1)}, font=\scriptsize, anchor=north east, legend cell align=left, align=left, draw=white!15!black}
]

\addlegendimage{empty legend}
\addlegendentry{\hspace{-6mm}$\SNR_\mathrm{LOS}$:}

\addplot [color=red, dashed, line width=1.0pt]
  table[row sep=crcr]{%
0.1	0.455710100148316\\
0.115478198468946	0.455037310764495\\
0.133352143216332	0.460003174900693\\
0.153992652605949	0.466834365290619\\
0.177827941003892	0.46334325336521\\
0.205352502645715	0.447329460223742\\
0.237137370566166	0.456883621335436\\
0.273841963426436	0.467852435755537\\
0.316227766016838	0.465376649413921\\
0.365174127254838	0.469147138113051\\
0.421696503428582	0.466956606563397\\
0.486967525165863	0.463628931812353\\
0.562341325190349	0.456240008556188\\
0.649381631576211	0.450797086650695\\
0.749894209332456	0.464626432255125\\
0.865964323360065	0.452713328032523\\
1	0.466779961745269\\
1.33352143216332	0.462253761433089\\
1.77827941003892	0.469912713860661\\
2.37137370566166	0.474120478746453\\
3.16227766016838	0.479768608018223\\
4.21696503428582	0.45805300721133\\
5.62341325190349	0.475972669841602\\
7.49894209332456	0.457918426064293\\
10	0.49961896327279\\
};
\addlegendentry{$\,\ 3\dB$}

\addplot [color=mycolor1, dotted, line width=2.0pt]
  table[row sep=crcr]{%
0.1	0.152310014621636\\
0.115478198468946	0.154159026592138\\
0.133352143216332	0.151622310143505\\
0.153992652605949	0.153309963989608\\
0.177827941003892	0.150689411007394\\
0.205352502645715	0.155074665809638\\
0.237137370566166	0.144879333384288\\
0.273841963426436	0.151314176572573\\
0.316227766016838	0.152960104596998\\
0.365174127254838	0.152927051565011\\
0.421696503428582	0.155121747521078\\
0.486967525165863	0.152143615829401\\
0.562341325190349	0.155558269001378\\
0.649381631576211	0.150238748224598\\
0.749894209332456	0.15335798288603\\
0.865964323360065	0.152540225222145\\
1	0.154421842639043\\
1.33352143216332	0.153579779599224\\
1.77827941003892	0.150418026826902\\
2.37137370566166	0.148179108907241\\
3.16227766016838	0.156004312206489\\
4.21696503428582	0.155499156967286\\
5.62341325190349	0.155448927311543\\
7.49894209332456	0.161655915856429\\
10	0.165840843945298\\
};
\addlegendentry{$13\dB$}

\addplot [color=black, line width=1.0pt]
  table[row sep=crcr]{%
0.1	0.0613262868672212\\
0.115478198468946	0.0594652950611334\\
0.133352143216332	0.0603901808474143\\
0.153992652605949	0.0593586454310692\\
0.177827941003892	0.0599532309328276\\
0.205352502645715	0.0604322581274668\\
0.237137370566166	0.0610209304204225\\
0.273841963426436	0.0617861914577341\\
0.316227766016838	0.0603656677415861\\
0.365174127254838	0.0613816268372409\\
0.421696503428582	0.0595182281542774\\
0.486967525165863	0.0605698083213133\\
0.562341325190349	0.0607192076681353\\
0.649381631576211	0.0603298863295345\\
0.749894209332456	0.061586328916487\\
0.865964323360065	0.0600131931319376\\
1	0.0587664343003127\\
1.33352143216332	0.0596825892794336\\
1.77827941003892	0.0607156169009126\\
2.37137370566166	0.0606399214402379\\
3.16227766016838	0.0607087752918074\\
4.21696503428582	0.0618552054925397\\
5.62341325190349	0.0625043999486475\\
7.49894209332456	0.0633625012123334\\
10	0.0692649049883226\\
};
\addlegendentry{$23\dB$}

\addplot [color=blue, dashdotted, line width=1.0pt]
  table[row sep=crcr]{%
0.1	0.04061663685434\\
0.115478198468946	0.0415021424166601\\
0.133352143216332	0.0405230892360247\\
0.153992652605949	0.04167660414968\\
0.177827941003892	0.0413033508891875\\
0.205352502645715	0.0414129506612948\\
0.237137370566166	0.0410187543451232\\
0.273841963426436	0.0411290615930635\\
0.316227766016838	0.0410624893233045\\
0.365174127254838	0.0410841563524294\\
0.421696503428582	0.041135199167756\\
0.486967525165863	0.0416044762163946\\
0.562341325190349	0.0412596008982035\\
0.649381631576211	0.0418169355151351\\
0.749894209332456	0.0417672762323761\\
0.865964323360065	0.0420344918668732\\
1	0.0424529126144965\\
1.33352143216332	0.0425379737075596\\
1.77827941003892	0.0419045505773756\\
2.37137370566166	0.0422259873710431\\
3.16227766016838	0.0424817830414511\\
4.21696503428582	0.0437520617455273\\
5.62341325190349	0.044033167291319\\
7.49894209332456	0.0459312319602698\\
10	0.0485174038776017\\
};
\addlegendentry{$33\dB$}

\end{axis}

\end{tikzpicture}%
}}
\caption{The effect of scaling the transmit power (i.e. scaling $E_1$), expressed in terms of the resulting LOS-path SNR. This evaluation uses $p_\mathrm{LOS} = 1$.}
\label{fig:EffectSNR}
\end{figure}


\Cref{fig:EffectSNR} visualizes the effect of the operating SNR. It has a significant effect on the distance estimation RMSE for small $d$ and on the position estimation RMSE throughout. This is in accordance with previous observations.


Finally, we study the performance implications of alien MPC occurrences. These can happen for any $d$ and are to be expected for larger $d$.
Our evaluation replaces a certain number of MPCs in the CIR $\cirB(\tau)$ with randomly sampled MPCs, using the aforementioned statistics, but in a fashion that does not alter the delay order of the CIR. Otherwise it would be unlikely that the alien MPCs would be erroneously selected and associated.
\Cref{fig:EffectAlienMPC} shows that the error of both distance estimation (MVUE) and position estimation (DD) deteriorates heavily, even from a single alien MPC. However, the DDN position estimator, which uses the association scheme from \Cref{sec:EstimateRelLocNoAssoc}, copes with alien MPCs very well because it discards MPCs that could not be associated with good fit. The error increases just slightly because $K$ decreases.

\renewcommand\SimPlotsInnerWidth{.8\columnwidth}  
\renewcommand\SimPlotsInnerHeight{.35\columnwidth} 
\begin{figure}[!ht]
\centering
\resizebox{.8\columnwidth}{!}{
\begin{tikzpicture}

\begin{axis}[%
width=\SimPlotsInnerWidth,
height=\SimPlotsInnerHeight,
at={(0,0)},
scale only axis,
xmin=0,
xmax=8,
xtick={0, 1, 2, 3, 4, 5, 6, 7, 8},
xlabel style={font=\color{white!15!black}},
xlabel={number of alien MPCs},
ymode=log,
ymin=0.03,
ymax=10,
yminorticks=true,
ylabel style={font=\color{white!15!black}},
ylabel={median absolute error [m]},
axis background/.style={fill=white},
legend style={at={(0.32,0.3)}, font=\scriptsize, anchor=south west, legend cell align=left, align=left, draw=white!15!black}
]
\addplot [color=black, \SimPlotsMVUELineStyle, line width=\SimPlotsLineWidth, mark size=\SimPlotsMVUEMarkSize, mark=\SimPlotsMVUEMarker, mark options={solid, rotate=\SimPlotsMVUERotate, black}]
  table[row sep=crcr]{%
0	0.164901736579933\\
1	0.76370821565319\\
2	2.38348062720697\\
3	4.32726479768095\\
4	5.66765706392106\\
5	6.57602569674849\\
6	8.00259025527167\\
7	8.83021400412085\\
8	9.99688644886892\\
};
\addlegendentry{distance \SimPlotsLegendMVUE}

\addplot [color=black, \SimPlotsDDLineStyle, line width=\SimPlotsLineWidth, mark size=\SimPlotsDDMarkSize, mark=\SimPlotsDDMarker, mark options={solid, black}]
  table[row sep=crcr]{%
0	0.0519440955310973\\
1	0.937392037789279\\
2	1.75642426100709\\
3	2.42988595734702\\
4	2.98484153744659\\
5	3.21841516948385\\
6	3.5170933899504\\
7	3.59638400823744\\
8	3.78655464217573\\
};
\addlegendentry{pos. \SimPlotsLegendDD{} scheme}

\addplot [color=black, \SimPlotsDDNLineStyle, line width=\SimPlotsLineWidth, mark size=\SimPlotsDDNMarkSize, mark=\SimPlotsDDNMarker, mark options={solid, black}]
  table[row sep=crcr]{%
0	0.0520209683318208\\
1	0.0563686889294816\\
2	0.065693533694913\\
3	0.0731825056641433\\
4	0.0860352981199923\\
5	0.108231850284318\\
6	0.152734236512406\\
7	0.292715333965909\\
8	1.50177545041629\\
};
\addlegendentry{pos. \SimPlotsLegendDDN{} scheme}

\end{axis}

\end{tikzpicture}%
}
\vspace{-2mm}
\caption{Accuracy deterioration due to alien MPCs, i.e. erroneously selected MPCs that can not be associated in terms of equal propagation paths.}
\label{fig:EffectAlienMPC}
\end{figure}
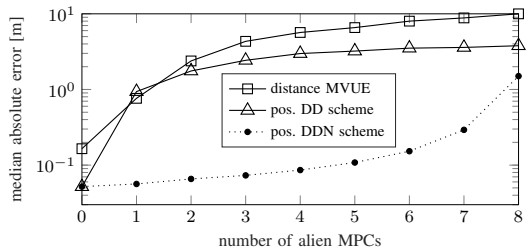


\section{Practical Feasibility Study}
\label{sec:EvalMeas}
This section gives a simple practical proof of concept of the proposed paradigm under compliant conditions. In a LOS scenario in a large empty room we evaluate one distance estimator and one position estimator, namely $\hat d\MVUE\AssAsyn$ from \Cref{eq:distMVUE} and $\hat\d\LSE\AssViaDiffPWA$ from \Cref{eq:displLSEAssViaDiffPWA}. 
The specific goals are as follows. For small A-B displacements we verify that the MPC differences behave as anticipated and, subsequently, we demonstrate that accurate estimation is indeed possible in practice.
The experiment does not demonstrate the paradigm's capabilities in terms of NLOS and time-varying channels.

\begin{figure}[!b]
\vspace{-3mm}
\centering
\subfloat[experiment setup and environment ]{\centering
\includegraphics[height=27.8mm]{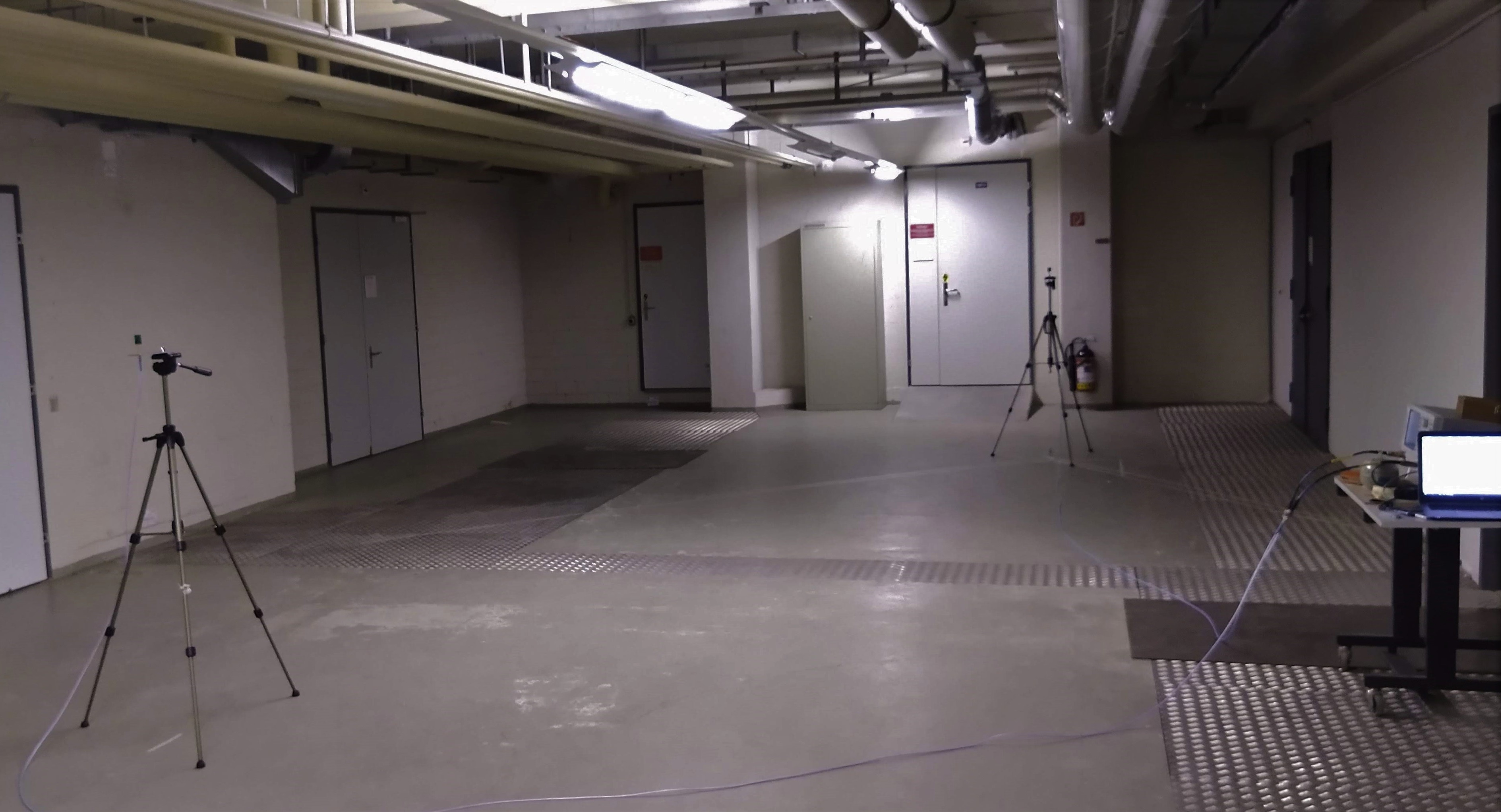}
\put(-140,52){\footnotesize{\color{white}\shortstack[c]{observer\\(static)
}}}
\put(-42,50){\footnotesize{\color{white}\shortstack[c]{node on\\trajectory}}}
\put(-21,17){\footnotesize{\color{white}VNA}}
\label{fig:PhotoSetupFull}
}
\subfloat[antenna detail]{\centering
\includegraphics[height=27.8mm]{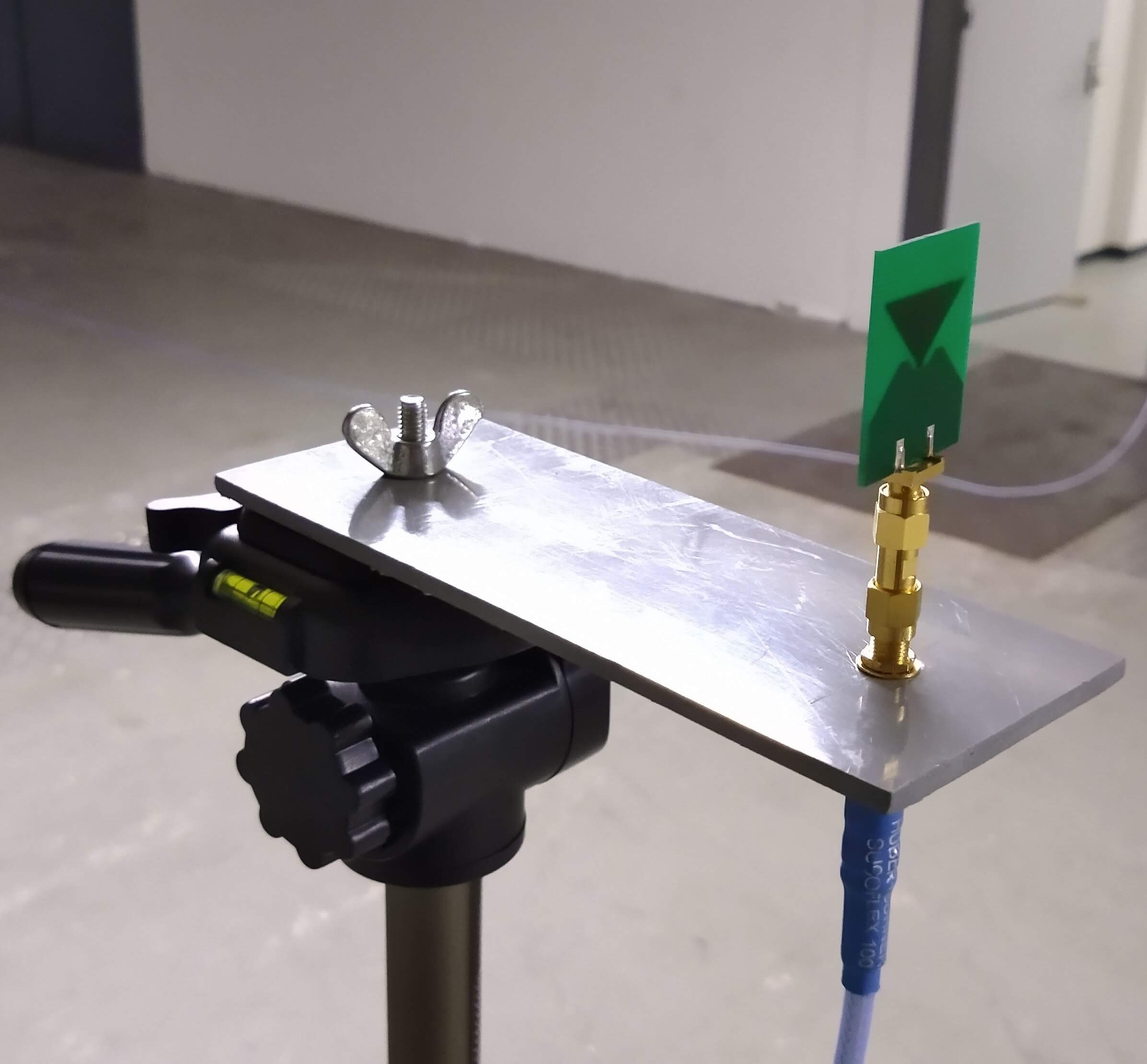}
\label{fig:PhotoSetupDetail}
}
\\
\subfloat[floor plan]{\centering
\resizebox{.62\columnwidth}{!}{\begin{tikzpicture}

\begin{axis}[%
width=.99\linewidth,
height=.5\linewidth,
at={(0,0)},
scale only axis,
xmin=-.5,
xmax=20,
xlabel={$x\ \mathrm{[m]}$},
xtick distance=2,
ymin=-.5,
ymax=11,
ytick distance=2,
label style={font=\large},
tick label style={font=\large},
ylabel={$y\ \mathrm{[m]}$},
axis background/.style={fill=white},
xmajorgrids,
ymajorgrids,
]
\addplot [color=black, line width=1.5pt, forget plot, fill=white!90!black,]
table[row sep=crcr]{%
	0	0\\
	0	10.5\\
	3.45	10.5\\
	3.45	7.5\\
	13.6	7.5\\
	13.6	8.66\\
	19.5	8.66\\
	19.5	0\\
	0	0\\
};
\addplot [only marks, mark=triangle*, mark options={color=black,fill=red,scale={4}}]
table[row sep=crcr]{%
	8.5	5\\
};

\addplot [only marks, mark=*,mark size=2.5pt]
table[row sep=crcr]{%
	15.5000    3.0000\\
};

\addplot [color=blue]
table[row sep=crcr]{%
	15.5000    3.0000\\
	15.4564    2.9981\\
	15.4132    2.9924\\
	15.3706    2.9830\\
	15.3290    2.9698\\
	15.2887    2.9532\\
	15.2500    2.9330\\
	15.2132    2.9096\\
	15.1786    2.8830\\
	15.1464    2.8536\\
	15.1170    2.8214\\
	15.0904    2.7868\\
	15.0670    2.7500\\
	15.0468    2.7113\\
	15.0302    2.6710\\
	15.0170    2.6294\\
	15.0076    2.5868\\
	15.0019    2.5436\\
	15.0000    2.5000\\
	15.0019    2.4564\\
	15.0076    2.4132\\
	15.0170    2.3706\\
	15.0302    2.3290\\
	15.0468    2.2887\\
	15.0670    2.2500\\
	15.0904    2.2132\\
	15.1170    2.1786\\
	15.1464    2.1464\\
	15.1786    2.1170\\
	15.2132    2.0904\\
	15.2500    2.0670\\
	15.2887    2.0468\\
	15.3290    2.0302\\
	15.3706    2.0170\\
	15.4132    2.0076\\
	15.4564    2.0019\\
	15.5000    2.0000\\
};

\node[font=\large,color=red] at (axis cs: 8.5,3.5) {observer};
\node[font=\large] at (axis cs: 16.4,4.3) {$\posA$ (static)};
\node[font=\large,color=blue] at (axis cs: 16,1) {$\posB$ trajectory};

\end{axis}

\end{tikzpicture}}
\label{fig:ZfloorTotal}
}\!\!\!\!\!%
\subfloat[trajectory detail]{\centering
\resizebox{.35\columnwidth}{!}{\begin{tikzpicture}

\begin{axis}[%
width=.4\linewidth,
height=.4\linewidth,
at={(0,0)},
scale only axis,
xmin=14.5,
xmax=16,
label style={font=\normalsize},
tick label style={font=\normalsize},
xlabel={$x\ \mathrm{[m]}$},
xtick distance=.5,
ymin=1.75,
ymax=3.25,
ytick distance=.5,
ylabel style={at={(-.13,.7)}},
ylabel={$y\ \mathrm{[m]}$},
axis background/.style={fill=white},
xmajorgrids,
ymajorgrids,
]
\addplot [color=black, line width=1.5pt, forget plot, ]
table[row sep=crcr]{%
	0	0\\
	0	10.5\\
	3.45	10.5\\
	3.45	7.5\\
	13.6	7.5\\
	13.6	8.66\\
	19.5	8.66\\
	19.5	0\\
	0	0\\
};
\addplot [only marks, mark=triangle*, mark options={color=black,fill=red,scale={2}},forget plot]
table[row sep=crcr]{%
	8.5	5\\
};

\addplot [only marks, mark=*,mark size=4pt]
table[row sep=crcr]{%
	15.5000    3.0000\\
};

\addplot [color=blue, mark=*, mark options={scale=.3}]
table[row sep=crcr]{%
	15.5000    3.0000\\
	15.4564    2.9981\\
	15.4132    2.9924\\
	15.3706    2.9830\\
	15.3290    2.9698\\
	15.2887    2.9532\\
	15.2500    2.9330\\
	15.2132    2.9096\\
	15.1786    2.8830\\
	15.1464    2.8536\\
	15.1170    2.8214\\
	15.0904    2.7868\\
	15.0670    2.7500\\
	15.0468    2.7113\\
	15.0302    2.6710\\
	15.0170    2.6294\\
	15.0076    2.5868\\
	15.0019    2.5436\\
	15.0000    2.5000\\
	15.0019    2.4564\\
	15.0076    2.4132\\
	15.0170    2.3706\\
	15.0302    2.3290\\
	15.0468    2.2887\\
	15.0670    2.2500\\
	15.0904    2.2132\\
	15.1170    2.1786\\
	15.1464    2.1464\\
	15.1786    2.1170\\
	15.2132    2.0904\\
	15.2500    2.0670\\
	15.2887    2.0468\\
	15.3290    2.0302\\
	15.3706    2.0170\\
	15.4132    2.0076\\
	15.4564    2.0019\\
	15.5000    2.0000\\
};

\node[font=\normalsize] at (axis cs: 15.29,3.145) {$\posA$, pos. $0$ (static)};
\node[font=\normalsize,color=blue] at (axis cs: 15.27,1.87) {$\posB$, pos. $1\,\ldots\,36$};

\draw[-latex,color=blue] (14.85,2.875) arc (150:210:0.75);

\end{axis}

\end{tikzpicture}}
\label{fig:ZfloorZoom}}
\\[1.1mm]
\subfloat[CIR measurements over trajectory]{\centering
\includegraphics[width=\columnwidth]{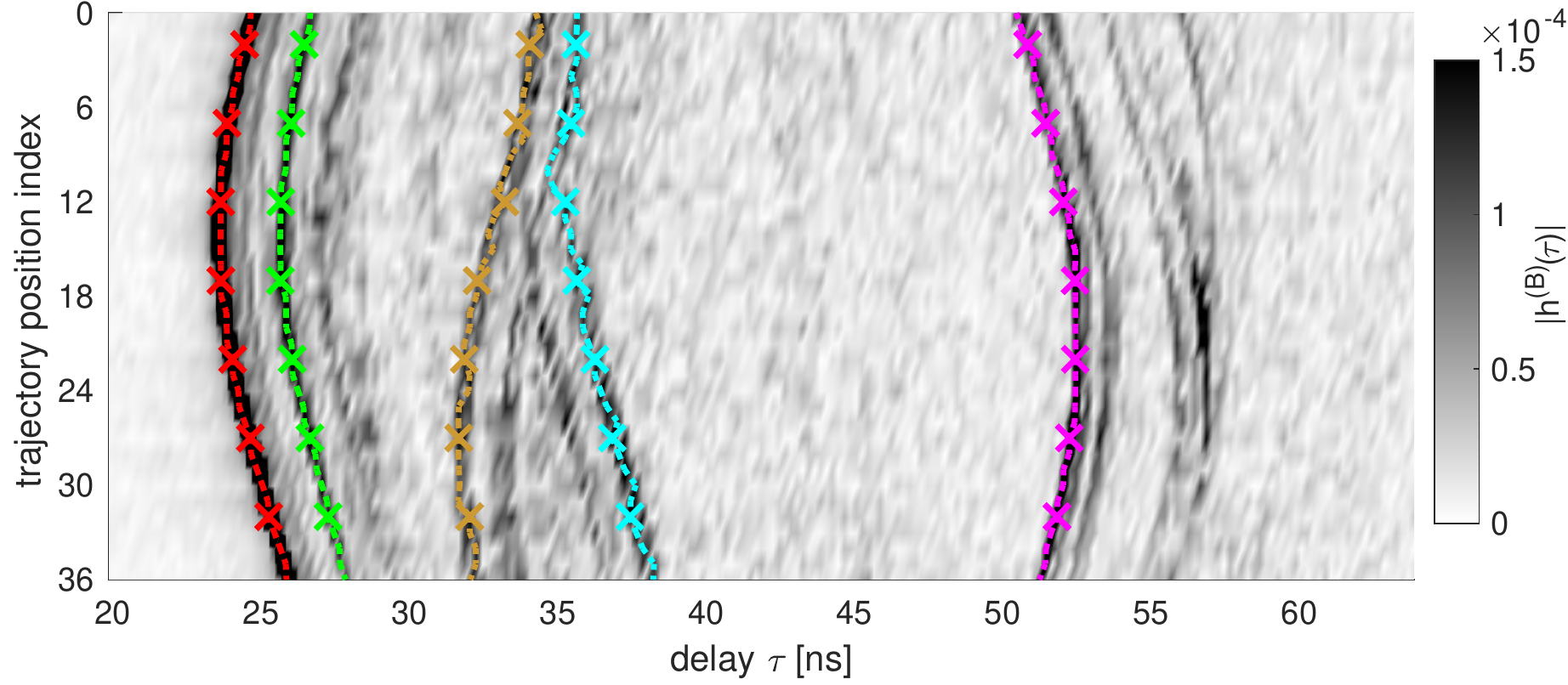}
\put(-239.5,98){\color[rgb]{.85,0,0}{\scriptsize{\begin{tabular}{l}MPC\\$k \!=\! 1$\end{tabular}}}}
\put(-208,61.5){\color[rgb]{0,1,0}{\scriptsize{\begin{tabular}{l}MPC\\$k \!=\! 2$\end{tabular}}}}
\put(-196,86){\color[rgb]{.6,.45,.2}{\scriptsize{\begin{tabular}{l}MPC\\$k \!=\! 3$\end{tabular}}}} 
\put(-157,48){\color[rgb]{0,.7,.7}{\scriptsize{\begin{tabular}{l}MPC\\$k \!=\! 4$\end{tabular}}}}
\put(-107,60){\color[rgb]{.7,0,.7}{\scriptsize{\begin{tabular}{l}MPC\\$k \!=\! 5$\end{tabular}}}}
\label{fig:mpcColorplot}}
\\[-.1mm]
\subfloat[example CIR measurements]{\centering
\resizebox{\columnwidth}{!}{\input{MeasuredCirTwoExamples.tex}}
\label{fig:measuredExampleCIRs}}
\caption{Experiment environment, measurement equipment, and obtained channel measurements.}
\label{fig:Zfloor}
\end{figure}

We conduct measurements in a large empty room (\Cref{fig:PhotoSetupFull}) with an approximate size of $20\unit{m} \times 10\unit{m} \times 3\unit{m}$. The floor plan is shown in \cref{fig:ZfloorTotal}.
The room exhibits mostly plain walls and a plain floor, which give rise to distinct reflections. The ceiling is however cluttered with pipes and other installations (cf. \Cref{fig:PhotoSetupFull}) which scatter the radio waves and prevent a clear ceiling reflection.
Our experiment uses a single observer node ($\NObs = 1$); the index $o = 1$ is discarded.
This observer node is static at the position $[8.5\unit{m}, 5\unit{m}, 1\unit{m}]\Tr$.

Node A is static at position $\posA = [15.5\unit{m}, 3 \unit{m}, 1\unit{m}]\Tr$. This is also the starting point of a half-circle trajectory on which the mobile node B is sequentially placed (see \cref{fig:ZfloorZoom}). The half-circle radius is $0.5\unit{m}$ and the center is at $[15.5\unit{m}, 2.5 \unit{m}, 1\unit{m}]\Tr$. The different positions $\posB$ are referred to by a trajectory position index $1,\ldots,36$ with increasing distance to $\posA$ (index $0$). We chose such a curved trajectory to ensure versatility of $\d$ compared to the MPC directions $\dirVectA[k]$.
The chosen node constellation and trajectory ensure that the A-B distance $d$ is much shorter than the observer distance at all times. This choice was made to support the plane-wave assumption in this proof-of-concept experiment, to allow us to estimate the MPC directions from the delay evolutions (without antenna arrays).

Each CIR measurement between the observer node and another node (A or B) was obtained as follows. With a vector network analyzer (VNA), we recorded the frequency response with a sweep from $5$ -- $10\unit{GHz}$ with $3.125\unit{MHz}$ resolution. An inverse Fourier transform then yielded the delay-domain CIR data. This measurement approach is rather slow but achieves decent SNR.
The evolution of the measured CIRs over the trajectory is shown in \Cref{fig:mpcColorplot}, with a $20 - 65\unit{ns}$ delay window that contains all major MPCs. Several distinct MPCs are clearly visible. \Cref{fig:measuredExampleCIRs} shows two examplary CIRs (position index $0$ and $18$) in detail. The SNR is approximately between $22\dB$ (LOS path) and $14\dB$ (later reflected paths).


As the propagation environment is time-invariant, simultaneous acquisition of $\cirA(\tau)$ and $\cirB(\tau)$ was not required. Instead, we measured the CIRs sequentially over the trajectory, for node positions with indices $0,\ldots,36$.
The CIR measurement at index $0$ was declared as $\cirA(\tau)$ while $\cirB(\tau)$ is drawn from the set of CIRs at index $1,\ldots,36$.

After applying a standard peak-extraction algorithm to the CIRs (Matlab function \texttt{findpeaks}), we handpicked a suitable selection of MPCs (see the colored graphs in \Cref{fig:mpcColorplot}). Thereby we omitted MPCs that do not occur distinctly over the entire trajectory. It is important to note that the chosen MPCs do not always have the largest amplitudes due to small-scale fading.
This shows the importance of appropriate MPC tracking and selection schemes for real-time systems, where careful offline processing would not be an option.

\begin{figure}[!ht]
\vspace{-2mm}
\centering
\subfloat[ranging accuracy]{
\resizebox{!}{39mm}{
%
%
\definecolor{colorSync}{rgb}{\SimPlotsSyncColor}%
\begin{tikzpicture}

\begin{axis}[%
width=.5\linewidth,
height=.5\linewidth,
at={(0,0)},
scale only axis,
xmin=0,
xmax=36,
xtick={ 0,  6, 12, 18, 24, 30, 36},
label style={font=\normalsize},
xlabel={trajectory position index},
ymin=0,
ymax=1.2,
ylabel={estimated distance [m]},
ylabel style={at={(-.1,.5)}},
axis background/.style={fill=white},
xmajorgrids,
ymajorgrids,
legend style={at={(1,0)}, anchor=south east, legend cell align=left, align=left, draw=white!15!black}
]

\addplot [color=blue, mark=*, mark options={scale=.3}, line width=0.5pt]
  table[row sep=crcr]{%
0	0\\
1	0.043619387365336\\
2	0.0871557427476582\\
3	0.130526192220052\\
4	0.17364817766693\\
5	0.216439613938103\\
6	0.258819045102521\\
7	0.300705799504273\\
8	0.342020143325669\\
9	0.38268343236509\\
10	0.422618261740699\\
11	0.461748613235034\\
12	0.5\\
13	0.537299608346824\\
14	0.573576436351046\\
15	0.608761429008721\\
16	0.642787609686539\\
17	0.67559020761566\\
18	0.707106781186547\\
19	0.737277336810124\\
20	0.766044443118978\\
21	0.793353340291235\\
22	0.819152044288992\\
23	0.843391445812886\\
24	0.866025403784439\\
25	0.887010833178222\\
26	0.90630778703665\\
27	0.923879532511287\\
28	0.939692620785908\\
29	0.953716950748227\\
30	0.965925826289068\\
31	0.976296007119933\\
32	0.984807753012208\\
33	0.99144486137381\\
34	0.996194698091746\\
35	0.999048221581858\\
36	1\\
};
\addlegendentry{ground truth}

\addplot [color=colorSync, dashed, line width=1.5pt, forget plot]
  table[row sep=crcr]{%
0	0\\
1	0.0676331785247981\\
2	0.137424862747199\\
3	0.208655550767999\\
4	0.214411565961598\\
5	0.282764246385599\\
6	0.3482389192128\\
7	0.356153440103999\\
8	0.4259451243264\\
9	0.429542633822399\\
10	0.4993343180448\\
11	0.567686998468799\\
12	0.571284507964798\\
13	0.638917686489598\\
14	0.644673701683198\\
15	0.7115873783088\\
16	0.7115873783088\\
17	0.7137458840064\\
18	0.720940902998399\\
19	0.8000861119104\\
20	0.800805613809599\\
21	0.872036301830398\\
22	0.867719290435198\\
23	0.802964119507199\\
24	0.815195651793601\\
25	0.9367914727584\\
26	0.9382304765568\\
27	0.935352468959999\\
28	0.939669480355199\\
29	0.933193963262399\\
30	0.925279442371199\\
31	0.943266989851201\\
32	0.802244617608\\
33	0.802964119507199\\
34	0.795049598615999\\
35	0.9324744613632\\
36	0.923120936673599\\
};

\addplot [color=black, dashed, line width=1.0pt]
  table[row sep=crcr]{%
%
0	0\\
1	0.0490160668829983\\
2	0.132208473978\\
3	0.219897767942999\\
4	0.227542475621998\\
5	0.311634260091\\
6	0.356603128791\\
7	0.402921063551999\\
8	0.448789309626\\
9	0.455534639931\\
10	0.539626424399999\\
11	0.630013850486999\\
12	0.590441246030998\\
13	0.717253455764999\\
14	0.766269522647999\\
15	0.766269522648\\
16	0.850361307117\\
17	0.890383600259999\\
18	0.897578619251999\\
19	0.946594686134999\\
20	0.948393440882998\\
21	0.994711375643999\\
22	0.989315111399998\\
23	0.949742506943998\\
24	0.956038148561999\\
25	1.030686470604\\
26	1.02978709323\\
27	0.987066667964999\\
28	0.987966045338998\\
29	0.942547487951999\\
30	1.023041762925\\
31	0.994261686957\\
32	0.902974883496\\
33	0.949742506943998\\
34	0.952440639066\\
35	1.038331178283\\
36	1.073856584556\\
};
\addlegendentry{distance estimator}

\addplot [color=colorSync, line width=2.0pt, forget plot]
  table[row sep=crcr]{%
%
0	0\\
1	0.025378469648685\\
2	0.0917657428904415\\
3	0.139742306017978\\
4	0.142708961606617\\
5	0.20454229690013\\
6	0.233039638769419\\
7	0.266504379006876\\
8	0.277032763455855\\
9	0.328179995230107\\
10	0.384183893246661\\
11	0.438387624922484\\
12	0.418425525572542\\
13	0.496676855864882\\
14	0.539298840478678\\
15	0.530301063560905\\
16	0.614138790449805\\
17	0.634491177226188\\
18	0.662884976312158\\
19	0.687564069534005\\
20	0.689022951909189\\
21	0.741579240578378\\
22	0.755421283815677\\
23	0.720144924943977\\
24	0.74577672078151\\
25	0.834418212129498\\
26	0.876537896515152\\
27	0.84270599449156\\
28	0.87278837386967\\
29	0.879140011935956\\
30	0.931444749243839\\
31	0.914189109471349\\
32	0.83587348481242\\
33	0.861731367016778\\
34	0.85538039755549\\
35	0.916147674224101\\
36	0.942875487919716\\
};

\addplot [color=black, line width=1.0pt]
  table[row sep=crcr]{%
0	0\\
1	0.050655208047914\\
2	0.0982824092500128\\
3	0.151958460807622\\
4	0.156910800603353\\
5	0.216834449774341\\
6	0.25201921812826\\
7	0.277473266090677\\
8	0.308935594219791\\
9	0.335251884125299\\
10	0.389697296579556\\
11	0.44540662893945\\
12	0.435709435279451\\
13	0.507405197377003\\
14	0.545075731320881\\
15	0.547929975621104\\
16	0.620565790380617\\
17	0.638065753564445\\
18	0.665948131870386\\
19	0.688733104937706\\
20	0.690237976214924\\
21	0.741705177467821\\
22	0.754783999411596\\
23	0.719575694152447\\
24	0.742337382318192\\
25	0.830616497056621\\
26	0.86911548496654\\
27	0.837116788265431\\
28	0.864432264545074\\
29	0.871916676184727\\
30	0.917004841636066\\
31	0.905767895067451\\
32	0.816493789102394\\
33	0.842852648161811\\
34	0.829245251053967\\
35	0.883166167713459\\
36	0.924948150471921\\
};
\addlegendentry{position estimator}

\end{axis}
\end{tikzpicture}%
}
\label{fig:MeasDistEstError}
}\!\!%
\subfloat[position accuracy]{
\resizebox{!}{39mm}{
\tikzstyle{densely dashed}=[dash pattern=on 3pt off 2pt]%

\definecolor{colorSync}{rgb}{\SimPlotsSyncColor}%
\definecolor{colorSyncDarker}{rgb}{\SimPlotsSyncColorDarker}
\begin{tikzpicture}

\begin{axis}[%
width=.5\linewidth,
height=.5\linewidth,
at={(0,0)},
scale only axis,
xmin=14.75,
xmax=16.25,
label style={font=\normalsize},
xlabel={$x$ [m]},
xtick distance=.5,
ymin=1.75,
ymax=3.25,
ytick distance=.5,
ylabel style={at={(-.09,.68)}},
ylabel={$y$ [m]},
axis background/.style={fill=white},
xmajorgrids,
ymajorgrids,
]

\addplot [only marks, mark=*,mark size=4pt, forget plot]
  table[row sep=crcr]{%
15.5	3\\
};

\addplot [color=blue, mark=*, mark options={scale=.3}]
  table[row sep=crcr]{%
15.5	3\\
15.4564221286262	2.99809734904587\\
15.4131759111665	2.9924038765061\\
15.3705904774487	2.98296291314453\\
15.3289899283372	2.96984631039295\\
15.2886908691297	2.95315389351833\\
15.25	2.93301270189222\\
15.2132117818245	2.9095760221445\\
15.1786061951567	2.88302222155949\\
15.1464466094067	2.85355339059327\\
15.1169777784405	2.82139380484327\\
15.0904239778555	2.78678821817552\\
15.0669872981078	2.75\\
15.0468461064817	2.71130913087035\\
15.030153689607	2.67101007166283\\
15.0170370868555	2.62940952255126\\
15.0075961234939	2.58682408883347\\
15.0019026509541	2.54357787137383\\
15	2.5\\
15.0019026509541	2.45642212862617\\
15.0075961234939	2.41317591116653\\
15.0170370868555	2.37059047744874\\
15.030153689607	2.32898992833717\\
15.0468461064817	2.28869086912965\\
15.0669872981078	2.25\\
15.0904239778555	2.21321178182448\\
15.1169777784405	2.17860619515673\\
15.1464466094067	2.14644660940673\\
15.1786061951567	2.11697777844051\\
15.2132117818245	2.0904239778555\\
15.25	2.06698729810778\\
15.2886908691297	2.04684610648167\\
15.3289899283372	2.03015368960705\\
15.3705904774487	2.01703708685547\\
15.4131759111665	2.0075961234939\\
15.4564221286262	2.00190265095413\\
15.5	2\\
};

\addplot [color=colorSync, line width=1.5] 
  table[row sep=crcr]{%
15.5	3\\
15.4925851888948	3.02427112065777\\
15.4173210143701	2.96018622219077\\
15.3689340600798	2.95152700252651\\
15.3584690215518	2.9817016431794\\
15.3080358155512	2.92937915952016\\
15.2861226363181	2.90745840642035\\
15.2478167086139	2.9138131589189\\
15.2241373467733	2.97456481604751\\
15.1750902798895	2.95378546714529\\
15.1207243425682	2.93878529988792\\
15.0817441545407	2.86867689683709\\
15.1113529260788	2.84497299466494\\
15.0628618432277	2.76420744064855\\
15.0536900373491	2.6972614055325\\
15.0501163625428	2.71924400137817\\
15.0382857972612	2.59504759671728\\
15.0161990898525	2.58949332122563\\
15.0427744128097	2.52004036186478\\
15.0315000830991	2.49675833084091\\
15.03254210227	2.49380266485539\\
15.0378359465786	2.42004814184114\\
15.0631328732911	2.38371400336951\\
15.078267446921	2.41626173108486\\
15.1109831515471	2.36372269498324\\
15.1077118069253	2.26354652128761\\
15.1382526095404	2.20159063787875\\
15.1760806777089	2.22203491993634\\
15.1990721167909	2.1807308411338\\
15.2436845125009	2.1590543825818\\
15.2866173556926	2.09332631669254\\
15.3196179883176	2.10378347608547\\
15.3556215027645	2.17669001454588\\
15.408940745413	2.14309326000161\\
15.4785401419843	2.144888838211\\
15.5110556245753	2.08391903515493\\
15.5283553877987	2.05755097872477\\
};

\addplot [color=black, line width=1] 
  table[row sep=crcr]{%
15.5	3\\
15.4707841577007	3.04138096979434\\
15.4069656876813	2.96831327250444\\
15.3522098606203	2.96465246100299\\
15.3432168583966	2.99367182056541\\
15.29082180946	2.94288902722477\\
15.2580439187985	2.92949509892496\\
15.2326564140213	2.92571123617205\\
15.1910648446288	3.00052072231733\\
15.167139925379	2.9600250511836\\
15.114381812405	2.94376303388911\\
15.0721399787261	2.87621443016782\\
15.0856970839875	2.8651081700952\\
15.0427873062855	2.77996230543032\\
15.0394663736316	2.70842439769592\\
15.0134454039984	2.74802404223042\\
15.014019368075	2.6140923354162\\
15.0029982202155	2.59985360583558\\
15.0218840121178	2.53643553148128\\
15.0216317038764	2.50450321585204\\
15.0219057644889	2.50215025782835\\
15.0305790582146	2.42574348092868\\
15.049768555618	2.39420256499599\\
15.0553790785864	2.43422494235229\\
15.0813710636213	2.38696280470639\\
15.0891721169067	2.278096810053\\
15.1116304384475	2.22248420603301\\
15.1588682127273	2.23554357814984\\
15.176666561082	2.19831513219721\\
15.227493097628	2.17176170194282\\
15.258551540226	2.11535288342935\\
15.3046177076282	2.11555597149257\\
15.3218787276116	2.20317196336631\\
15.3801138507357	2.16571713566204\\
15.4430912250973	2.17270979835693\\
15.4683954723131	2.11739950508548\\
15.5060310705441	2.07507151235919\\
};

\node[font=\normalsize] at (axis cs: 15.7,3.1) {$\posA$};
\node[font=\normalsize,color=blue] at (axis cs: 15.55,1.87) {true $\posB$ trajectory};
\node[font=\normalsize] at (axis cs: 15.59,2.61) {$\posA\! + \hat\d\LSE\AssViaDiffPWA$};
\node[font=\normalsize,color=colorSyncDarker] at (axis cs: 15.68,2.37) {$\posA\! + \hat\d\LSE\AssViaDiffPWASync$};

\end{axis}

\end{tikzpicture}%
}
\label{fig:MeasTrajectoryEst}
}
\caption{Estimation accuracy resulting from measured delays. The light-colored (cyan) graphs relate to cases with precise time-synchronization.}
\label{fig:MeasEstResults}
\end{figure}
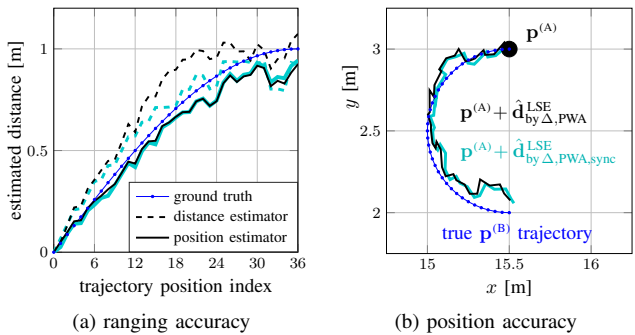

For each $\posB$ on the trajectory and the associated CIR, we evaluate the considered estimators in their synchronous and asynchronous versions in \Cref{fig:MeasEstResults}. 
We note that the relative position estimator $\hat\d\LSE\AssViaDiffPWA$ from \Cref{eq:displLSEAssViaDiffPWA} requires knowledge of the MPC directions $\dirVectA[k]$. We estimated these directions with a least-squares approach applied to the set of measured MPC delays over the entire trajectory.


The observed estimation errors are impressively small, even for the farthest positions. 
The distance estimates exhibit a limited relative error, but no additional absolute error can be observed. As expected, this results in especially accurate estimates at small $d$. Furthermore, the estimated trajectory in \Cref{fig:MeasTrajectoryEst} 
has a maximum position error of $19.2\unit{cm}$.

We conclude that the proposal has the potential for great practical accuracy under compliant conditions. An extensive evaluation of different node arrangements in different environments and a thorough study of the various expected problems (e.g., alien MPCs at large $d$, low SNR and diffuse propagation in cluttered NLOS environments) are out of the scope of this paper. They are left for future work.

\section{Technological Aspects}
\label{sec:TechComparison}
We shall discuss the proposal's enabling features and its advantages, disadvantages, and possible interfaces to state-of-the-art wireless ranging and localization. We note that the conceptual and technological uniqueness of the proposed paradigm prevents a direct performance comparison.


\subsection{Applicability and Limitations}
\label{sec:Applicability}

As introduced in \Cref{sec:SystemModel}, the presented estimators require rich multipath propagation with distinct MPCs from diverse directions (one exception is \Cref{eq:EstimateRelLocViaTauSync}). 
Such conditions can be found in indoor, urban, or industrial settings. Outdoor channels can be eligible if several scatterers are near the nodes and/or the observer(s). Free-space propagation is however unsuitable. This is a key difference to conventional schemes which are designed for free space but suffer major NLOS problems \cite{GeziciSPM2005}. Our estimators on the other hand suffer only minor accuracy losses from NLOS situations, as shown by \Cref{sec:EvalSim}. They furthermore perform best at small distances. We see this as an important advantage over TOA ranging, which is prone to large relative errors at small distances (as described in \Cref{sec:Intro}).

Very diffuse propagation environments like forests or very cluttered rooms are also unsuitable. Because there, even multiple $\mathrm{GHz}$ of bandwidth will not allow for reliable MPC resolution \cite{MolischPIEEE2009}.
This is a disadvantage compared to RSS schemes which allow for coarse distance estimates in such settings, even with minimal bandwidth requirements \cite{GeziciSPM2005}.

Besides IEEE 802.15.4a, candidate wideband technologies are 802.11ad and 5G NR FR2, where CIRs could be computed from OFDM channel state information with an inverse Fourier transform. Acoustic technology could also be suitable \cite{DokmanicPNAS2013}.



%

\subsection{Advantages}

The key advantages of the paradigm are revealed by the following observation. Certain technical quantities and conditions usually play a crucial role in localization algorithms but simply do not occur in our estimator formulations. In particular, \textit{major advantages} arise from the following \textit{absent requirements}:
\begin{enumerate}
\item \label{li:advLOS}
No line-of-sight conditions required, neither from A to B, from observers to A or B, nor between observers.
\item \label{li:advTI}
No time-invariant propagation environment required.
\item \label{li:advEnv}
No knowledge of the environment required.
\item \label{li:advObsPos}
No knowledge of the observer positions required.
\item \label{li:advInteract}
No interaction between A and B required.
\end{enumerate}
A subset of the estimators exhibits the following advantages:
\begin{enumerate}
\setcounter{enumi}{5}
\item \label{li:advSync}
No precise synchronization required between A and B, between observers, or between an observer and A or B.
\item \label{li:advAssoc}
No knowledge of the MPC association required.
\item \label{li:advDir}
No knowledge of the MPC directions required.
\end{enumerate}
%

\subsection{Opportunities and Use Cases}

The advantage \ref{li:advLOS} makes the proposed paradigm \textit{suitable for dense and crowded environments}, where LOS obstruction is typical \cite{WitrisalSPM2016}. The advantages \ref{li:advTI}, \ref{li:advEnv}, \ref{li:advObsPos} make the paradigm \textit{qualified for dynamic settings} with time-variant channels. This is in contrary to fingerprinting \cite{SteinerTSP2010,HePC2016} and estimators based on calibrated models \cite{HeynVTC2019}, which rely on up-to-date training data.

The proposed paradigm enables \textit{interaction-free distance estimation between mobile users} by evaluating their channels only \textit{at the infrastructure} end (e.g., anchor nodes, cellular base stations, WiFi routers). This supersedes the perception \cite{WinPIEEE2018} that inter-node estimates always require communication among the pair. And it provides a new way of acquiring pairwise estimates for \textit{cooperative localization} \cite{WinPIEEE2018,BuehrerPIEEE2018}. The surveillance potential could however prompt ethical questions.

Another interesting use case is \textit{proximity testing} of a mobile (node B) to some \textit{point of interest} (node A), e.g., an access gate.
The CIR $\cirA$ could be a pre-recorded fingerprint or updated periodically. For example, node A could be a stationary ultra-low-complexity beacon whose only task is the periodical transmission of a training sequence. The resultant live updates qualify the approach for time-varying channels.

Magnificently, \textit{mobiles can act as observers} due to advantage \ref{li:advObsPos}.
Thus, pairwise estimates can be obtained without any infrastructure, e.g. for \textit{self-localization of ad-hoc networks} \cite{IhlerJSAC2005}. Thereby, $M$ mobiles allow for $\NObs = M-2$ observers and thus large $K$, which promises high accuracy.



\subsection{Relationship with Other Approaches}
\label{sec:TechRelation}






The proposed paradigm provides novel acquisition methods for inter-node location information for the use in cooperative network localization. They can complement or replace the traditional  RSS- or TOA-based range measurements in the following established Bayesian frameworks: the network localization and navigation (NLN) framework \cite{WinPIEEE2018,ShenJSAC2012,WinCM2011} and the collaborative localization (CL)  framework \cite{BuehrerPIEEE2018}. 
These frameworks could integrate the proposed paradigm in a systematic fashion, for which we identify the following suitable mathematical interfaces (a detailed formal description is relayed to future work). The distance likelihood functions from the paper at hand can be inserted into the inter-node measurements PDF of the NLN framework in \cite[Eq.~(3),(5)]{WinPIEEE2018}. Likewise, the position likelihood functions can be inserted into the associated measurement model for node relative positions \cite[Eq.~(6)]{WinPIEEE2018}. Likewise, the likelihood functions could be inserted into the CL framework measurement model \cite[Eq.~(1)]{BuehrerPIEEE2018}. 
These interfaces constitute a straightforward method for integrating our proposal into belief-propagation algorithms like \cite[Sec.~VI-B]{WinPIEEE2018}, \cite{LeitingerTWC2019,YuPLANS2020,PrietoTSP2016,MeyerIOT2018,MeyerPIEEE2018}, which use Bayesian frameworks like the aforementioned, or into related algorithms \cite{VaghefiCL2015,PatwariSPM2005,IhlerJSAC2005}.

The proposal also provides a novel way of estimating the node velocity $v$. Consider the positions $\posA, \posB$ of the same node at different times $t\AnnotateNodeA , t\AnnotateNodeB$. With $T := t\AnnotateNodeB - t\AnnotateNodeA$ and \Cref{eq:ApdxDistLhf} we can formulate a velocity likelihood function
$L(\Hypo{v},\Hypo{\errClock}) = f(\delayDiff \, | \, \Hypo{d} = \Hypo{v}T, \Hypo{\errClock})$. This allows for velocity estimation analogous to \Cref{sec:EstimateDist} and for integration into Bayesian localization frameworks, where it could complement or replace intra-node measurements of inertia or Doppler \cite{BuehrerPIEEE2018}.



Our proposal has no direct implications for localization systems that rely exclusively on estimates to far-away anchors.

The proposal has the potential to supplement or replace machine-learning-based technology for wireless location fingerprinting as presented in \cite{SteinerTSP2010,VieiraPIMRC2017,LiSENS2021}. 

\subsection{Comparison to RSS and TOA Ranging Accuracy}

\begin{figure}[!ht]
\vspace{-2mm}
\centering
\resizebox{!}{34mm}{\begin{tikzpicture}
\begin{axis}[%
width=93mm,
height=38mm,
at={(0,0)},
scale only axis,
xmin=0,
xmax=7,
label style={font=\normalsize},
xlabel={shadowing standard deviation $\sigma_\mathrm{sh}$ (RSS scheme) [dB]},
ymin=0,
ymax=16,
ytick={ 0,  2,  4,  6,  8, 10, 12, 14, 16, 18, 20},
ylabel={\begin{minipage}[c]{40mm}\centering required $K$\\to beat RSS scheme\end{minipage}},
ylabel style={at={(-.06,.5)}},
xlabel style={at={(.5,-.095)}},
axis background/.style={fill=white},
xmajorgrids,
ymajorgrids,
legend style={at={(.96,1)}, anchor=north east, legend cell align=left, align=left, draw=white!15!black}
]

\addlegendimage{empty legend}
\addlegendentry{\hspace{-6.5mm}RSS path-loss exponent:}

\addplot [color=red, line width=1.5pt]
  table[row sep=crcr]{%
7.77135338803061	3\\
7.76643990636829	4\\
5.79195826969964	4\\
5.78922855778634	5\\
4.64368087676474	5\\
4.64192606199815	6\\
3.88506221742782	6\\
3.88383384710437	7\\
3.34365549433864	7\\
3.34274559040124	8\\
2.93671763837909	8\\
2.93601571248750	9\\
2.61917346845692	9\\
2.61861511831751	10\\
2.36422722846969	10\\
2.36377227650520	11\\
2.15489302847824	11\\
2.15451506838526	12\\
1.97985752190304	12\\
1.97953846468207	13\\
1.83128281317128	13\\
1.83100984199360	14\\
1.70356122677246	14\\
1.70332500171503	15\\
1.59257162639925	15\\
1.59236517760969	16\\
1.49521570327307	16\\
1.49503372248828	17\\
1.40911817934218	17\\
1.40895655167155	18\\
1.33242691087255	18\\
1.33228239672004	19\\
1.26367600424105	19\\
1.26354601796635	20\\
1.20168987541188	20\\
1.20157232801517	21\\
1.14551461417646	21\\
1.14540779936816	22\\
1.09436798024677	22\\
1.09427049054080	23\\
1.04760237536620	23\\
1.04751303934475	24\\
1.00467701931932	24\\
1.00459485408221	25\\
0.965136762826658	25\\
0.965060937499828	26\\
0.928595758334052	26\\
0.928525565745795	27\\
0.894724735294831	27\\
0.894659569761973	28\\
0.863240983453058	28\\
0.863180323191613	29\\
0.833900394011748	29\\
0.833843787085749	30\\
0.806491081221271	30\\
0.806438134225837	31\\
};
\addlegendentry{$\alpha = 4$}

\addplot [color=white!60!black, line width=1pt]
  table[row sep=crcr]{%
9.21738588851489	2\\
9.20817310901612	3\\
5.82851504102296	3\\
5.82482992977622	4\\
4.34396870227473	4\\
4.34192141833976	5\\
3.48276065757356	5\\
3.48144454649862	6\\
2.91379666307086	6\\
2.91287538532828	7\\
2.50774162075398	7\\
2.50705919280093	8\\
2.20253822878432	8\\
2.20201178436562	9\\
1.96438010134269	9\\
1.96396133873813	10\\
1.77317042135226	10\\
1.77282920737890	11\\
1.61616977135868	11\\
1.61588630128895	12\\
1.48489314142728	12\\
1.48465384851156	13\\
1.37346210987846	13\\
1.37325738149520	14\\
1.27767092007935	14\\
1.27749375128627	15\\
1.19442871979944	15\\
1.19427388320727	16\\
1.12141177745480	16\\
1.12127529186621	17\\
1.05683863450664	17\\
1.05671741375366	18\\
0.999320183154416	18\\
0.999211797540032	19\\
0.947757003180790	19\\
0.947659513474759	20\\
0.901267406558913	20\\
0.901179246011376	21\\
0.859135960632344	21\\
0.859055849526121	22\\
0.820775985185078	22\\
0.820702867905603	23\\
0.785701781524649	23\\
0.785634779508559	24\\
0.753507764489489	24\\
0.753446140561656	25\\
0.723852572119993	25\\
0.723795703124871	26\\
0.696446818750539	26\\
0.696394174309346	27\\
0.671043551471123	27\\
0.67099467732148	28\\
0.647430737589793	28\\
0.647385242393709	29\\
0.625425295508811	29\\
0.625382840314312	30\\
0.604868310915953	30\\
0.604828600669378	31\\
};
\addlegendentry{$\alpha = 3$}

\addplot [color=black, line width=1.5pt]
  table[row sep=crcr]{%
7.00000000000000	2\\
6.14492392567659	2\\
6.13878207267742	3\\
3.8856766940153	    3\\
3.88321995318415	4\\
2.89597913484982	4\\
2.89461427889317	5\\
2.32184043838237	5\\
2.32096303099908	6\\
1.94253110871391	6\\
1.94191692355218	7\\
1.67182774716932	7\\
1.67137279520062	8\\
1.46835881918955	8\\
1.46800785624375	9\\
1.30958673422846	9\\
1.30930755915876	10\\
1.18211361423484	10\\
1.18188613825260	11\\
1.07744651423912	11\\
1.07725753419263	12\\
0.98992876095152	12\\
0.989769232341037	13\\
0.915641406585640	13\\
0.915504920996800	14\\
0.851780613386231	14\\
0.851662500857515	15\\
0.796285813199626	15\\
0.796182588804844	16\\
0.747607851636533	16\\
0.747516861244141	17\\
0.704559089671090	17\\
0.704478275835776	18\\
0.666213455436277	18\\
0.666141198360021	19\\
0.631838002120527	19\\
0.631773008983173	20\\
};
\addlegendentry{$\alpha = 2$}

\end{axis}
\end{tikzpicture}%
}
\vspace{-2mm}
\caption{Minimum number of MPCs $K$ such that the proposed delay-difference-based distance MVUE \Cref{eq:distMVUE} has lower RMSE than RSS-based distance estimation between A and B in log-normal shadowing.}
\label{eq:DistEstimation_ProposedVersusRSS}
\end{figure}
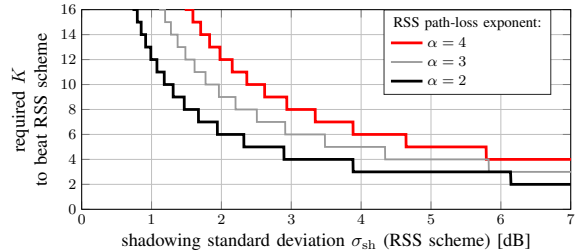

First, we address the question of whether the novel $\hat d\MVUE\AssAsyn$ from \Cref{eq:distMVUE} can beat the accuracy of RSS-based distance estimation between A and B. We conduct a comparison based on analytical RMSE expressions: on the one hand
$d\,(\f{2}{(K-1)(K+2)})^{0.5}$
from \Cref{eq:stdMVUE}
and on the other hand
$\f{\log(10)}{10\,\alpha} \sigma_\mathrm{sh} \, d$
from the CRLB on the RSS-based distance RMSE \cite{GeziciSPM2005}. This CRLB assumes a log-normal shadowing model with standard deviation $\sigma_\mathrm{sh}$ in $\mathrm{dB}$ and path-loss exponent $\alpha$. We find that the asynchronous $\hat d \MVUE\AssAsyn$ has lower RMSE than the RSS scheme if
$K \geq \lceil \sqrt{x^2 + {\scriptstyle\f{9}{4}}} - \f{1}{2} \rceil$ with
$x = \f{10\sqrt{2}}{\log(10)} \f{\alpha}{\sigma_\mathrm{sh}}$. 
This threshold is plotted in \Cref{eq:DistEstimation_ProposedVersusRSS} as a function of $\sigma_\mathrm{sh}$ and typical values of $\alpha$. We conclude that with large $K$ (e.g., by using many observers), the proposed $\hat d\MVUE\AssAsyn$ outperforms an RSS scheme, especially in dense propagation environments where $\sigma_\mathrm{sh}$ is large.

Likewise, we compare the proposed $\hat\d\LSE\AssViaDiff$ to TOA ranging (with RMSE denoted $\cd \sigma_\mathrm{TOA}$). We recall the position RMSE approximation
$3\cd\sigma / \sqrt{K}$
from \Cref{eq:RMSEdisplLSEViaDiff}, which holds for diverse MPC directions and dominant delay-measurement errors.
We require that $3\cd\sigma / \sqrt{K} < \cd \sigma_\mathrm{TOA}$.
For simplicity we assume $\sigma\AnnotateNodeA_{k,o} = \sigma\AnnotateNodeB_{k,o} = \sigma_\mathrm{TOA}$ $\forall k,o$, resulting in $\cd\sigma = \cd\sigma_\mathrm{TOA}\sqrt{2}$. We obtain
the criterion $K > 18$ for the RMSE of $\hat\d\LSE\AssViaDiff$
(an asynchronous, NLOS-compatible position estimator)
to actually beat the distance RMSE of TOA in LOS.
Having $K = \sum_{o=1}^\NObs K_o > 18$ MPCs is realistic for large $\NObs$.

The accuracy witnessed in \Cref{sec:EvalSim,sec:EvalMeas} beats accounts of practical TOA ranging accuracy \cite{LiJSAC2015,Chen2017,AlaviCL2006}, although the different circumstances make a comparison difficult.


\section{Summary \& Outlook}
\label{sec:Summary}
We proposed a novel paradigm for obtaining pairwise distance or position information between wireless users by comparing their UWB channels to observers. It is applicable in the spatial and temporal domains and opens up exciting new technological opportunities. We derived various estimators and studied their accuracy in theory and  practice.
Open topics for future research are the integration into real-time localization algorithms (e.g, for cooperative network localization),
extensive field trials (e.g., with mmWave massive-MIMO systems),
a detailed comparison to machine learning approaches,
an analytic study of the error caused by inaccurately measured MPC directions,
and a study of suitable schemes for MPC selection and association (e.g., with belief propagation and possibly incorporating the RSS between A and B) in the light of path overlap and selective MPC shadowing.


\begin{appendices}
\crefalias{section}{appendix}



\section{Formal Definition of MPC Terms}
\label[appendix]{apdx:MpcSetTheory}
An MPC (here modeled as object $\mpc$) is characterized by familiar quantities like the path delay $\tau(\mpc)$.
We also consider a propagation path identifier $P(\mpc)$; example values are
$\texttt{"LOS}$ $\texttt{path"}$,
$\texttt{"reflection}$ $\texttt{(ground)"}$,
$\texttt{"reflection}$ $\texttt{(east wall)"}$, or
$\texttt{"scattered}$ $\texttt{path}$ $\texttt{(lamp)"}$.
Relating to an observer $o$, let $\mpcSetAllA_o$ and $\mpcSetAllB_o$ be the sets of MPCs that were resolved in the CIRs $\cirA_o$ and $\cirB_o$, respectively.
\textit{MPC selection} reduces them to subsets
$\mpcSetA_o \subseteq \mpcSetAllA_o$ and
$\mpcSetB_o \subseteq \mpcSetAllB_o$
of equal size
$|\mpcSetA_o| = |\mpcSetB_o| =: K_o$.
We require that the path identifiers are unique within either set. 

Two MPCs
$\mpcA \in \mpcSetA_o$,
$\mpcB \in \mpcSetB_o$
have \textit{matching paths} if $P(\mpcA) = P(\mpcB)$.
An \textit{alien MPC} does not have matching paths with any MPC of the other set. We note that $\mpcSetA_o$ and $\mpcSetB_o$ always hold the same number of alien MPCs.

\textit{MPC association} establishes the same MPC index $k \in \{1,\ldots,K_o\}$ for both sets $\mpcSetA_o , \mpcSetB_o$. The indexed elements
$\mpcA_{k,o} \in \mpcSetA_o$,
$\mpcB_{k,o} \in \mpcSetB_o$
are consequently associated in pairs $( \mpcA_{k,o}\, , \mpcB_{k,o} \,)$.
A \textit{correct association} exhibits matching paths for all pairs $k,o$.
A pairing of MPCs with mismatching paths we call an \textit{association error}, unless both MPCs are alien.

\section{Proof of Propagation-Geometric Properties}
\begin{figure}[!ht]
\vspace{-5mm}
\centering
\subfloat[distances]{
\includegraphics[height=32mm,trim=0 -9 0 0]{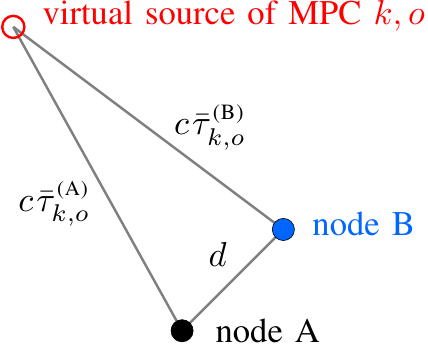}
\label{fig:ApdxTriangleDistances}}
\hspace{3mm}
\subfloat[positions and vectors]{
\includegraphics[height=32mm,trim=0 -9 0 0]{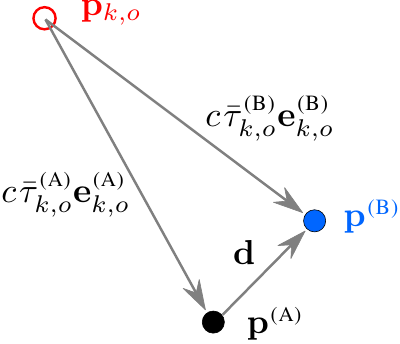}
\label{fig:ApdxTriangleVectors}}
\caption{Triangle formed by the positions of node A, node B, and the virtual source representation of the $k$-th MPC associated with observer $o$.}
\label{fig:ApdxTriangle}
\end{figure}

\label[appendix]{apdx:PropagationGeometry}
This appendix derives the geometric properties \Cref{eq:DelayDiffBounds,eq:VectorEquality,eq:ProjectionEquality} which hold for each MPC $k,o$. Without loss of generality we assume that the observer is transmitting. We consider a specific MPC $k,o$ in terms of its virtual source \cite{Borish1984,LeitingerJSAC2015,WitrisalSPM2016}. The virtual source position is denoted ${\bf p}_{k,o} \in \bbR^3$.
The model is shown in \Cref{fig:ApdxTriangle}. It is an equivalent description of the MPC delays $\delayTrueA[k,o], \delayTrueB[k,o]$ and directions of arrival $\dirVectA[k,o] , \dirVectB[k,o]$.

The delay-difference bounds 
in \Cref{eq:DelayDiffBounds} are obtained by applying the triangle inequality to \Cref{fig:ApdxTriangleDistances} twice in order to obtain
$\cd\delayTrueA[k,o] \leq \cd\delayTrueB[k,o] + d$
and
$\cd\delayTrueB[k,o] \leq \cd\delayTrueA[k,o] + d$.
With simple rearrangements we obtain the lower bound
$-d \leq c(\delayTrueB[k,o] - \delayTrueA[k,o])$
and the upper bound
$c(\delayTrueB[k,o] - \delayTrueA[k,o]) \leq d$, respectively.

The vector equality
$\d = \cd\delayTrueB[k,o] \dirVectB[k,o] - \cd\delayTrueA[k,o] \dirVectA[k,o]$
in \Cref{eq:VectorEquality} follows from
$\d = \posB - \posA$
with
$\posB = {\bf p}_{k,o} + \cd\delayTrueB[k,o] \dirVectB[k,o]$ and
$\posA = {\bf p}_{k,o} + \cd\delayTrueA[k,o] \dirVectA[k,o]$,
which are deduced from \Cref{fig:ApdxTriangleVectors}.


To prove the projection equality \Cref{eq:ProjectionEquality} we consider the above vector equality. On both sides we form the inner product with $\dirVectA[k,o]$ to obtain
$\d\Tr \dirVectA[k,o]
= \cd\delayTrueB[k,o] a - \cd\delayTrueA[k,o]$
with
$a := ( \dirVectA[k,o] )\Tr \dirVectB[k,o]$.
Likewise,
${\bf d}\Tr \dirVectB[k,o]
= \cd\delayTrueB[k,o] - \cd\delayTrueA[k,o] a$.
The sum of the equations is
$\d\Tr ( \dirVectA[k,o] \! + \dirVectB[k,o] )$
$= c( \delayTrueB[k,o] \! - \delayTrueA[k,o] ) ( 1 + a )$
$= c \delayDiffTrue[k,o] ( 1 + a )$.
\qed

\section{Derivation of Distance Estimators for Known MPC Association}
\label[appendix]{apdx:DeriveEstDistanceWithAssoc}

\subsubsection{General Case MLE}
\label[appendix]{apdx:DeriveEstDistanceWithAssoc_GeneralMLE}
We derive the distance MLE rule \Cref{eq:distMLE_general}.
We recall that the true delay differences have uniform distribution $\delayDiffTrue[k o] \overset{\text{i.i.d.}}{\sim} \mathcal{U}(-d/c,d/c)$ under the employed assumptions.
Thus the PDF
$f_{\delayDiffTrue[k o]|d}(\delayDiffTrue[k o]|d) = \f{c}{2d} \IndFunc_{[-d/c,d/c]}(\delayDiffTrue[k o])$.
We consider $\delayDiff[k o] = \delayDiffTrue[k o] + \errMeas[k o] + \errClock$ where
$\errClock$ is non-random and $\delayDiffTrue[k o], \errMeas[k o]$ are statistically independent. The PDF of $\delayDiff[k o]$ is thus given by the convolution of PDFs
\begin{align}
& f(\delayDiff[k o]\,|\,d,\errClock) 
=
\int_\mathbb{R} f_{\errMeas[k o]}(z) \,
f_{\delayDiffTrue[k o] | d}(\delayDiff[k o] \!-\! \errClock \!-\! z \, | \,d) \, dz \\
&=
\int_\mathbb{R} f_{\errMeas[k o]}(z) \,
\f{c}{2d}\,\IndFunc_{[-d/c,d/c]}(\delayDiff[k o] \!-\! \errClock \!-\! z) \, dz \\
&=
\f{c}{2d} \int_{\delayDiff[k o] - \errClock - d/c}^{\delayDiff[k o] - \errClock + d/c} 
f_{\errMeas[k o]}(z) \, dz
=
\f{c}{2d} \, I_{k o}(\delayDiff[k o]\! - \errClock ,d)
\label{eq:ApdxDistLhfPrep}
\end{align}
with the definition of $I_{k o}$ in \Cref{eq:SoftIndicFunc}. The $\delayDiff[k o]$ are statistically independent for different $k,o$ and their joint PDF is thus the product of PDFs \Cref{eq:ApdxDistLhfPrep}.
We replace the true values $d, \errClock$ with free variables $\Hypo{d},\Hypo{\errClock}$ to express the likelihood function
\begin{align}
L\big( \Hypo{d} , \Hypo{\errClock} \big)
&=
f\big( \delayDiff[1,1] \, , \, \delayDiff[2,1] \ \ldots \ ,\delayDiff[K_\NObs,\NObs] \,|\, d=\Hypo{d},\errClock=\Hypo{\errClock}\, \big) \nonumber\\
&=
\bigg( \f{c}{2} \bigg)^{\!K} \Hypo{d}^{-K} \prod_{o=1}^\NObs \prod_{k=1}^{K_o}
I_{k o} \big( \delayDiff[k o]\! - \Hypo{\errClock}, \Hypo{d} \,\big)
. \label{eq:ApdxDistLhf}
\end{align}
Any value pair $(\Hypo{d},\Hypo{\errClock})$ that maximizes $L(\Hypo{d},\Hypo{\errClock})$ is an MLE. We discard the constant prefactor $(\f{c}{2})^K$ and obtain \Cref{eq:distMLE_general}.

\subsubsection{No-Measurement-Error Case MLE}
\label[appendix]{apdx:DeriveEstDistanceWithAssoc_NoErrorMLE}
Here $\errMeas[k o] \equiv 0 \ \Rightarrow \ f_{\errMeas[k o]}(x) = \delta(x)$ and thus the soft indicator function in \Cref{eq:ApdxDistLhfPrep} becomes the actual indicator function
$I_{k o}(\delayDiff[k o]\! - \errClock ,d)
= \IndFunc_{[-d/c,d/c]}(\delayDiff[k o]\! - \errClock)$.
We use it in \Cref{eq:ApdxDistLhf} and note that the likelihood scales like
$L \propto \Hypo{d}^{-K}$ on the LHF support $\mathrm{supp}(L)$. The MLE is thus given by the pair $(\Hypo{d},\Hypo{\errClock}) \in \mathrm{supp}(L)$ with minimum $\Hypo{d}$. We note that $(\Hypo{d},\Hypo{\errClock}) \in \mathrm{supp}(L)$ iff $\Hypo{\errClock}-\Hypo{d}/c \leq \delayDiff[k o] \leq \Hypo{\errClock}+\Hypo{d}/c \ \forall k,o$ and find the MLE by requiring
$\Hypo{\errClock}+\Hypo{d}/c = \max_{k o} \delayDiff[k o]$ as well as
$\Hypo{\errClock}-\Hypo{d}/c = \min_{k o} \delayDiff[k o]$.
The sum of those equations yields the $\hat\errClock\MLE\AssAsyn$ formula in \Cref{eq:EpsMVUE},
the difference yields $\hat d\MLE\AssAsyn$ in \Cref{eq:distMLE}.

\subsubsection{MLE Bias and MVUE Property}
\label[appendix]{apdx:DeriveEstDistanceWithAssoc_NoErrorMVUE}
We use an index $i \in \{1,\ldots,K\}$ for the MPCs across all observers.
The uniform distribution
$\delayDiff[i] \iid \calU(-\f{d}{c}+\errClock,\f{d}{c}+\errClock)$
follows from  the assumptions in \Cref{sec:EstimateDist} with $\errMeas[i] \equiv 0$.
The problem of estimating $d,\errClock$ (or just $d$ given $\errClock$) from all $\delayDiff[i]$ is equivalent to the problem of estimating the parameters of a uniform distribution from iid samples. Hence, the MLE results \eqref{eq:distMLE},\eqref{eq:distMLESync},\eqref{eq:EpsMVUE}  and MVUE results \eqref{eq:distMVUE},\eqref{eq:EpsMVUE},\eqref{eq:distMVUESync} follow directly from respective statements in the mathematical literature, e.g. \cite[Cpt.~8]{David1970}.

Since these estimators are central to the paper, we provide more detail. 
Let
$x_i := \f{c|\delayDiff[i]-\errClock|}{d}$
and
$y_i := \f{c(\delayDiff[i]-\errClock)}{2d} + \f{1}{2}$;
they fulfill $x_i \sim \calU(0,1)$, $y_i \sim \calU(0,1)$. With their order statistics $x_{(i)}$, $y_{(i)}$ we can express 
$\hat d\MLE\AssSync = x_{(K)} \cdot d$
and
$\hat d\MLE\AssAsyn
= ( y_{(K)} - y_{(1)} ) \cdot d$.
From \cite[Eq.~(1.146)]{Gentle2009} we find that $x_{(i)}, y_{(i)} \sim \mathrm{Beta}(i,K-i+1)$ with
$\EVSymb[x_{(i)}] = \EVSymb[y_{(i)}] = \f{i}{K+1}$.
We obtain
$\EVSymb[\hat d\MLE\AssSync] = \f{K}{K+1} d$
and
$\EVSymb[\hat d\MLE\AssAsyn] = \f{K-1}{K+1} d$,  i.e. the estimators are biased, however
$\hat d\MVUE\AssSync = \f{K+1}{K} \hat d\MLE\AssSync$
and
$\hat d\MVUE\AssAsyn = \f{K+1}{K-1} \hat d\MLE\AssAsyn$
are unbiased.
Their minimum-variance property is due to the Lehmann–Scheffé theorem \cite{Casella2002} as the determining $x_{(K)}$ and $y_{(K)},y_{(1)}$ are complete sufficient statistics of the respective uniform distribution parameters.

\section{Proof: Distance MLE, Unknown MPC Association}
\label[appendix]{apdx:DeriveEstDistanceNoAssoc}
\subsubsection{Derivation of MLE rule \Cref{eq:distMLENoAss_} for $\NObs=1$}
We consider only one specific observer $o$ and discard the index $o$.
We recall $\delayMeasB[k]\! = \delayMeasA[k]\! + \delayDiff[k]$ from \Cref{eq:SignalModel},
$\delayDiff[k] \iid \calU(-d/c + \errClock,d/c + \errClock)$ from \Cref{sec:EstimateDist}, and
$f(\delayMeasB[k] | \delayMeasA[k]\! , d , \errClock) =
\f{c}{2d} \, I_k( \delayMeasB[k]\! - \delayMeasA[k]\! - \errClock, d )$
with
$I_k(x, d)
= F_{\errMeas[k]}(x + d/c) - F_{\errMeas[k]}(x - d/c)$
from \Cref{eq:SoftIndicFunc,eq:ApdxDistLhfPrep}.
We introduce the shorthand notation
$f(\delayMeasB[k] | \delayMeasA[k]\! , d, \errClock) = g_k(\delayMeasB[k])$
with
$g_k(\bullet) := \f{c}{2d} \, I_k( \bullet - \delayMeasA[k]\! - \errClock, d )$.

The observed delays $\delayMeasA[k]$ are considered as non-random parameters. Without loss of generality we assume that the MPC indexation is such that $\delayMeasA[1] \leq \ldots \leq \delayMeasA[K]$. The corresponding delays $\delayMeasB[k]$ may or may not have the same order; this circumstance is unknown and unobserved. Given the non-random parameters, the random variables $\delayMeasB[1] \ldots \delayMeasB[K]$ are statistically independent but have non-identical distributions (PDFs $g_k$). As random observations we consider the order statistics $\delayMeasB[(k)]$ which are observable and fulfill $\delayMeasB[(1)] \leq \ldots \leq \delayMeasB[(K)]$ by definition.
According to \cite[Eq.~(6)]{VaughanVenables1972}, the joint PDF of the order statistics $\delayMeasB[(k)]$ is given by the permanent 
of a $K \times K$ matrix $(${\bf A}$)_{k,l} = g_k(\delayMeasB[(l)])$.
By a property of matrix permanents \cite{VaughanVenables1972} we obtain a sum over all length-$K$ permutations $\pi(\bullet)$,
\begin{multline}
f\big(\delayMeasB[(1)] , \ldots , \delayMeasB[(K)] \, | \, \delayMeasA[1]\! , \ldots , \delayMeasA[K]\! , d, \errClock \big) \\[-1.5mm]
= \mathrm{perm}({\bf A}) 
= \sum_{\perm \in \permSet[K]} \prod_{k=1}^K g_k(\delayMeasB[(\pi(k))])
\, . \label{eq:ApdxOrderStatsJointPDF}
\end{multline}
%
This sum does not depend on the actual order or the observed order of $\delayMeasB[1], \ldots , \delayMeasB[K]$. Thus $\delayMeasB[(\pi(k))]$ can be replaced by $\delayMeasB[\pi(k)]$.
By fixing all $\tau$-values in \Cref{eq:ApdxOrderStatsJointPDF} upon observation and replacing the true values $d,\errClock$ with variables $\Hypo{d},\Hypo{\errClock}$ we obtain the LHF
\begin{align}
L(\Hypo{d},\Hypo{\errClock})
= \left( \f{c}{2\Hypo{d}} \right)^K \sum_{\perm \in \permSet[K]} \prod_{k=1}^K
I_k(\delayMeasB[\pi(k)] - \delayMeasA[k]\! - \Hypo{\errClock}, \Hypo{d} ) \, .
\label{eq:ApdxLhfPermSum}
\end{align}
Furthermore discarding the constant factor $(\f{c}{2})^K$ yields the MLE rule \Cref{eq:distMLENoAss_} for the case $\NObs=1$. \qed

Finally, we note that the resulting $\hat d\MLE$ is symmetric and $\hat\errClock\MLE$ is antisymmetric in the arguments $\delayMeasB[k], \delayMeasA[k]$ (if $f_{\errMeas[k]}$ is symmetric).
This is a desirable property because the A/B labeling is arbitrary. It justifies our seemingly arbitrary choice of considering only the distribution of $\delayMeasB[k]$ but not of $\delayMeasA[k]$. 

\subsubsection{Extension to multiple observers}
The LHF is now the product
$L(\Hypo{d},\Hypo{\errClock}) = \prod_{o=1}^\NObs L_o(\Hypo{d},\Hypo{\errClock})$
of terms \Cref{eq:ApdxLhfPermSum} written as $L_o$. This is due to the statistical independence of the order statistics
$\delayMeasB[(1),o] \, , \ldots , \delayMeasB[(K_o),o]$
between different observers $o$, as a result of the known MPC-observer association (as argued in \Cref{sec:SystemModel}).
\qed

\subsubsection{MLE candidates for the case $\errMeas[k o] \equiv 0$}
\label[appendix]{app:lhfEvalNoAssDerviation}

We consider only one specific observer $o$ and discard the index. Analogous to \Cref{apdx:DeriveEstDistanceWithAssoc_NoErrorMLE},
$\errMeas[k] \equiv 0$
transforms the LHF from \Cref{eq:distMLENoAss_} into
$(\frac{c}{2\Hypo{d}})^K
\sum_{\perm \in \permSet[K]}
\prod_{k=1}^{K}
\IndFunc_{[-\Hypo{d}/c,\Hypo{d}/c]}( \delayMeasB[\perm(k)] - \delayMeasA[k] - \Hypo{\errClock})$.
This LHF is a superposition of wedges (see \cref{fig:LHF_Asyn_NoAssoc}), each determined by the inequalities
$\errClock \leq d/c + S_\perm$
and
$\errClock \geq - d/c + L_\perm$
with
$L_\perm = \max_k \delayMeasB[\perm(k)] - \delayMeasA[k]$
and
$S_\perm = \min_k \delayMeasB[\perm(k)] - \delayMeasA[k]$.
The points of interest for evaluation are then the peaks of the wedges, located at 
$(\f{c}{2}(L_\perm - S_\perm), \f{1}{2}(L_\perm + S_\perm))$, 
and intersections between the wedges-borders located either at 
$(\f{c}{2}(L_{\perm_1} - S_{\perm_2}), \f{1}{2}(L_{\perm_1} + S_{\perm_2}) )$
or at
$(\f{1}{2}(L_{\perm_2} - S_{\perm_1}), \f{1}{2}(L_{\perm_2} + S_{\perm_1}) )$
if they exist.
Thus we can write all candidate points as
$(\Hypo{d},\Hypo{\errClock}) =
(\f{c}{2}(L - S), \f{1}{2}(L + S))$
with
$(L,S) \in \calL \times \calS$
and
$\calL= \bigcup_{\perm \in \permSet[K]} \max_k \delayMeasB[\perm(k)] -\delayMeasA[k]$
and $\calS$ equivalently. 
The proof generalizes to the multi-observer case by forming the union of candidate points of each observer.

\section{Derivation of Relative Position Estimators}
\label[appendix]{apdx:DeriveRelLocMLE}
\subsubsection{General MLE from $\delayDiff[k o]$}
\label[appendix]{apdx:DeriveRelLocMLE_GeneralMLE}
Using standard tools from estimation theory \cite{Kay1993} we derive the joint MLE $(\hat \d \MLE\AssAsyn, \hat \errClock \MLE\AssAsyn)$ in \Cref{eq:displMLEViaDiff}, based on observed delay differences
$\delayDiff \in \bbR^K$
with
$\delayDiff
= \delayDiffTrue + \mathbf{1}\errClock + \errMeas
= \f{1}{c} \E\Tr \boldsymbol\theta + \errMeas$ cf. \Cref{eq:SignalModel} and
\Cref{eq:DelayShiftStackVector,eq:ErrorStackVector,eq:EMatrix,eq:sVectorForEMatrix}.
Here $\boldsymbol\theta := [\d\Tr,\cd\errClock]\Tr \in \bbR^4$ is the estimation parameter.
The only randomness is constituted by the error vector with PDF $f_{\errMeas}(\errMeas)$.
Now
$\errMeas = \delayDiff - \f{1}{c} \E\Tr \boldsymbol\theta$
implies
$f(\delayDiff | \boldsymbol\theta)
= f_{\errMeas}(\delayDiff - \f{1}{c} \E\Tr \boldsymbol\theta)$.
Fixing the observed $\delayDiff$ and replacing the true value $\boldsymbol\theta$ with a free variable $\Hypo{\boldsymbol\theta}$ yields the likelihood function 
$L(\Hypo{\boldsymbol\theta})
= f_{\errMeas}(\delayDiff - \f{1}{c} \E\Tr \Hypo{\boldsymbol\theta})$.
Any $\hat{\boldsymbol\theta} \in \argmax L(\Hypo{\boldsymbol\theta})$ is an MLE.
\qed

\subsubsection{Approximate RMSE of LSE from $\delayDiff[k o]$}
\label[appendix]{apdx:DeriveApproxRMSE}
$(\E\E\Tr)^{-1} \E\, (\cd\delayDiff)$ from \Cref{eq:displLSEAssViaDiff} exhibits a random estimation error $(\E\E{}\Tr)^{-1} \E\, (\cd\errMeas)$ for $K \geq 4$ under the employed assumptions ($\errMeas$ is random, $\E$ is not).
The estimation error has the covariance matrix
${\bf C} = c^2 (\E\E\Tr)^{-1} \E\,\EVSymb[\errMeas \errMeas\Tr] \E\Tr (\E\E\Tr)^{-1}$.
The assumption
$\errMeas \sim \mathcal{N}({\bf 0}, \sigma^2 \eye_K)$
gives
$\EVSymb[\errMeas \errMeas\Tr] = \sigma^2 \eye_K$
and we obtain
${\bf C} = (c\sigma)^2 (\E\E\Tr)^{-1} \E\E\Tr (\E\E\Tr)^{-1}
= (c\sigma)^2 (\E\E\Tr)^{-1}$.
We use the plane-wave assumption ${\bf s}_{ko} \approx \dirVectA[k o]$. 
For $\dirVectA[k o]$ with i.i.d. uniform distributions on the 3D unit sphere, $\f{1}{K} \E\E\Tr \in \bbR^{4 \times 4}$ converges to
$\diag(\f{1}{3},\f{1}{3},\f{1}{3},1)$ for $K \rightarrow \infty$  \cite[Lemma 4.3]{Dumphart2020}.
Thus ${\bf C} \approx \f{(c\sigma)^2}{K} \diag(3,3,3,1)$ with
$\EVSymb[ | \hat\errClock\LSE\AssViaDiff \! - \errClock|^2 ]
= C_{4,4} \approx \f{\sigma^2}{K}$
and
$\EVSymb[ \| \hat\d\LSE\AssViaDiff \! - \d\|^2 ]
= \sum_{i=1}^3 C_{i,i}
\approx \f{(3c\sigma)^2}{K}$.
\qed

\subsubsection{LSE from $\delayMeasA[k o], \delayMeasB[k o]$}
\label[appendix]{apdx:DeriveRelLocLSE_FromTau}

We recall $\d
= \cd\delayTrueB[k o] \dirVectB[k o] - \cd\delayTrueA[k o] \dirVectA[k o]$
from \Cref{eq:VectorEquality} and reformulate to
$\d
= c(\delayMeasB[k o] \!- \errClockA[o] \!- \errClock \!- \errMeasB[k o]) \dirVectB[k o] - 
c(\delayMeasA[k o] \!- \errClockA[o] \!- \errMeasA[k o]) \dirVectA[k o]$
with \Cref{eq:SignalModelTauA,eq:SignalModelTauB,eq:QuickChannelEstimationAssumption}.
We express this as a linear equation in terms of the unknown parameters,
$[ \eye_3, \dirVectB[k o]\!, \dirVectB[k o]\!-\dirVectA[k o] ]\Tr \cdot
[\,\d\Tr, \cd\errClock, \cd\errClockA[o] ]\Tr
=
  \cd\delayMeasB[k o]\dirVectB[k o]
- \cd\delayMeasA[k o]\dirVectA[k o]
+ \cd\errMeasA[k o]  \dirVectA[k o]
- \cd\errMeasB[k o]  \dirVectB[k o]$.
In the least-squares sense we discard the random error terms (the two rightmost summands) and require that the equation holds for all MPCs $k,o$. We obtain
${\bf G} \cdot [\,\d\Tr, \cd\errClock, \cd\errClockA[1],\ldots, \cd\errClockA[\NObs]]\Tr = {\bf t}$ with ${\bf G}$ and ${\bf t}$ from \Cref{eq:EstimateRelLocViaTauDetails}. 
The Moore-Penrose inverse $({\bf G}\Tr{\bf G})^{-1} {\bf G}\Tr$ yields the LSE.
\qed

\section{Estimators for the Fully Asynchronous Case}
\label[appendix]{apdx:EstimatorsAsyncObs}
\newcommand{\obsOnesVectorBlockMatrix}{\bm{\mathcal{O}}}
This appendix states the estimators adapted to the case that assumption \Cref{eq:QuickChannelEstimationAssumption} does not hold, e.g., when clock drift is significant over the duration of acquiring different CIR measurements. Then 
the inter-node clock offset $\errClock$ is replaced by individual offsets $\errClock_o$ for the different observers $o = 1 , \ldots , \NObs$.

The distance MLE for unknown MPC directions, previously the two-dimensional problem \Cref{eq:distMLE_general}, is now given by
$( \hat d \MLE\AssAsyn , \hat{\errClockO}\MLE\AssAsyn )
= \argmax \,\f{1}{\Hypo{d}^K} \prod_{o=1}^{\NObs} \prod_{k=1}^{K_o}
I_{k,o}( \delayDiff[k,o] - \Hypo{\errClock}_o, \Hypo{d} )$,
$\Hypo{d} \in \bbR_+$,
$\Hypo{\errClockO} \in \bbR^\NObs$.
For the zero-measurement-error case, cf. \Cref{eq:distMLE}, 
$\hat d\MLE =
\f{c}{2} \cdot \max_{o} ( \max_{k} \delayDiff[k,o] - \min_{k} \delayDiff[k,o] )$
applies.
Now any clock offset in the interval
$\hat \errClock\MLE_o \in [
-\tfrac{1}{c} \hat d\MLE \! + \max_{k} \delayDiff[k,o] \, , \,
 \tfrac{1}{c} \hat d\MLE \! + \min_{k} \delayDiff[k,o] ]$
has maximum likelihood; the interval midpoint is
$\f{1}{2}( \max_{k} \delayDiff[k,o] + \min_{k} \delayDiff[k,o] )$. The $o$ with the largest spread
$\max_{k} \delayDiff[k,o] - \min_{k} \delayDiff[k,o]$ determines $\hat d\MLE$ and its $\hat\errClock\MLE_o$ interval reduces to the stated midpoint value.

Now $\hat\d\MLE\AssViaDiff$, previously the four-dimensional problem \Cref{eq:displMLEViaDiff}, becomes a $(3+\NObs)$-dimensional problem:
$( \hat\d\MLE\AssViaDiff , \hat{\errClockO}\MLE\AssViaDiff )
\in \argmax
f_{\errMeas}( \, \delayDiff - \tfrac{1}{c} \tilde\E{}\Tr [ \Hypo{\d}{}\Tr , \cd\Hypo{\errClockO}{}\Tr ]{}\Tr )$ with
$\Hypo{\d} \in \bbR^3$,
$\Hypo{\errClockO} \in \bbR^\NObs$.
Here $\tilde\E{}\Tr = [[{\bf s}_{1,1} \,\ldots\, {\bf s}_{K_\NObs, \NObs}]{}\Tr, \obsOnesVectorBlockMatrix] \in \bbR^{K \times (3+\NObs)}$
and
$\obsOnesVectorBlockMatrix \in \bbR^{K \times \NObs}$ is a block-diagonal matrix of the all-ones vector blocks ${\bf 1}_{K_1 \times 1}, \ldots , {\bf 1}_{K_\NObs \times 1}$.
The Gaussian-error-case MLE is
$\big( \tilde\E\,\boldsymbol\Sigma^{-1} \tilde\E\Tr \big)^{-1} \tilde\E \, \boldsymbol\Sigma^{-1} (\cd\delayDiff - c\boldsymbol\mu)$
and the LSE is
$( \tilde\E \tilde\E\Tr )^{-1} \tilde\E (\cd\delayDiff)$,
analogous to \Cref{eq:displLSEAssViaDiff}.

To adapt $\hat\d\LSE\AssViaTau$ in \Cref{eq:EstimateRelLocViaTau,eq:EstimateRelLocViaTauDetails}, expand $\bf G$ from size $(3K)\times(4+\NObs)$ to $(3K)\times(3+2\NObs)$ by separating the entries $\dirVectB[k o]$ of the fourth column into $\NObs$ separate columns for $o = 1,\ldots,\NObs$.

\end{appendices}
\section*{Acknowledgment}
We would like to thank Florian Tr\"osch of Schindler Aufz\"uge AG for innovative ideas,
the group of Klaus Witrisal at Graz University of Technology for providing the ray tracer used in \cite{LeitingerJSAC2015},
Erik Leitinger for valuable discussions,
Malte G\"oller for help with initial simulations, 
Ninad Chitnis for help with the measurement acquisition and processing, and the reviewers for valuable feedback that improved the paper.


\bibliographystyle{IEEEtran}
\bibliography{IEEEabrv,ref}


\ifdefined\ForIEEE
\input{biographies}
\vfill
\fi

\flushend
	
\pdfinfo{
/Author(Gregor Dumphart, Robin Kramer, Robert Heyn, Marc Kuhn, Armin Wittneben)
/Title(\OurPaperTitle)
/Subject(Wireless Ranging and Localization)
/Keywords(ultra-wideband ranging, distance estimation, indoor localization)
}

\end{document}